\documentclass{aa}  
\usepackage{graphicx}
\usepackage{txfonts}
\usepackage{gensymb}
\usepackage{natbib,twoopt}
\usepackage[breaklinks=true]{hyperref} 
\newcommand{\kms}{ km\,s$^{-1}$}
\newcommand{\teff}{T_\mathrm{eff}}
\newcommand{\vsini}{v \sin i}

\begin{document}

\title{Starspot activity of HD 199178}
\subtitle{Doppler images from 1994--2017
 \thanks{Based on observations made with the Nordic Optical Telescope,
    operated by the Nordic Optical Telescope Scientific Association at
    the Observatorio del Roque de los Muchachos, La Palma, Spain, of
    the Instituto de Astrofisica de Canarias.}}

\author{T. Hackman \inst{1}
\and I. Ilyin \inst{2}
\and J.J. Lehtinen \inst{3,} \inst{4}
\and O. Kochukhov \inst{5}
\and M.J. K{\"a}pyl{\"a} \inst{3,} \inst{4}
\and N. Piskunov \inst{5}
\and T. Willamo \inst{1}
}

\institute{Department of Physics, P.O. Box 64, FI-00014 University of Helsinki, 
Finland
 \and 
Leibniz Institute for Astrophysics Potsdam (AIP), An der Sternwarte 16, 
D-14482 Potsdam, Germany
\and Max Planck Institute for Solar System Research, Justus-von-Liebig-Weg 3, 
D-37077 G\"ottingen, Germany
\and
ReSoLVE Centre of Excellence, Department of Computer Science, 
Aalto University, PO Box 15400, 
FI-00076 Aalto, Finland
\and
Department of Physics and Astronomy, Uppsala University, Box 516, 
S-75120 Uppsala, Sweden
}


\abstract
{
Studying the spots of late-type stars is crucial for distinguishing between the
various proposed
dynamo mechanisms believed to be the main cause of starspot activity.
For this research it is important to collect observation time series that are 
long enough  to unravel both long- and short-term spot evolution. Doppler
imaging is a very efficient method for studying spots of stars that
cannot be angularly resolved.}
{High-resolution spectral observations during 1994--2017 are analysed in order 
to reveal long- and short-term changes in the spot activity of the FK Comae-type
subgiant \object{HD 199178}.}
{Most of the observations were collected with the Nordic Optical Telescope. The
Doppler imaging temperature maps were calculated using an inversion technique
based on Tikhonov regularisation and utilising multiple spectral lines.} 
{We present a unique series of 41 temperature maps spanning more than 23 years. 
All reliable images 
show a large cool spot region centred near the visible rotation pole. 
Some lower latitude cool features are also recovered, although the reliability
of these is questionable. There is an expected anti-correlation between
the mean surface temperature and the spot coverage.
Using the Doppler images, we construct the equivalent of a solar 
butterfly diagram for HD 199178.}
{\object{HD 199178} clearly has a long-term large and cool spot structure
at the rotational pole. This spot structure dominated the spot activity
during the years 1994--2017. The size and position of the 
structure has evolved with time, with a gradual increase during the last
years. 
The lack of lower latitude features
prevents the determination of a possible differential rotation. 
}

\keywords{stars: activity -- starspots -- late-type -- imaging -- individual: 
HD 199178}

\maketitle

\section{Introduction}
\label{intro}

\object{HD 199178} (also known as \object{V1794 Cyg}) is one of the original members of the
FK Comae-type stars originally defined by \cite{BoppRuc81}. This very small
group consists of stars that are single, rapidly rotating, and extremely active 
G--K-type 
subgiants or giants. The magnetic activity is observed both as strong emission 
from the corona, transition region, and chromosphere and as large cool spots
causing rotational modulation of the brightness and photospheric absorption 
line profiles. 

There are very few confirmed FK Comae-type stars (only three or four).
The reason for this is that the class
probably represents a very short evolutionary stage.
\cite{BoppRuc81} postulated that they are coalesced W Uma-type systems. Thus, 
they would be observed in the transitionary stage when they have formed a 
single star, but still rotate rapidly. Eventually the rapid rotation will slow 
down due to magnetic braking.
Even though these stars are rare, they still provide useful information on
stellar magnetic activity. It seems that most rapidly rotating late-type stars
show similar spot activity: large non-axisymmetric spot structures, often
on high latitudes and even forming polar caps \citep[see e.g.][]{Strass2009}.
This is true  for evolved stars, e.g. RS CVn and FK Comae -type stars, and for young solar-type stars \citep[see e.g.][]{Willamo2018}.
These  high-latitude structures
have been confirmed, not only by Doppler imaging, but also using long-baseline 
infrared interferometry \citep{Roett2016,Roett2017}. The reason behind the 
similarities
of these late-type stars in very different evolutionary stages is apparently
a common kind of dynamo mechanism operating in their convection zone.
The general notion is that rapid rotation suppresses the differential rotation,
which could lead to dynamos of $\alpha^2$- or $\alpha^2 \Omega$-type, in 
contrast to the $\alpha \Omega$ dynamo operating in the Sun 
\citep[see e.g.][and references therein]{Oss2003}. 

There is no doubt that the photometric rotation period of HD 199178 is around 
$3\fd3$ \citep[see e.g.][]{Jetsu1999}. Furthermore, \cite{Jetsu1999} showed 
that the light curve of the star cannot be explained with a single, unique 
period. \cite{Jetsu1999} applied the Kuiper method on the times of the 
photometric minima and retrieved the period $P \approx 3\fd3175$ implying
a long-lived active longitude.
\cite{Panov2007} used a mean period of $P\approx 3\fd30025$ to minimise 
the systematic drift of the phase of the photometric minimum.
The variations in the photometric rotation period of HD 199178 were
interpreted by \cite{Jetsu1999} as signs of differential rotation. 
In a recent study using long-term photometry of \object{FK Comae}, 
\cite{Jetsu2018} suggested that the light curve is better explained with a 
combination of two periods. The interference between these two periods would 
cause apparent period variations, when fitting a single period model to the 
data. Thus estimating differential rotation from period variations in a model that is  too
simple  may be misleading. However, two co-existing periods may
itself be a consequence of differential rotation. As HD 199178 belongs to
the same class as FK Comae, there are strong reasons to doubt that
its differential rotation could be estimated by studying variations of
a single period fit. 

\citet{Oneal1996,Oneal1998} estimated starspot 
parameters of  HD 199178, among other stars, using photometry and molecular bands. They 
reported the values $T_Q = 5350\,$K and $T_S = 3800\,$K for the `quiet' 
surface and spots, respectively.
Previous Doppler imaging studies of HD 199178 include the papers by 
\cite{Vogt1988}, \cite{Strass1999}, \cite{Hackman2001}, and \cite{Petit2004}.
All of these revealed large polar or high-latitude spot structures.
\cite{Petit2004} also calculated surface magnetic field maps for HD 199178 
applying Zeeman-Doppler imaging (ZDI) on Stokes V spectropolarimetry obtained
in 1998--2003. The common features in the ZDI maps presented in their study 
were mixed polarity radial fields ranging from -500 to +500 G in the polar 
region, and azimuthal fields forming rings or arches around the polar region.

The attempts to determine the differential rotation of HD 199178 using 
Doppler imaging have yielded 
conflicting results. \cite{Petit2004} employed differential rotation in
the ZDI inversion and derived the value $\alpha \approx 0.041$ 
based on the goodness of the Doppler imaging solution. However, 
\cite{Hackman2001} tried different values of $\alpha$ and concluded that a 
negative $\alpha$ gave the best fit and smallest systematic error. As
the general notion is that rapidly rotating single stars should not have
strong anti-solar differential rotation, the latter result has been questioned (e.g. \citealt{Rice2002}).

Several studies have aimed at determining the possible cyclic behaviour
in the spot activity of HD 199178.  \citet{Jetsu1990} 
reported a 9.07-year cycle in $UBVRI$-photometry, but failed to confirm it in 
a later study \citep{Jetsu1999}. \cite{Panov2007} reported a cycle of
4.2 years in spot longitudes, while \cite{Savanov2009} derived an eight-year 
brightness cycle.
 
In this study we present new Doppler images for HD 199178 using high-resolution
spectroscopy collected with three different instruments in 1994--2017. 
Despite some gaps, this data set constitutes one of the longest series
of Doppler images for any star. 
This unique series of Doppler images
enables us to  construct the equivalent of a solar butterfly diagram.
Such diagrams are important for studying stellar dynamos,
and have previously been constructed using three different approaches:
by modelling light curves \citep[see e.g.][]{Livshits2003}, based on Doppler 
imaging \citep[see e.g.][]{Hackman2011,Willamo2018}, and utilising 
asteroseismology \citep{Bazot2018,Nielsen2018}.

\begin{table*}
      \caption[]{Summary of spectral observations.}
\centering
\begin{tabular}{lcrrrrccc}
\hline \hline
Month & $\langle$HJD$\rangle - 2400000$ & $R$ & $n_\phi$ & $\Delta t [$d$]$ & $f_\phi$  & Wavelength regions [\AA] & 
$\langle S/N \rangle $ & $\sigma$ [\%]\\

            \hline
July 1994      & 49558.5 & 35 000  & 19 & 14 & 98\% & 6431, 6439 & 239 & 0.529 \\
August 1994    & 49583.5 & 80 000  & 11 & 10 & 90\% & 6411, 6431, 6439, 7511 & 265 & 0.455 \\
November 1994  & 49673.8 & 140 000 & 9  &  9 & 86\% & 6431, 6439 & 206 & 0.540\\
July 1995      & 49916.5 & 80 000  & 11 & 11 & 95\% & 6411, 6431, 6439, 7511 & 333 & 0.381 \\
October 1996   & 50384.8 & 80 000  & 8  &  7 & 75\% & 6411, 6431, 6439, 7511 & 326 & 0.408 \\
June 1997      & 50623.0 & 80 000  & 12 & 10 & 97\% & 6411, 6431, 6439, 7511 & 278 & 0.435 \\
July 1998      & 51003.8 & 80 000  & 13 & 13 & 96\% & 6411, 6431, 6439, 7511 & 320 & 0.480 \\
October 1998   & 51092.1 & 80 000  & 10 &  8 & 85\% & 6411, 6431, 6439, 7511 & 244 & 0.546 \\
November 1998  & 51124.4 & 80 000  &  7 &  6 & 69\% & 6411, 6431, 6439, 7511 & 230 & 0.520 \\
May 1999       & 51328.2 & 80 000  & 14 & 10 & 93\% & 6411, 6431, 6439, 7511 & 203 & 0.617 \\
July 1999      & 51388.5 & 80 000  & 11 & 10 & 97\% & 6411, 6431, 6439, 7511 & 174 & 0.694 \\
October 1999   & 51474.0 & 80 000  &  9 &  9 & 64\% & 6411, 6431, 6439, 7511 & 192 & 0.780 \\
August 2000    & 51768.6 & 80 000  & 21 & 11 & 100\%&  6431, 6439, 7511 & 222 & 0.618 \\
June 2001      & 52068.4 & 80 000  & 15 &  7 & 89\% & 6431, 6439, 7511 & 201 & 0.502 \\
August 2002    & 52511.8 & 80 000  & 23 &  9 & 100\%& 6411, 6431, 6439 & 259 & 0.524 \\
November 2002  & 52594.3 & 80 000  &  8 & 12 & 68\% & 6411, 6431, 6439 & 285 & 0.399 \\
June 2003      & 52803.4 & 80 000  & 16 & 19 & 95\% & 6411, 6431, 6439 & 268 & 0.462 \\
November 2003   & 52950.6 & 80 000 & 18 & 20 & 100\%& 6411, 6431, 6439 & 254& 0.515 \\
August 2004    & 53220.2 & 80 000  & 16 & 13 & 89\% & 6411, 6431, 6439 & 257 & 0.527 \\
July 2005      & 53571.4 & 80 000  & 11 & 16 & 75\% & 6411, 6431, 6439 & 270 & 0.500 \\
November 2005  & 53690.8 & 80 000  &  6 & 10 & 50\% & 6411, 6431, 6439 & 242 & 0.525 \\
September 2006 & 53982.6 & 80 000  & 13 &  9 & 91\% &6411, 6431, 6439 & 280 & 0.514 \\
December 2006  & 54074.8 & 80 000  & 8  &  7 & 75\% & 6265, 6439, 6644 & 211 & 0.728 \\
July 2007      & 54305.9 & 80 000  & 12 & 10 & 94\% & 6411, 6431, 6439 & 238 & 0.540 \\
September 2008 & 54721.2 & 80 000  & 7  &  6 & 60\% & 6265, 6431, 6439, 6644 & 337 & 0.494 \\
December 2008  & 54811.8 & 80 000  & 6  &  6 & 49\% & 6265, 6431, 6439, 6644 & 213 & 0.835 \\
August 2009    & 55074.7 & 80 000  & 9  & 11 & 78\% & 6439, 6644, 6663 & 297 & 0.663 \\
September 2009 & 55075.7 & 80 000  & 11 & 12 & 80\% & 6411, 6439, 6644 & 259 & 0.524 \\
December 2009  & 55197.1 & 80 000  & 5  &  6 & 47\% & 6265, 6439, 6644 & 205 & 0.536 \\
July 2010      & 55402.2 & 80 000  & 10 & 13 & 89\% & 6265, 6439, 6644 & 287 & 0.551 \\
December 2010  & 55552.6 & 80 000  & 4  & 12 & 31\% & 6265, 6439, 6644 & 266 & 0.422 \\
December 2011  & 55908.8 & 80 000  & 4  & 4  & 40\% & 6265, 6439, 6644 & 355 & 0.469 \\ 
August 2012    & 56170.5 & 80 000  & 9  &  7 & 87\% & 6265, 6439, 6644 & 349 & 0.382 \\
November 2012  & 56259.6 & 80 000  & 4  & 11 & 40\% & 6265, 6439, 6644 & 232 & 0.540 \\
August 2014    & 56886.0 & 67 000  & 10 & 9  & 84\% & 6265, 6411, 6431, 6439, 6644 & 352 & 0.382\\ 
December 2014  & 56996.2 & 67 000  &  6 & 6  & 58\% & 6265, 6411, 6431, 6439, 6644 & 293 & 0.431 \\
July 2015      & 57209.1 & 67 000  &  9 & 9  & 78\% & 6265, 6411, 6431, 6439, 6644 & 380 & 0.380\\
November 2015  & 57355.9 & 67 000  &  7 & 8  & 69\% & 6265, 6411, 6431, 6439, 6644 & 246 & 0.481 \\
June 2016      & 57560.0 & 67 000  &  8 & 8  & 75\% & 6265, 6411, 6431, 6439, 6644 & 318 & 0.663 \\
June 2017      & 57914.4 & 80 000  & 7  & 10 & 59\% & 6265, 6431, 6439, 6644 & 240 & 0.451 \\
December 2017  & 58108.3 & 67 000  & 5  &  4 & 49\% & 6265, 6411, 6431, 6439, 6644 & 245 & 0.636 \\

\hline
 \label{tobs}
\end{tabular}
\tablefoot{$\langle$HJD$\rangle$ is the mean heliocentric Julian date,
$R$ the spectral resolution, $n_\phi$ the number of observations, 
$\Delta t$ the time span of the observations, $f_\phi$ the phase coverage of 
the season (see section \ref{secdi} for a definition), 
$\langle S/N \rangle$ the mean signal-to-noise ratio, and $\sigma$
the mean deviation between the observations and DI solution.}
\end{table*}

\section{Observations}
\label{obs}

The first observational set, July 1994, was collected with the 2 m 
Ritchey-Chreti\'en telescope
of the National Astronomical Observatory (Rozhen), Bulgaria. The rest of the
data was obtained with the Nordic Optical Telescope (NOT) from August
1994 to December 2017. The NOT observations were obtained with the SOFIN 
spectrograph 
during 1994--2012 and in June 2017. During 2013--2016 and in December 2017 the FIES 
spectrograph was used.

Seven spectral regions were used for Doppler imaging: 6263.0 -- 6267.0\,{\AA} 
(strongest lines \ion{Fe}{I} 6265.13\,{\AA} and \ion{V}{I} 6266.31\,{\AA}), 
6409.6 -- 6413.4\,{\AA} 
(\ion{Fe}{I} 6411.65\,{\AA}), 6428.5 -- 6434.5\,{\AA} (\ion{Fe}{I} 6430.84\,{\AA} 
and \ion{Fe}{II} 6432.68\,{\AA}), 6437.0 -- 6441.0\,{\AA} (\ion{Ca}{I} 
6439.08\,{\AA}), 6641.6 -- 6645.4\,{\AA} (\ion{Ni}{I} 6643.63\,{\AA}), 6661.3 -- 
6665.3\,{\AA} (\ion{Fe}{I} 6663.23\,{\AA} and \ion{Fe}{I} 6663.44\,{\AA}), and 
7509.0 --7513.4\,{\AA} (\ion{Fe}{I} 7511.02\,{\AA}). A summary of all 
the observations is given in Table \ref{tobs}.
As can be seen from the summary, different seasons covered different wavelength 
regions. However, the
region with the line \ion{Ca}{I} 6539.08\,{\AA} was included in every data set.
The signal-to-noise ratio was usually around 200--300. Spectra with $S/N$ as
poor as $\sim$100 were used in a few cases. The average $S/N$ is listed in 
Table \ref{tobs}.

The observations from July 1994 to July 1998 were already published 
\citep{Hackman2001,Hackman2004}. They were reduced 
using the 3A package \citep{ilyin2000}. 
The rest of the SOFIN data was reduced
using a new reduction pipeline developed by the same author. A more detailed
description of the new reduction pipeline can be 
found in \cite{Willamo2018} and will also be presented by Ilyin (2019, in 
prep.). The FIES data was reduced using the FIEStool \citep{Telting2014}.
The reduction covered 
all standard reduction procedures for high-resolution CCD spectra.
The final continuum normalisation was done separately
for each wavelength region (i.e. line blend) by comparing the observed spectrum
to the synthetic spectrum during the Doppler imaging inversion (see section
\ref{secdi}). Plots of all spectral data together with the Doppler imaging
modelled spectra, calculated using the adopted stellar parameters
(Table \ref{spar}), are shown in Appendix \ref{spectra}.

In general, a minimum number of $\sim$10 evenly distributed rotational phases
is considered  optimal for a fully reliable Doppler image. However, 
\cite{Hackman2011} demonstrated, that some useful information can be retrieved
with a much more limited phase coverage. Therefore, we calculated
Doppler images for all sets  with $n_\phi \ge 4$.

\begin{table}
\setlength{\tabcolsep}{3pt}

      \caption[]{Adopted stellar parameters for HD 199178.}
         \label{spar}

\centering
\begin{tabular}{lcl}
\hline \hline
Parameter & Adopted & Reference \tablefootmark{a} \\
\hline
Gravity $\lg (g)$ & 3.5  & \cite{Hackman2001} \\
Inclination $i$ & 60 $\degree$ & \cite{Hackman2001}\\
Rotation velocity & 72\,\kms & \cite{Hackman2001} \\
Rotation period & $3\fd3175$  & \cite{Jetsu1999} \\
Differential rotation $\alpha$ & 0 & \\
Micro turbulence $\xi$ & 1.7\,\kms & \\ 
Macro turbulence $\zeta_\mathrm{RT}$ & 5.0\,\kms & \\
Element abundances & solar & \cite{Strass1999} \\  

\hline
\end{tabular}
\tablefoot{
\tablefoottext{a}{In case of no reference,
a new parameter value was
derived/adopted for this study.}}
\end{table}

\section{Spectral and stellar parameters}

The spectral parameters were obtained from the Vienna Atomic Line Database 
\citep{VALD1,VALD2}. Some adjustments were made in the $gf$-values
in order to get a satisfactory fit for all spectral regions. This is a standard 
procedure in Doppler imaging since the spectra of real stars always show some 
deviations from calculations based on LTE-models. 
As we did not intend to derive exact element abundances,
it is safe to correct for all discrepancies between the model
and the real star by adjusting the $gf$-values. Furthermore,
to the first order all photospheric 
absorption lines will follow the same
behaviour (i.e. a cool spot will cause a `bump' in the spectral line).

The adopted parameters
for the strongest lines are listed in Table \ref{spepar}. The largest
adjustment was for the line \ion{Ca}{I} 6439.075\,{\AA}, were VALD listed
$\lg (gf)=0.370$ and we adopted the value 0.450. A total number of 917 lines
were included in the line synthesis of the seven wavelength regions. Most of these
were molecular lines. For the unspotted surface a smaller number, i.e. only the atomic lines,
would have been sufficient. For the cool spots molecular lines, 
in particular TiO-bands, are important. The local line profiles were 
calculated using the code LOCPRF7 developed at the University of Uppsala. 
Recent changes in the code include a
molecular equilibrium solver \citep{Piskunov2017}.

\begin{table}
      \caption[]{Wavelengths, lower excitation potentials, and adopted
$\lg(gf)$-values for the strongest lines.}

\centering
\begin{tabular}{lcrr}
\hline \hline
Line & $\lambda$ (\AA) & $\chi_{_\mathrm{low}}$ (eV) & $\lg (gf)$ \\
\hline
\ion{Fe}{I} &       6265.1319 & 2.1759 & -2.550 \\
\ion{V}{I} &        6266.3069 & 0.2753 & -2.290 \\
\ion{Fe}{I} &       6411.6476 & 3.6537 & -0.705 \\
\ion{Fe}{I} &       6430.8446 & 2.1759 & -2.206 \\
\ion{Fe}{II} &      6432.6757 & 2.8910 & -3.320 \\
\ion{Ca}{I} &       6439.0750 & 2.5257 &  0.450 \\
\ion{Ni}{I} &       6643.6304 & 1.6764 & -1.920 \\
\ion{Fe}{I} &       6663.2311 & 4.5585 & -1.239 \\
\ion{Fe}{I} &       6663.4407 & 2.4242 & -2.479 \\
\ion{Fe}{I} &       7511.0181 & 4.1777 & 0.199 \\

\hline

\label{spepar}
\end{tabular}
\end{table}

HD 199178 is a rapidly rotating late-type star, which was classified as 
G 5 {\sc III - IV}  by \cite{Herbig1958}. \cite{Strass1999} calculated
its luminosity and mass to be 11 $L_\odot$ and $1.65 M_\odot$. \cite{Hackman2001}
estimated its radius to be $\sim$5$R_\odot$ and adopted the rotation velocity
$v \sin i = 72$\,\kms, inclination angle $i = 60 \degree$, and gravity 
$\lg (g) = 3.5$.
With such a high rotation velocity, the macroturbulence is hard to determine,
but in any case should  not strongly influence Doppler imaging. In this
study we used a radial-tangential macroturbulence of $\zeta_\mathrm{RT} = 5.0$
\kms. We adopted an effective temperature $\teff = 5300$\,K for the unspotted 
surface as a starting point. The microturbulence was determined by comparing 
the synthetic spectrum of an unspotted surface to the observations. This 
yielded  $\xi = 1.7$\,\kms.

\cite{Jetsu1999} and \cite{Panov2007} concluded, that the photometric light
curve cannot be modelled  using a constant rotation period. However,
in order to compare temperature maps of different seasons, all observations 
have to be phased by a single rotation period. This is particularly important
because of our intention to study possible active longitudes and/or
azimuthal dynamo waves. Thus, we adopted the period
$P_\mathrm{al} = 3\fd3175$, which was reported by \cite{Jetsu1999} as the
most significant one in the analysis of possible long-term active longitudes.
Even though this may not be fully correct for all seasons, each
observation run is short enough that a small error in the period will
not distort the image. The phases for the observations were obtained with the 
ephemeris

\begin{equation}
\mathrm{HJD}_{\phi=0} = 2449557.664 + 3.3175 \times E.
\label{ephem}
\end{equation}

The surface differential rotation parameter $\alpha$ is usually expressed as

\begin{equation}
\alpha = {\Omega_\mathrm{eq} - \Omega_\mathrm{pole} \over \Omega_\mathrm{eq}},$$
\label{alpha}
\end{equation}
\noindent
where $\Omega_\mathrm{eq}$ and $\Omega_\mathrm{pole}$ are the angular rotation 
velocities at the equator and approaching the pole, respectively. In the Sun the 
differential rotation follows the approximate profile

\begin{equation}
\label{soldif}
 \Omega(\theta) \approx \Omega_\mathrm{eq} (1 - \alpha \sin^2\theta).$$
\end{equation}

\noindent
Here $\theta$ is the stellar latitude. It should be noted that Eq. 
\ref{soldif}
has only been proven valid for the Sun. Numerical simulations show that
faster rotating stars could have completely different differential rotation 
curves \citep[see e.g.][]{Kitchatinov1999,Kitchatinov2004}. Still the solar 
differential rotation curve is, for lack of a better option, used as a model 
for other stars.

\begin{figure}

\includegraphics[width=8cm,clip]{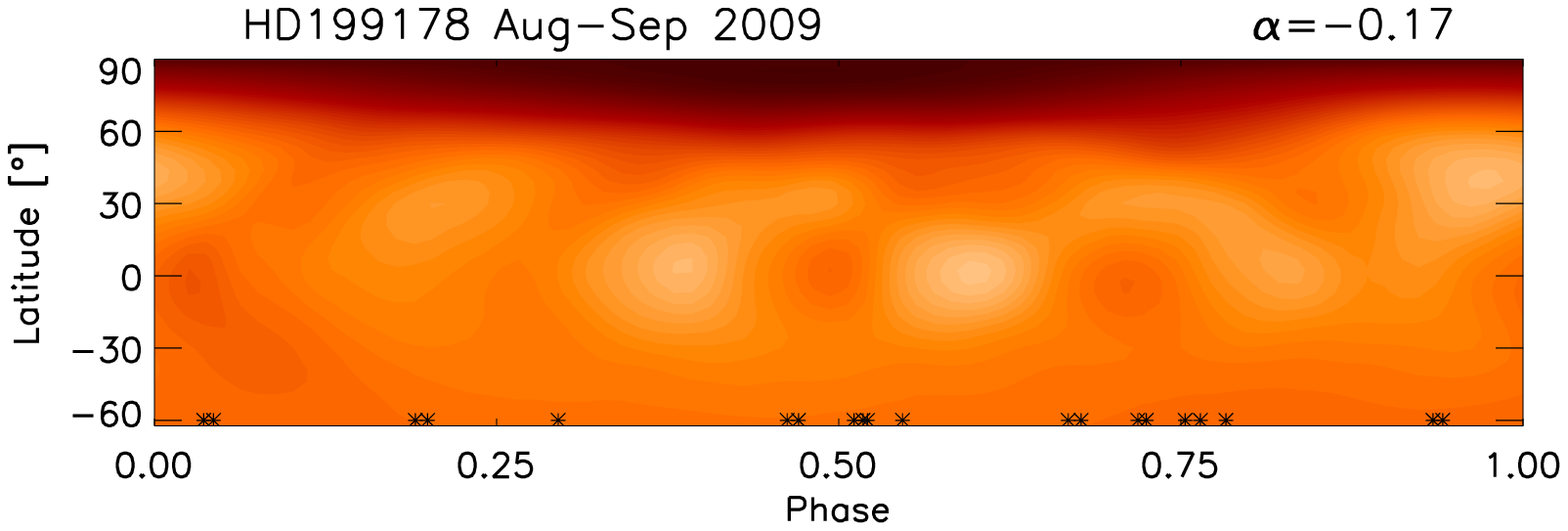}

\vspace{-3cm}

\includegraphics[width=8cm,clip]{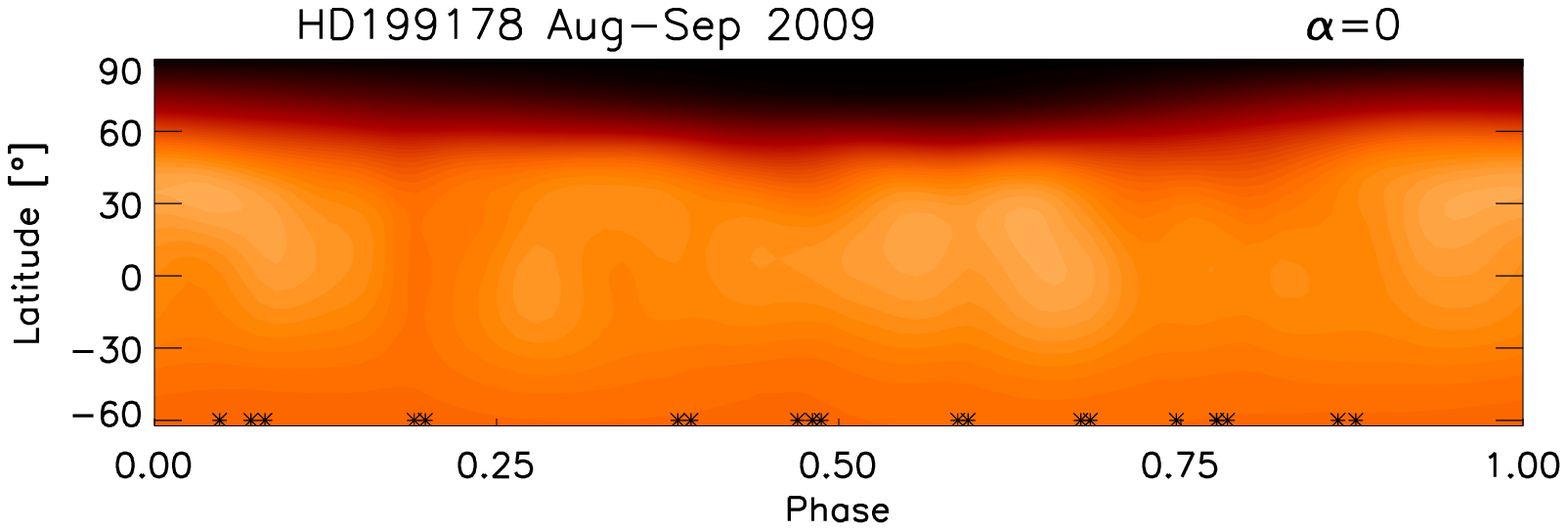}

\vspace{-3cm}

\includegraphics[width=8cm,clip]{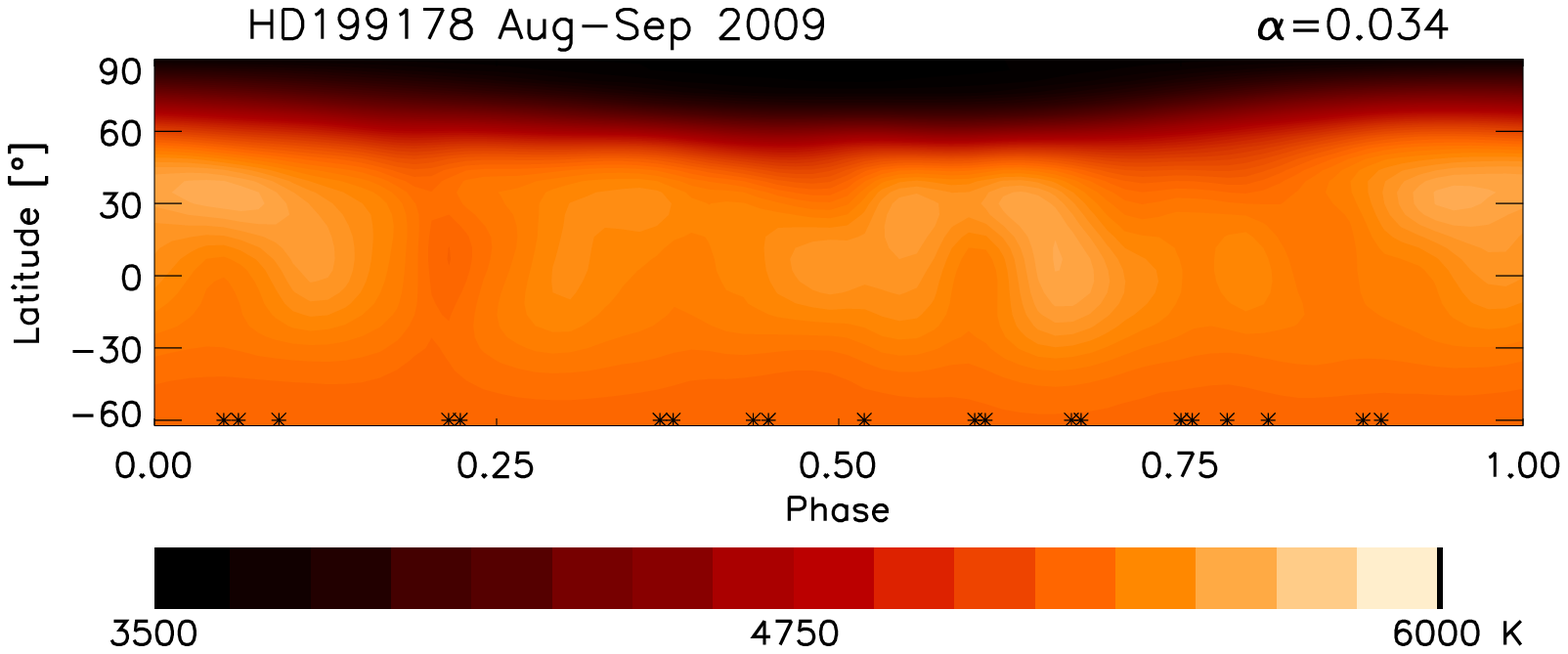}

\vspace{-2.5cm}

\caption{DI surface temperature maps of TEST2.
The observed phases are 
indicated with `*' and the same temperature scale is used in the three images.}
         \label{didr}
   \end{figure}

In the study by \cite{Hackman2001} the best fit for the Doppler images 
was achieved using the combination $\vsini= 70$\,\kms, $\alpha = -0.17$ and 
$P_\mathrm{eq}=3.81$. For the present study, we tested the cases of $\alpha=0$, 
$\alpha=-0.17$, and 
$\alpha=0.034$, the last value corresponding to the differential 
rotation derived 
for the epoch 2001.97 by \cite{Petit2004}. The reason for choosing this value
instead of $\alpha = 0.041$ is that the latter was a weighted mean and each 
value of $\alpha$ had a specific corresponding $\Omega_\mathrm{eq}$-value. Using
weighted means for both of these parameters would not necessarily have formed 
a consistent pair.
In these tests we accounted for both the surface shearing effect and
change in the rotational profile caused by differential rotation. For the test
we chose 
two sets. We can assume that any differential rotation will be
 detected most easily when the observation set covers a long period and the 
phase coverage is dense. Thus, we chose the longest set, November 2003 (TEST1), 
and combined the August and September 2009 sets into one set with better phase 
coverage (TEST2). In the latter case we could only use the common wavelength 
regions 6439 {\AA} and 6644 \AA.
Before the tests we chose 
the optimal $\vsini$ for each 
$\alpha$ value by testing values with a grid of 1\,\kms. The resulting 
deviation 
between the observations and 
the Doppler imaging (DI) solution after 0, 20, and 150 iterations ($\sigma_{0}$, $\sigma_{20}$, and $\sigma_{150}$) are shown in Table \ref{diffrot}.
When running the inversions for a sufficient number of iterations 
(150 iterations) the
differences in the deviation of the fits were small. The DI temperature 
maps for the August-September 2009 set are shown in Fig. \ref{didr}.

\begin{table}
\setlength{\tabcolsep}{3pt}

\caption[]{Convergence of DI solutions for different values of $\alpha$ 
in TEST1 and TEST2.}
         \label{diffrot}

\centering
\begin{tabular}{lclc|ccc}
\hline \hline
& $\alpha$ & $P_\mathrm{eq}$ & $\vsini$ & $\sigma_{0}$ & $\sigma_{20}$ & $\sigma_{150}$ \\
\hline
TEST1 & -0.17 & 3\fd81   & 70\kms  & 0.645\% & 0.539\% & 0.502\% \\
      &  0.0   & 3\fd3175 & 72\kms  & 0.737\% & 0.564\% & 0.515\% \\
      & 0.034 & 3\fd2488 & 73\kms  & 0.721\% & 0.567\% & 0.508\% \\
\hline
TEST2 & -0.17 & 3\fd81   & 70\kms  & 0.619\% & 0.548\% & 0.531\% \\
      &  0.0   & 3\fd3175 & 72\kms  & 0.701\% & 0.596\% & 0.544\% \\
      & 0.034 & 3\fd2488 & 73\kms  & 0.710\% & 0.599\% & 0.547\% \\
\hline
\end{tabular}

\end{table}

We thus confirmed the previous result by \cite{Hackman2001} that using an
anti-solar differential rotation will lead to a faster convergence and
slightly smaller deviation
of the Doppler imaging solution. 
An anti-solar differential rotation as strong as $\alpha=-0.17$ would be 
surprising taking into account
theoretical and observational results. Firstly, only slowly
rotating stars should have anti-solar differential rotation 
\citep{Gastine2014}, 
although there have been studies reporting weak anti-solar differential
rotation for some RS CVn binary components \citep[see e.g][]{Haru2016}.
Secondly,
rapidly rotating stars should not have such a strong differential rotation 
\citep[see e.g.][]{Kuker2011,Reinhold2015}. 
However, HD 199178 differs
significantly from most rapidly rotating late-type stars in that it has likely 
undergone a merger in the transit from a W Uma-type binary to a FK Comae-type star \citep[see e.g.][]{BoppRuc81}. Furthermore, 
asteroseismologic studies report strong internal shears as a result of core 
contraction in stars of similar luminosity class \citep[e.g.][]{Deheuvels2012}.
It is unclear what imprints 
a merger and possibly a strong radial shear would
leave at the surface.

Nevertheless, we
believe that the approach of fastest convergence and/or smallest
deviation is not the correct way to 
determine stellar rotational parameters like $\alpha$. One
reason for this is the regularisation function (e.g. maximum entropy or
minimum local gradient), which will in practice favour 
solutions with less deviation from a constant surface. By manipulating
stellar rotation parameters, axisymmetric spot structures can be minimised and 
the inversion procedure will thus favour this set of parameters. As has been 
demonstrated in several studies \citep[see e.g.][]{Hackman2001}, an anti-solar
differential rotation will have an effect on rotationally broadened
spectral line profiles similar to the effect  of a large polar spot.
Such high-latitude structures have been confirmed independently using 
interferometric observations \citep{Roett2016}.

The conclusion is that the rotational parameters cannot be determined
merely by optimising the Doppler imaging solution. The best way to determine
the differential rotation would be to trace the movement of spot structures
versus their latitude. The problem with this approach in our case is that the 
reliable cool spot structures in HD 199178 are found in a limited latitude 
range near the rotation pole \citep[see e.g.][and Fig. \ref{di} in the present 
study]{Petit2004}.

In conclusion: The differential rotation of HD 199178 is expected to be 
small and the determination of it is not possible with the present 
spectroscopic data. Thus, we chose to adopt the value $\alpha=0$ as this 
allows for unambiguous rotation phases for the whole set of images.

\section{Doppler imaging}
\label{secdi}

We used the Doppler imaging  code INVERS7DR originally written 
at the 
University of Uppsala, with some changes made at 
the University of Helsinki \citep[see e.g.][]{pisk1991,Hackman2012}. A total
of 41 new images were calculated for both the previously used spectra from
1994 to  1998 and all new spectra.
A table of synthetic spectra was first calculated
for a set of temperatures (3500 -- 6000\,K) and a number of
limb angles ($n_\mu = 40$) using stellar model atmospheres supplied by MARCS 
\citep{marcs}. This table was then used to solve the inverse problem, i.e.
searching for the surface temperature distribution that best reproduces
the observed photospheric spectra. In order to regularise this ill-posed
inversion problem, Tikhonov regularisation was employed. This means adding
the additional constraint of a minimum surface gradient of the solution. 
Furthermore, the solution was limited to the temperature range of the used
model atmospheres (i.e. 3500 -- 6000\,K) by an additional penalty function
similar to the one described by \cite{Hackman2001}.

In general, the method of regularisation is of less importance than the
calculation of the synthetic line profiles. However, with a high inclination
angle ($ \ge 60 \degree$) the region at the rotational pole will always be
near  the stellar limb and have a limited effect on the observed line 
profiles. Therefore, the Tikhonov regularisation may cause the convergence
of the solution to proceed slowly just at the pole. Thus, a large number
of iterations are needed to map this region correctly. In this study
we used 150 iterations and a regularisation parameter of $\Lambda = 1 \cdot 
10^{-9}$. All DI surface temperature maps are displayed in equirectangular 
projection in Fig. \ref{di}. The spectra calculated from the DI solutions
are displayed in Appendix \ref{spectra}.

Based on the assumption that ten evenly distributed rotational phases would
cover the whole star, we estimated a phase coverage $f_\phi$ for each DI 
(listed in Table \ref{tobs}). The mean phase coverage was 
$\langle f_\phi \rangle \approx 77$\%.
In six cases (December 2008, December 2009, December 2010,
December 2011, November 2012, and December 2017) $f_\phi$ was less than 50\%.

\begin{figure*}
\includegraphics[width=5.5cm,clip]
{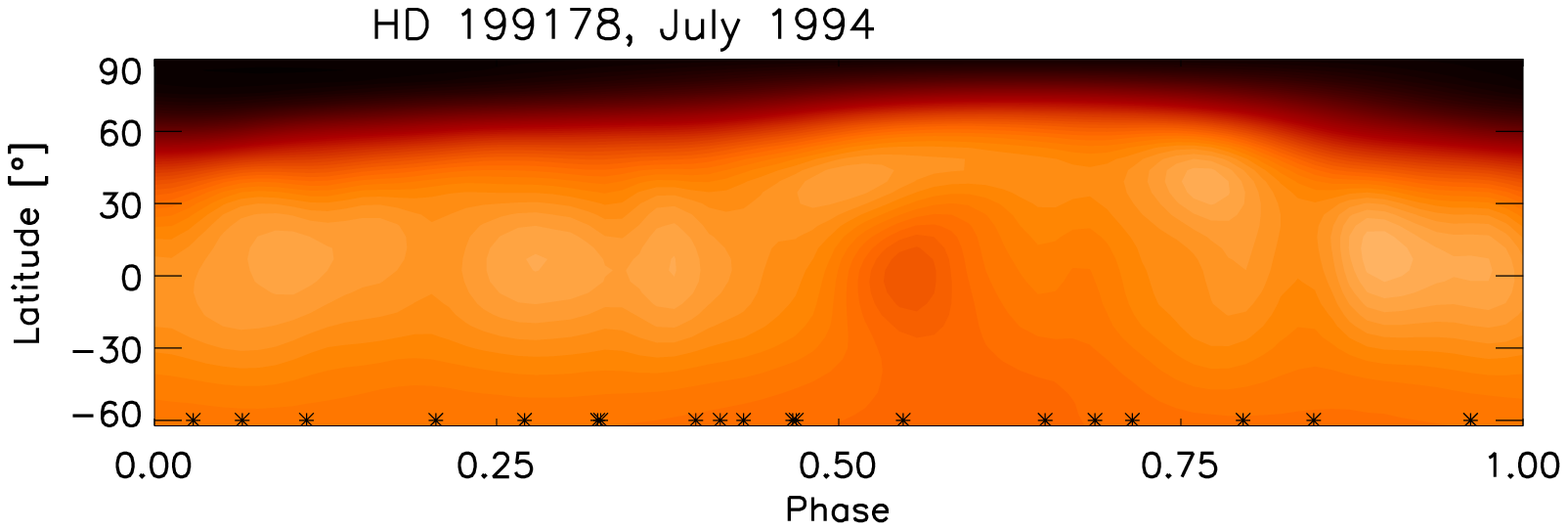}
\includegraphics[width=5.5cm,clip]
{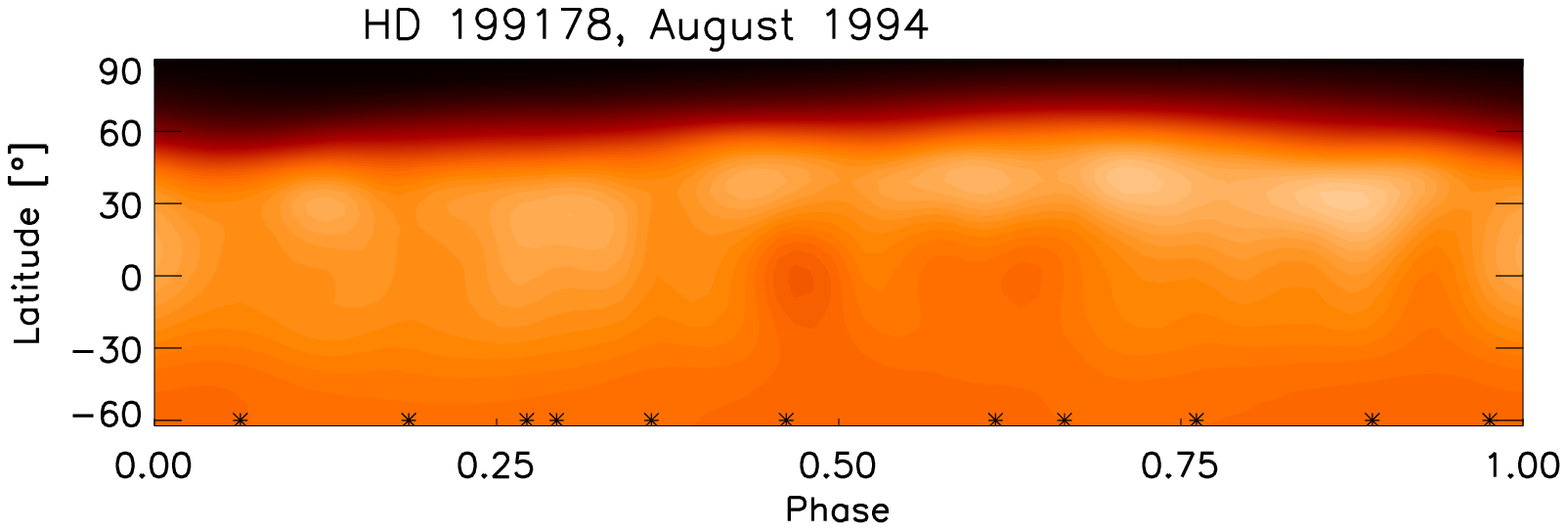}
\includegraphics[width=5.5cm,clip]
{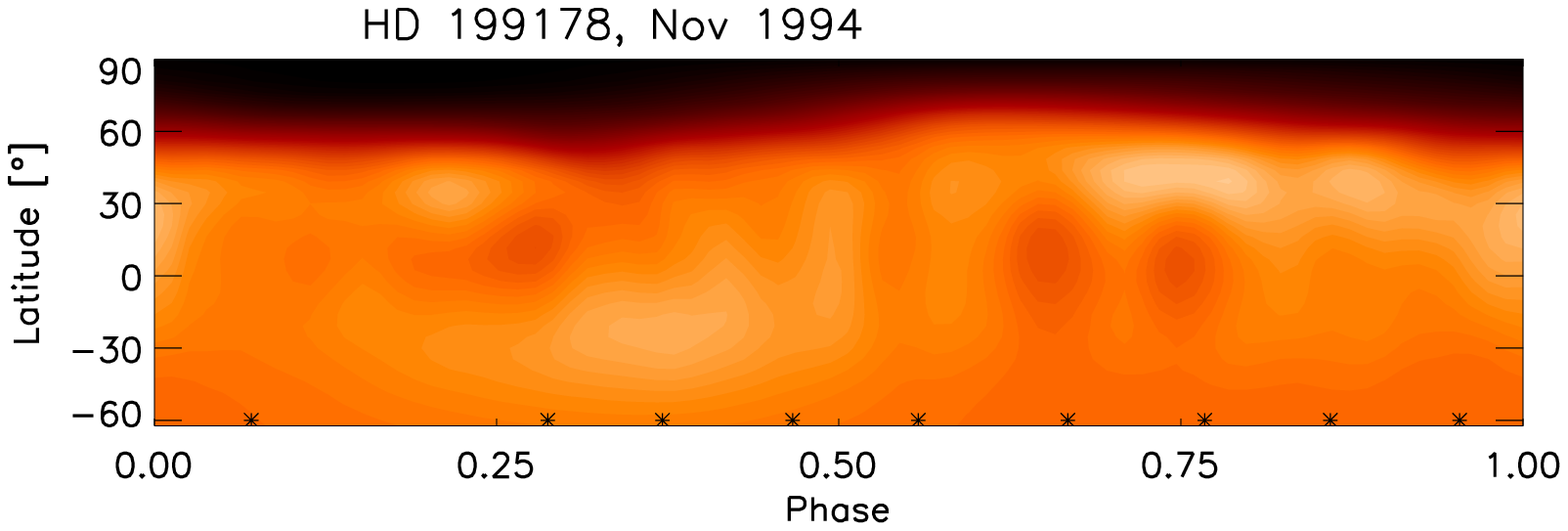}

\vspace{-2.25cm}

\includegraphics[width=5.5cm,clip]
{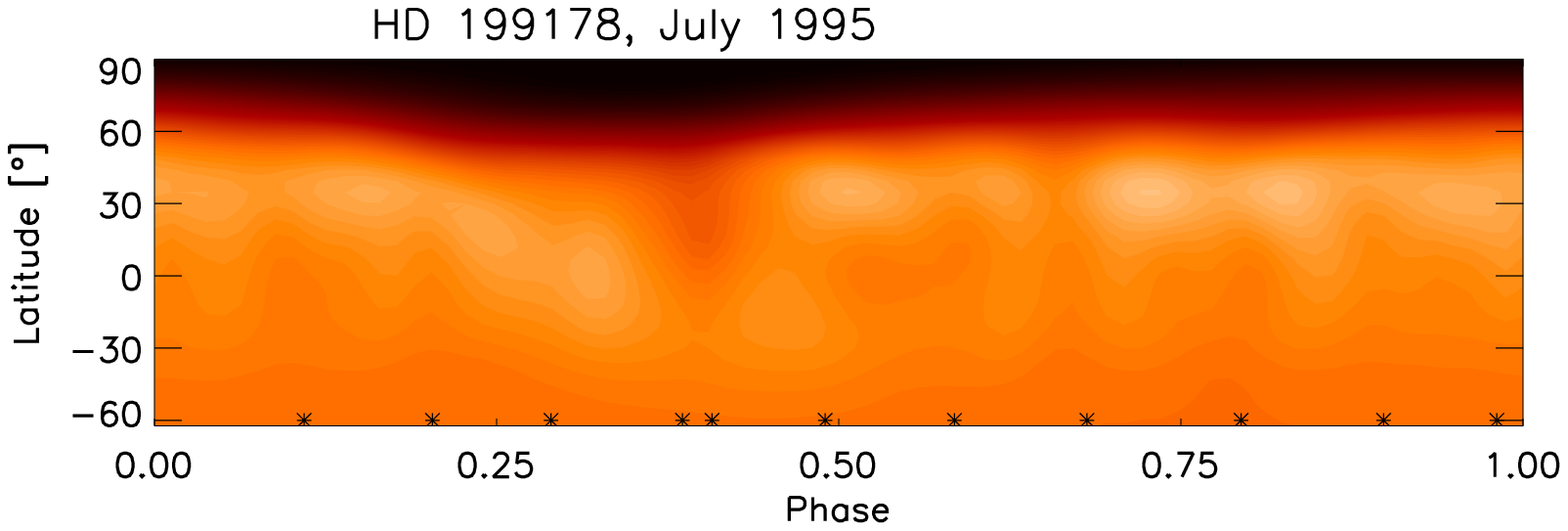}
\includegraphics[width=5.5cm,clip]
{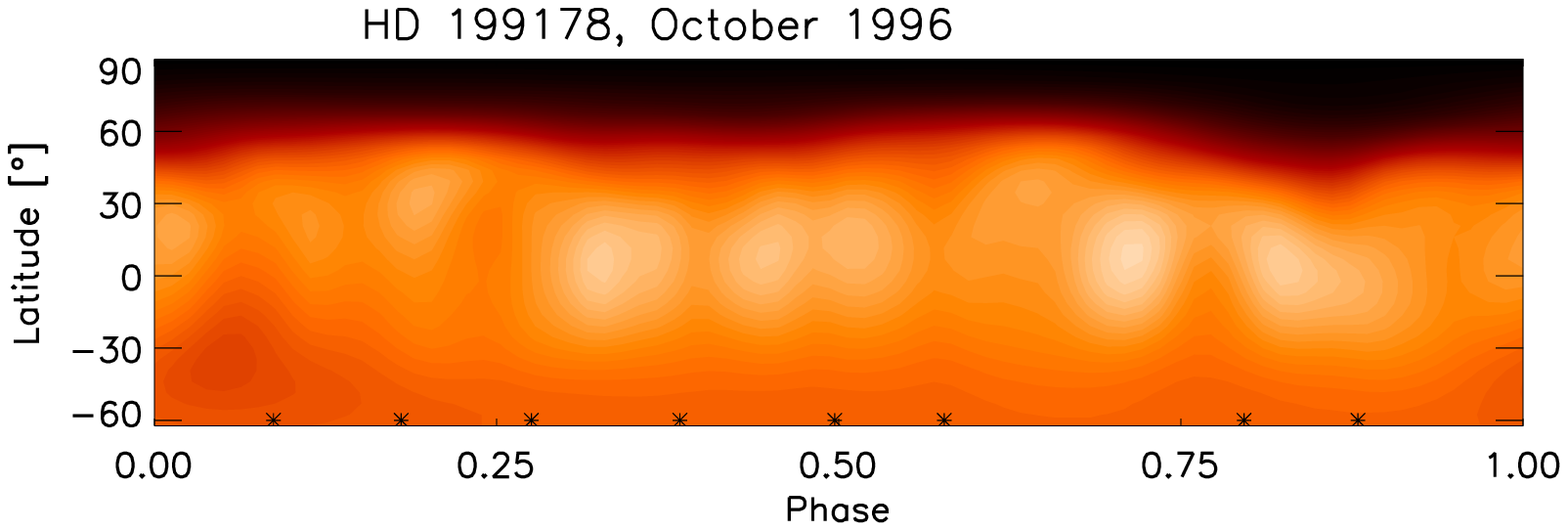}
\includegraphics[width=5.5cm,clip]
{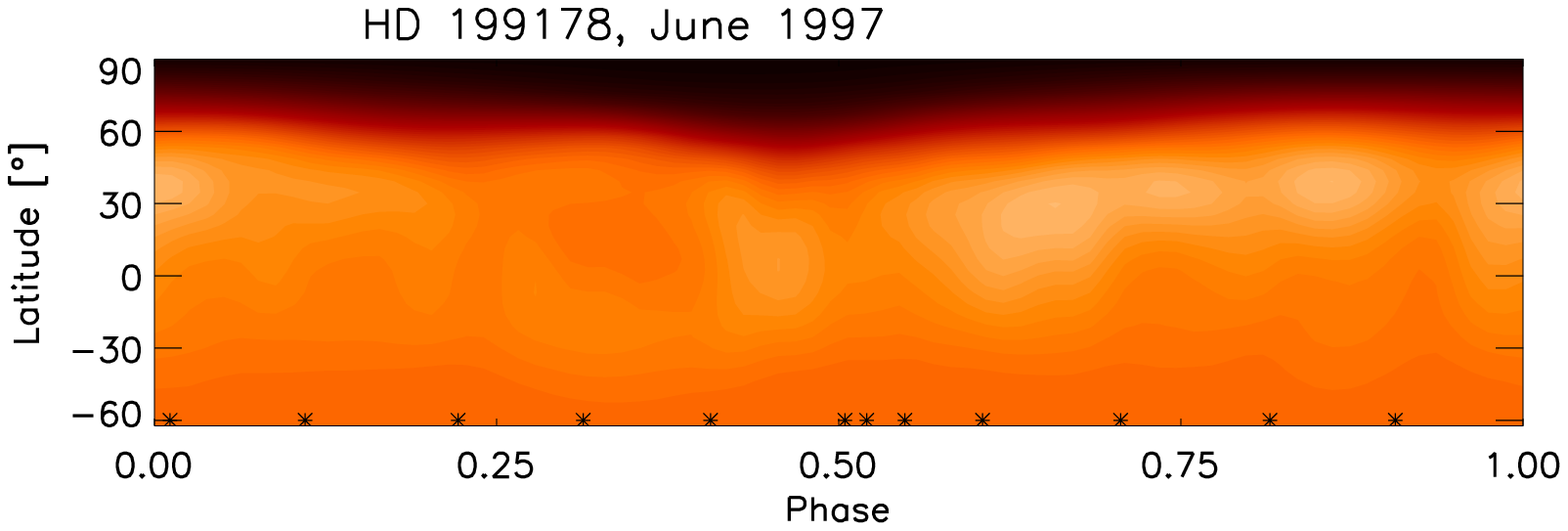}

\vspace{-2.25cm}

\includegraphics[width=5.5cm,clip]
{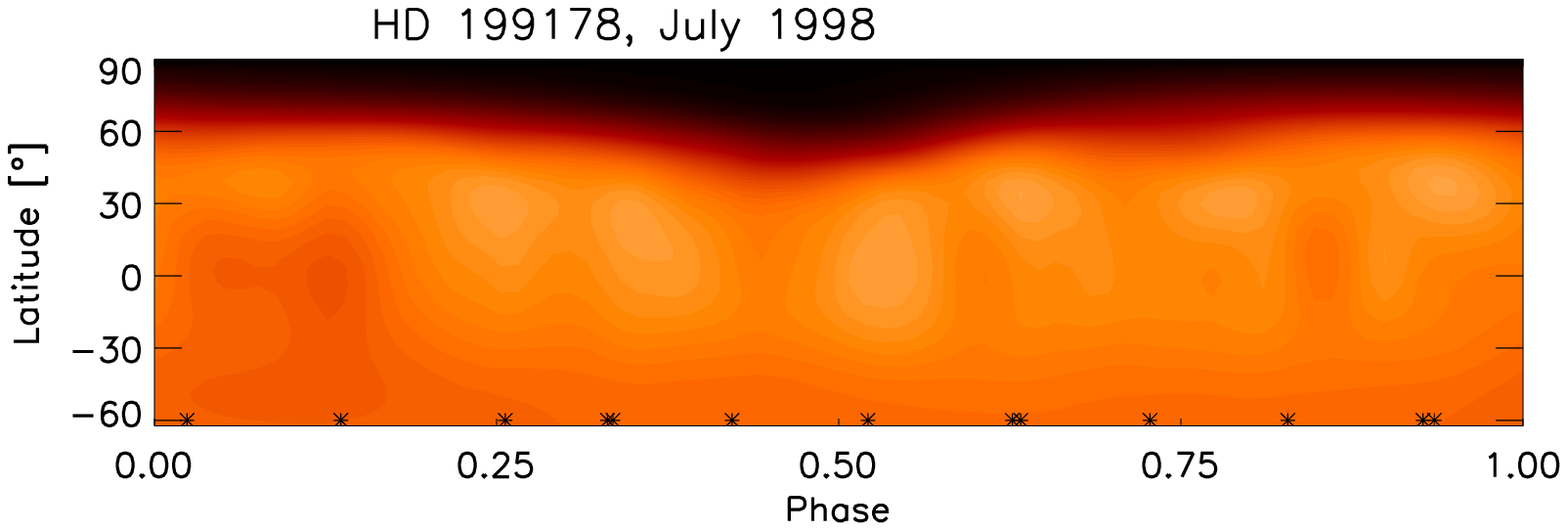}
\includegraphics[width=5.5cm,clip]
{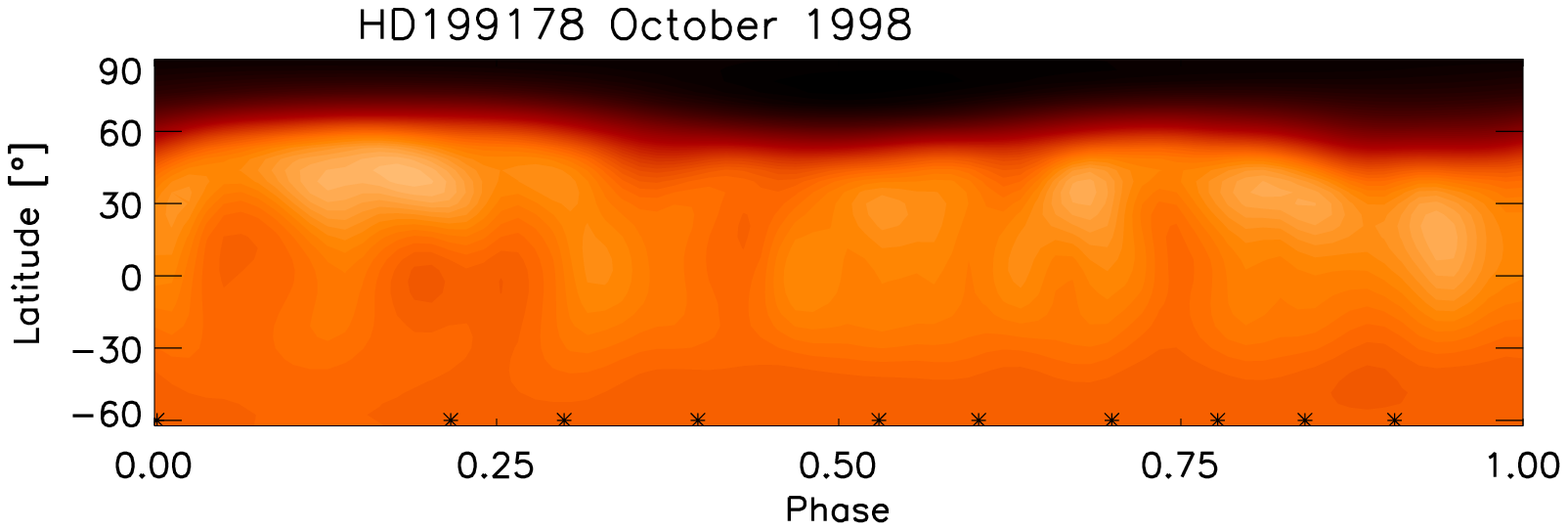}
\includegraphics[width=5.5cm,clip]
{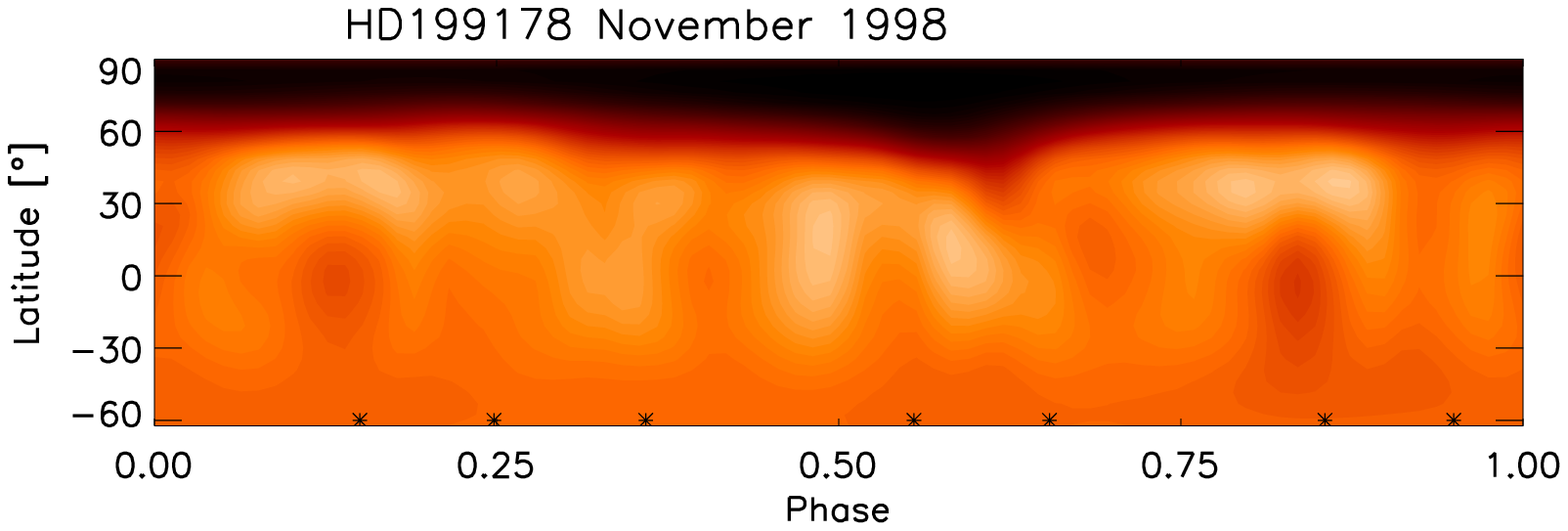}

\vspace{-2.25cm}

\includegraphics[width=5.5cm,clip]
{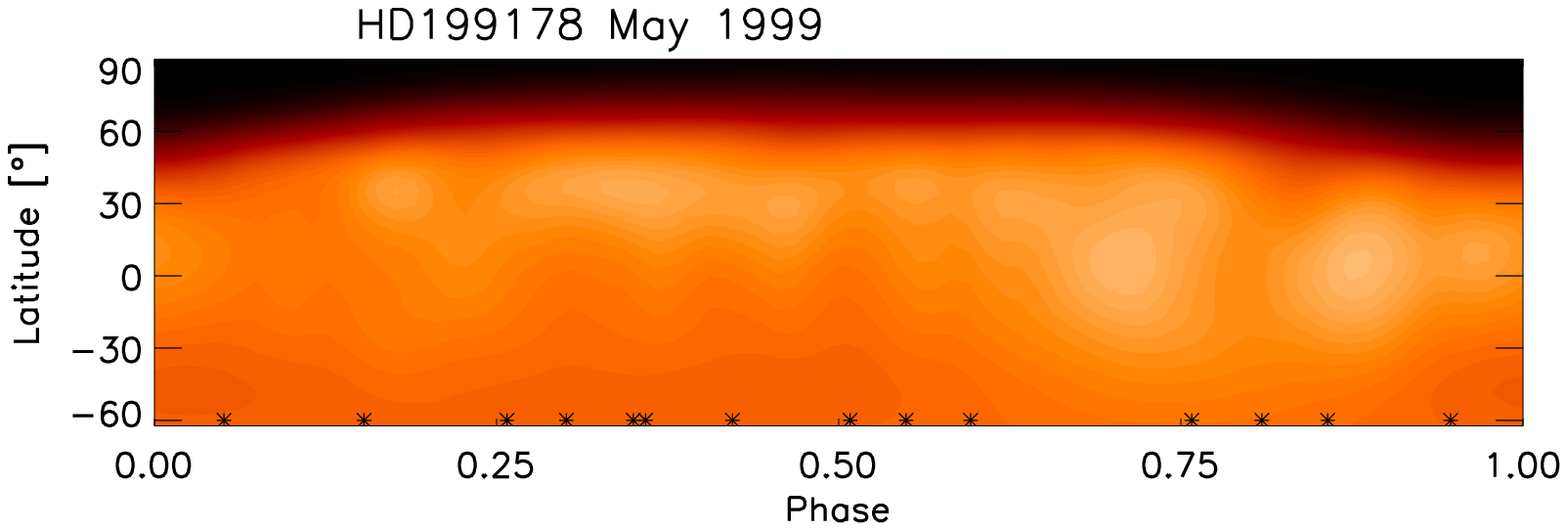}
\includegraphics[width=5.5cm,clip]
{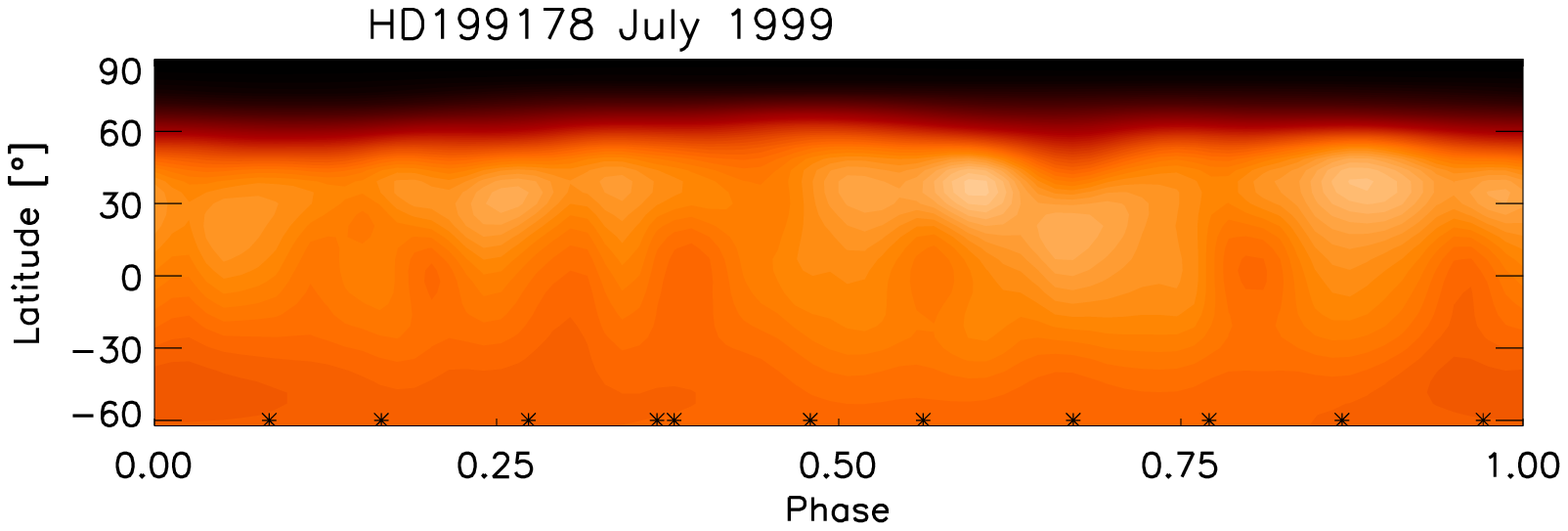}
\includegraphics[width=5.5cm,clip]
{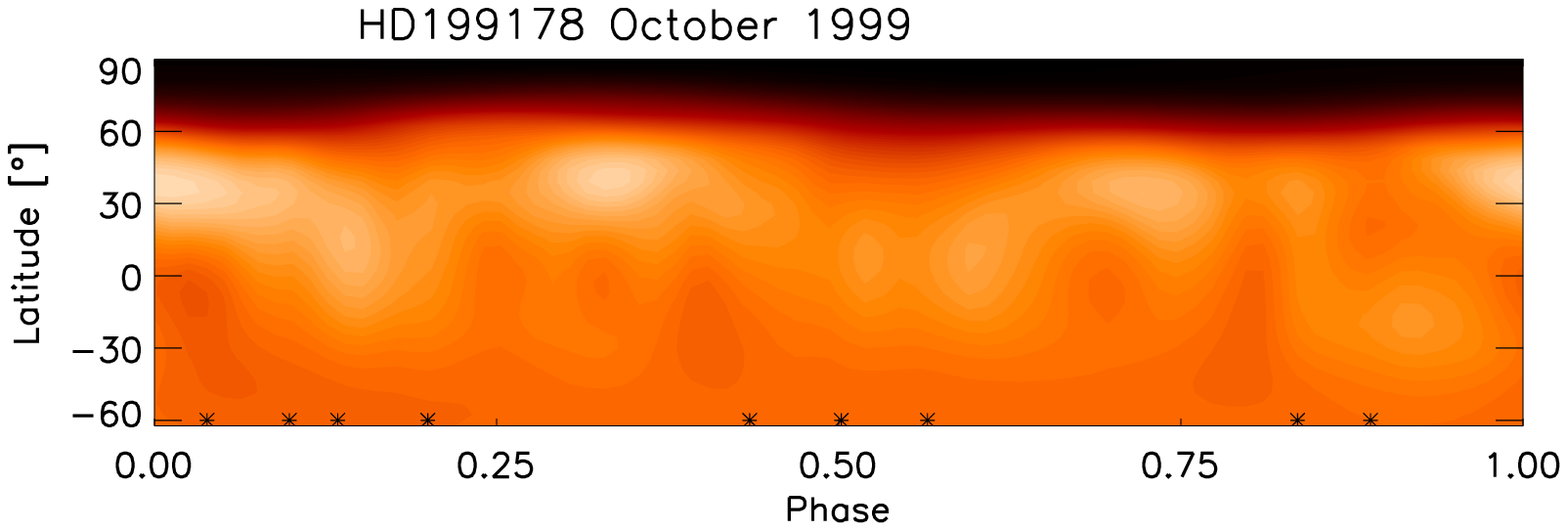}

\vspace{-2.25cm}

\includegraphics[width=5.5cm,clip]
{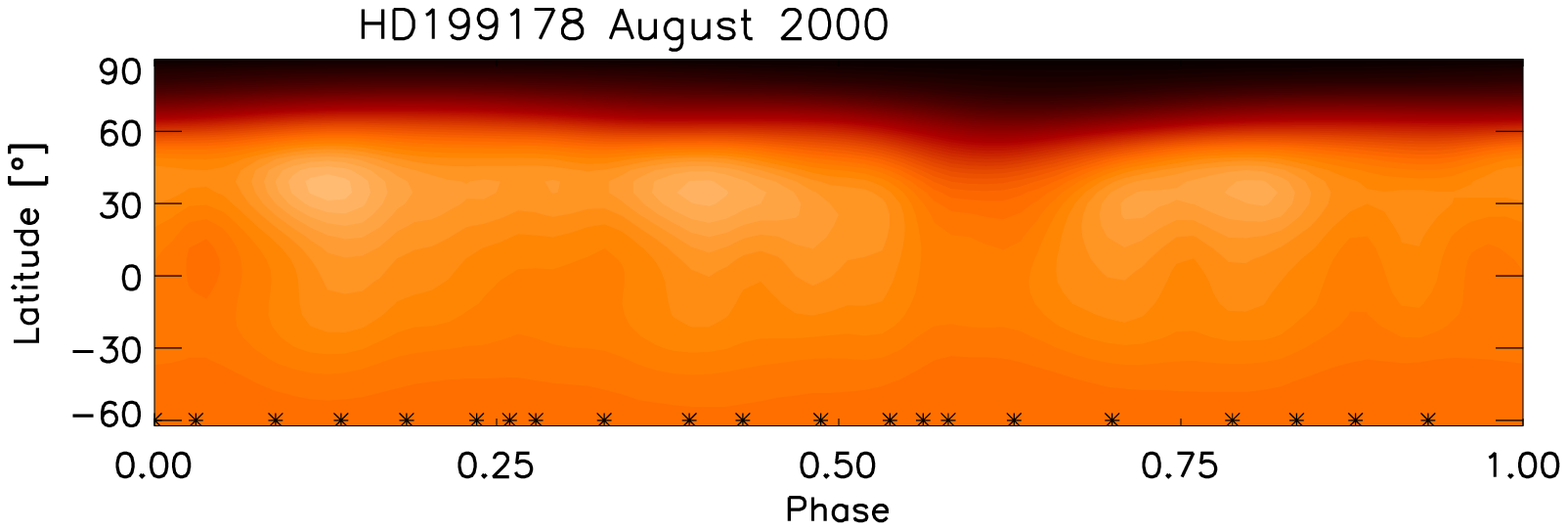}
\includegraphics[width=5.5cm,clip]
{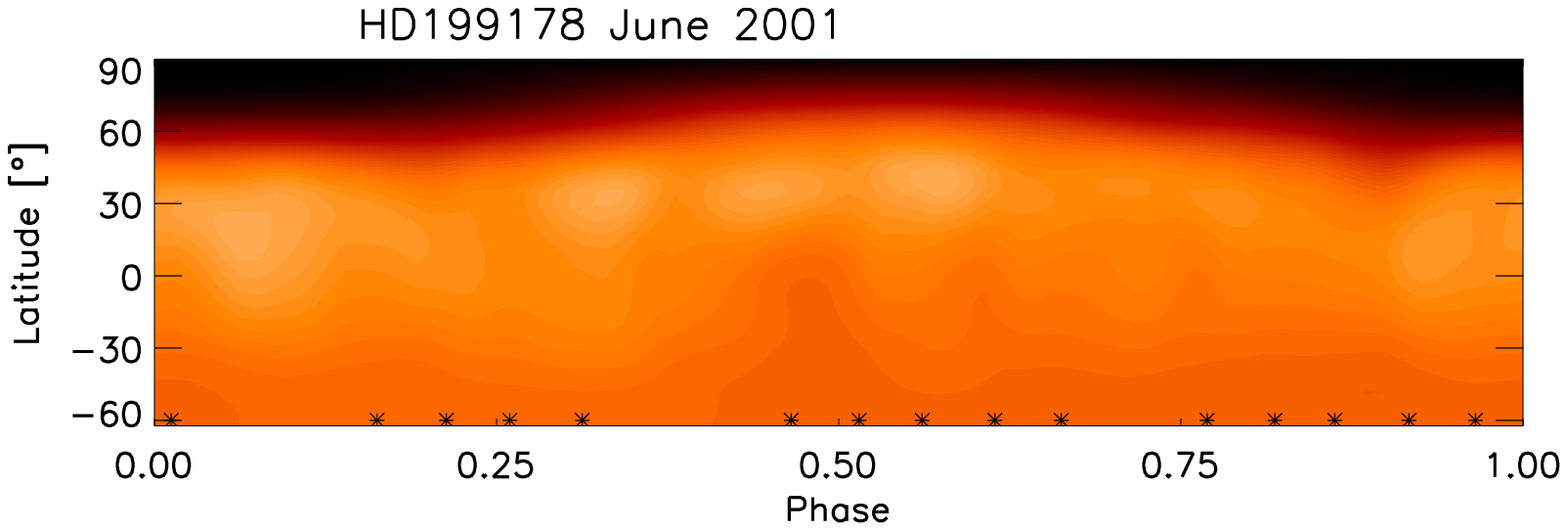}
\includegraphics[width=5.5cm,clip]
{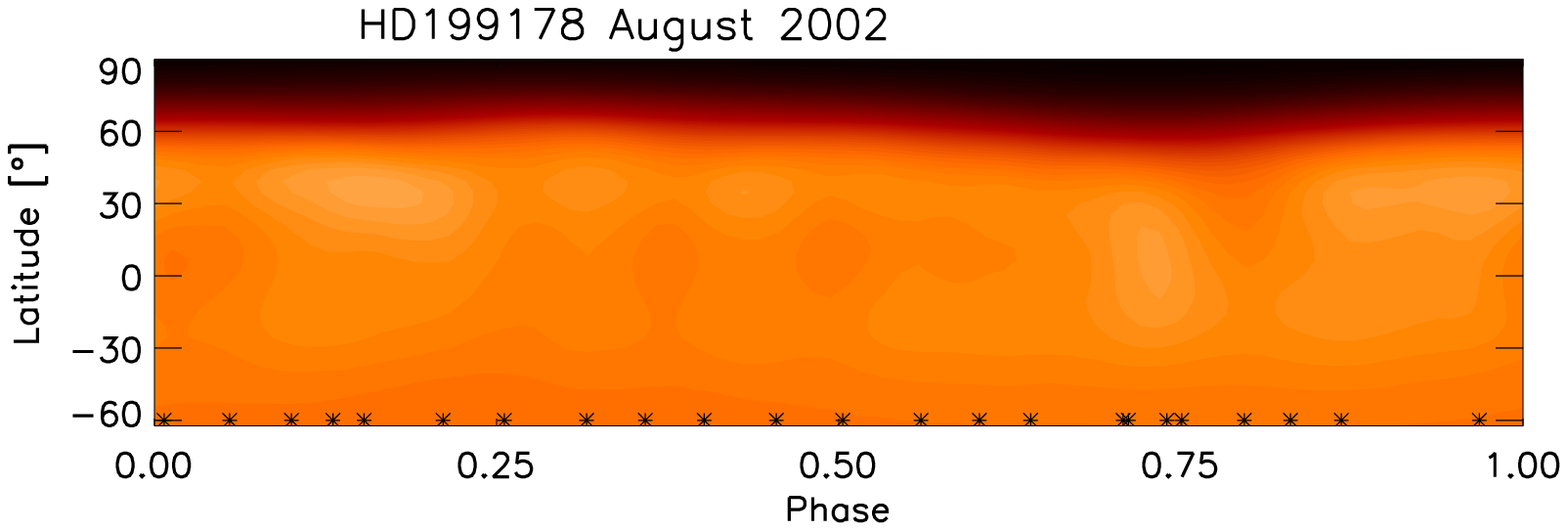}

\vspace{-2.25cm}

\includegraphics[width=5.5cm,clip]
{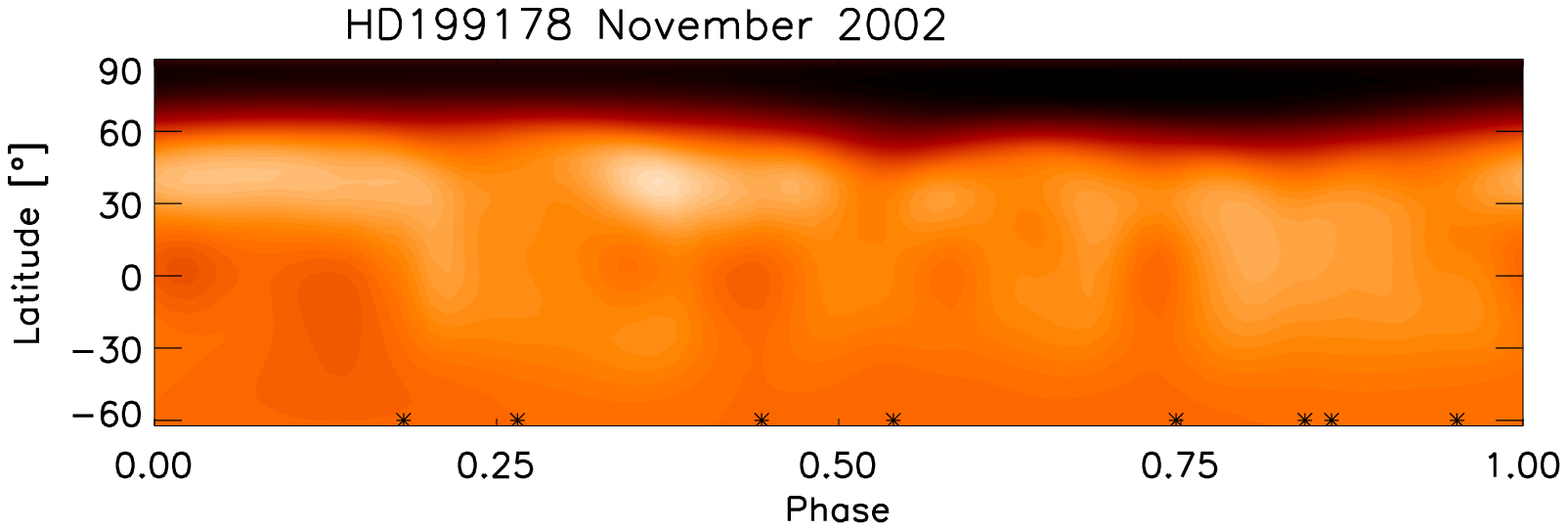}
\includegraphics[width=5.5cm,clip]
{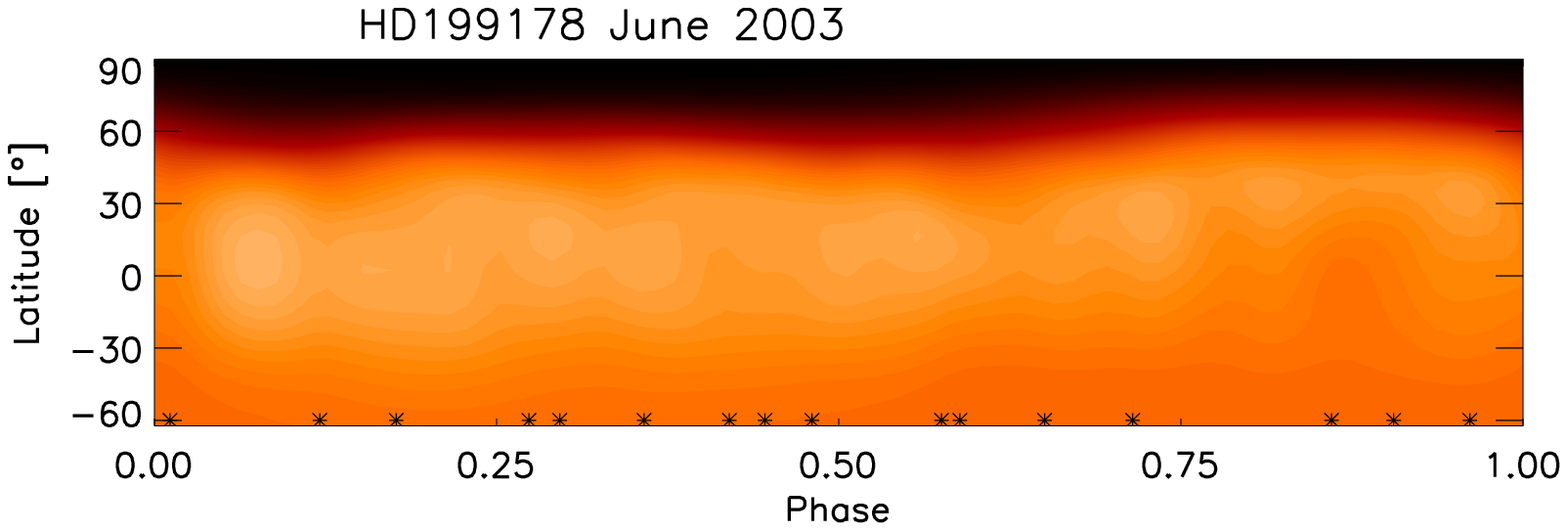}
\includegraphics[width=5.5cm,clip]
{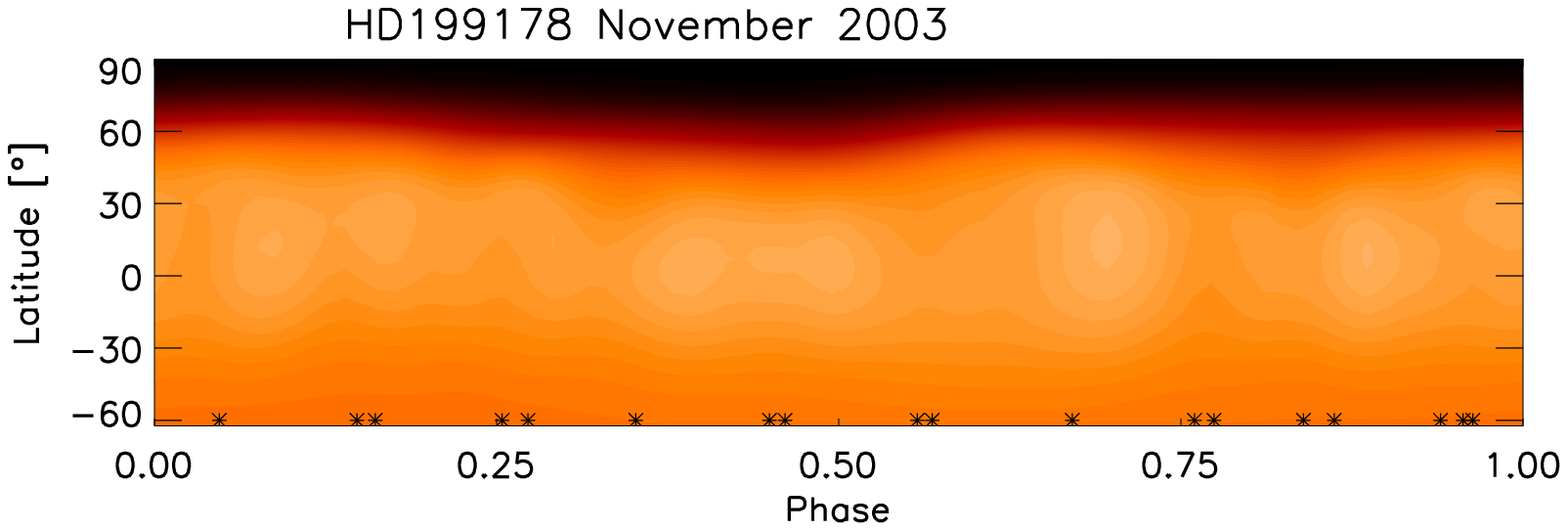}

\vspace{-2.25cm}

\includegraphics[width=5.5cm,clip]
{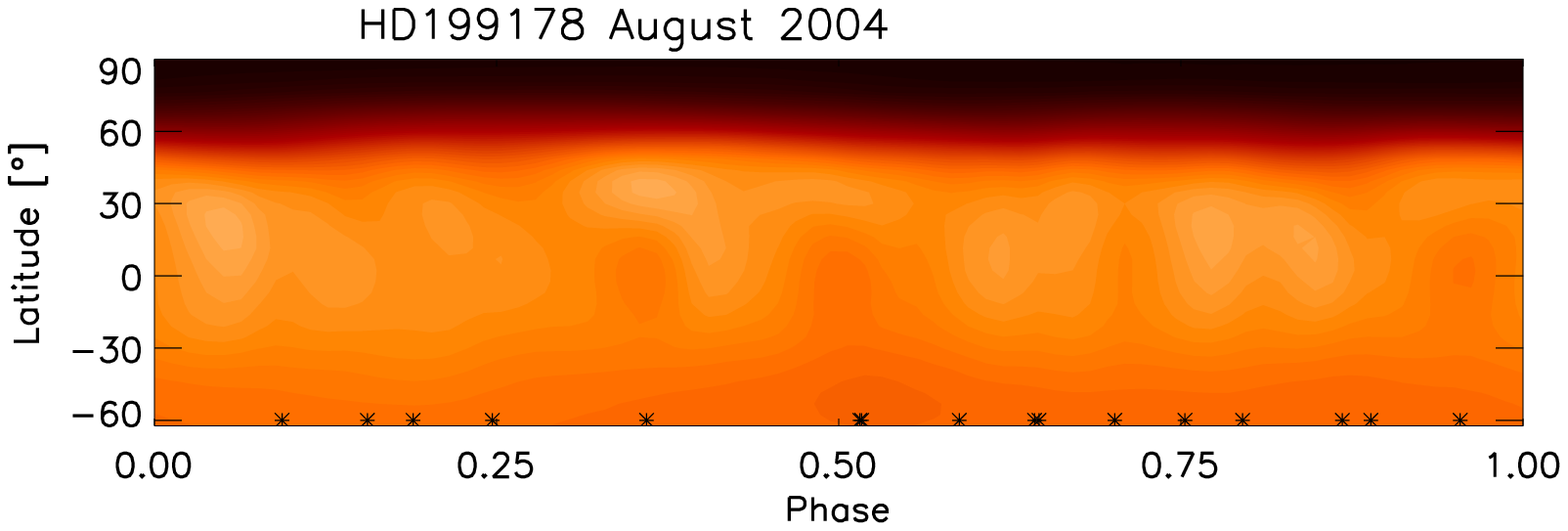}
\includegraphics[width=5.5cm,clip]
{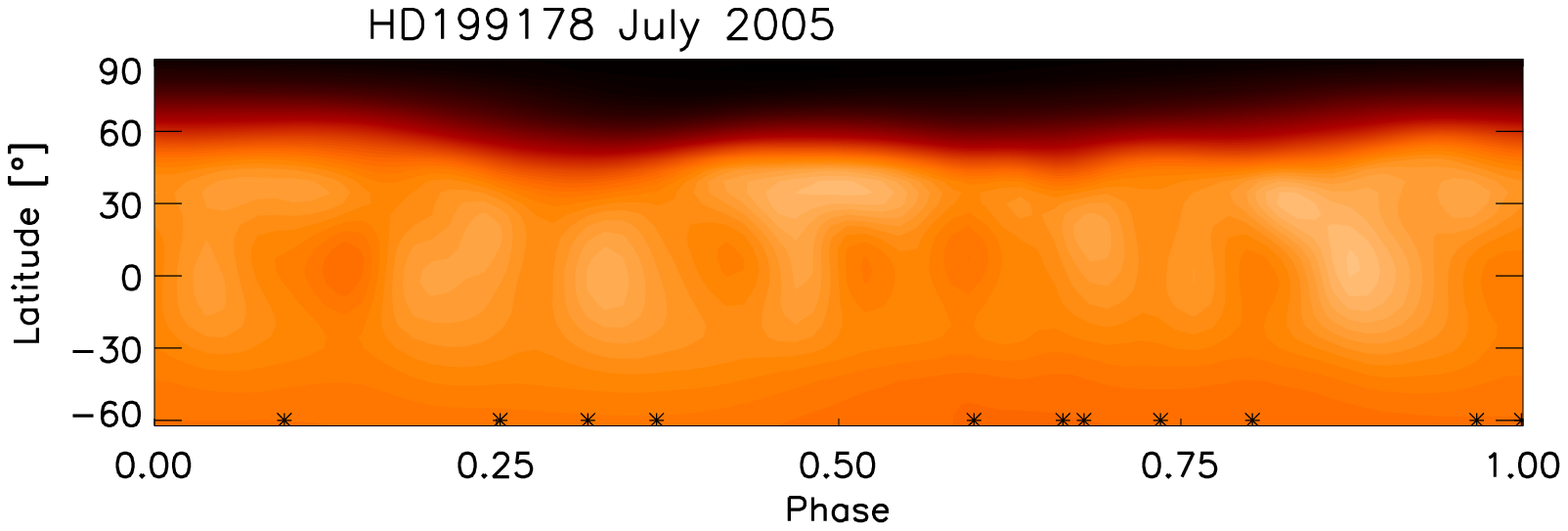}
\includegraphics[width=5.5cm,clip]
{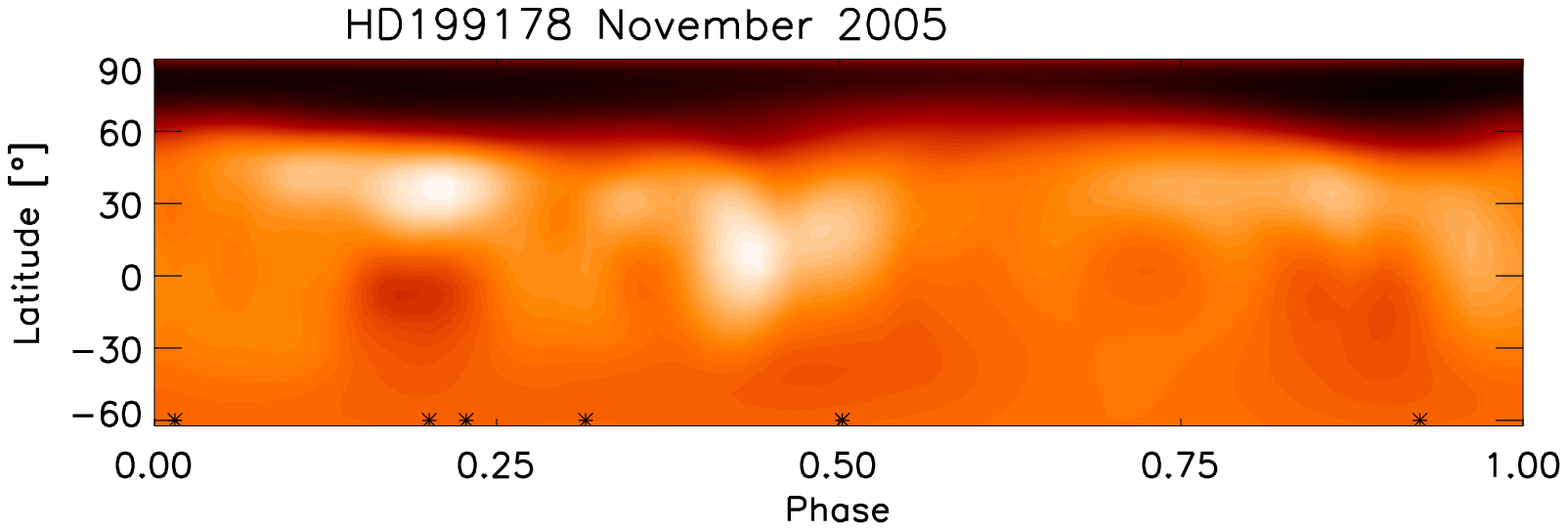}

\vspace{-2.25cm}

\includegraphics[width=5.5cm,clip]
{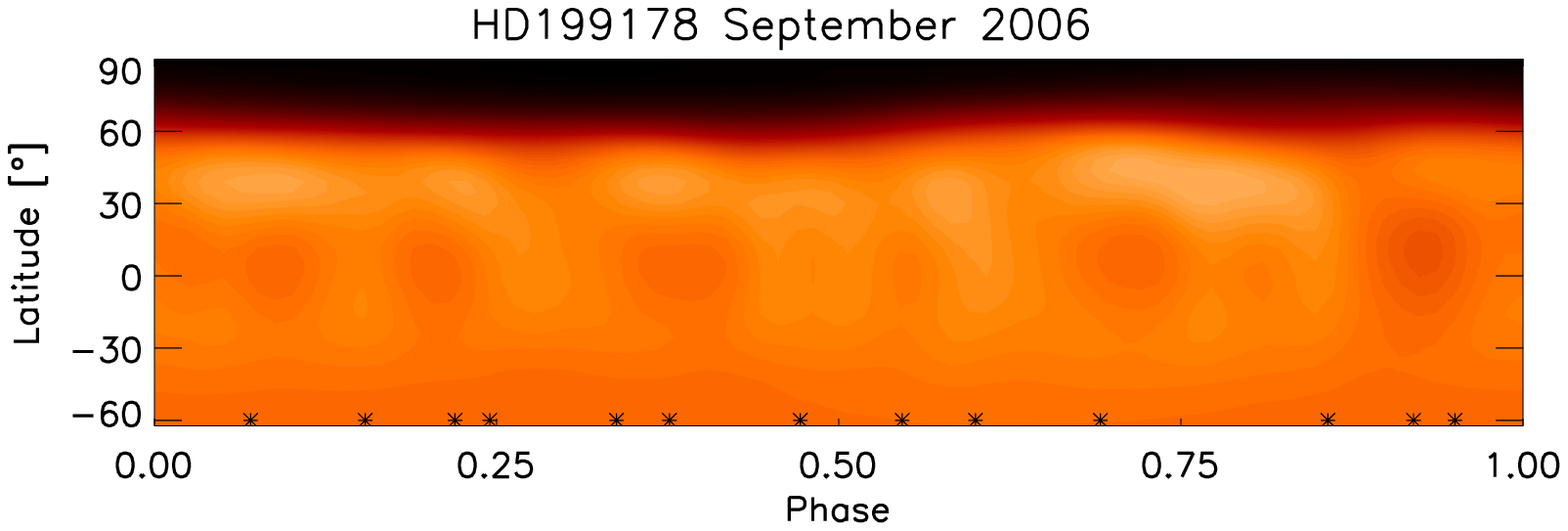}
\includegraphics[width=5.5cm,clip]
{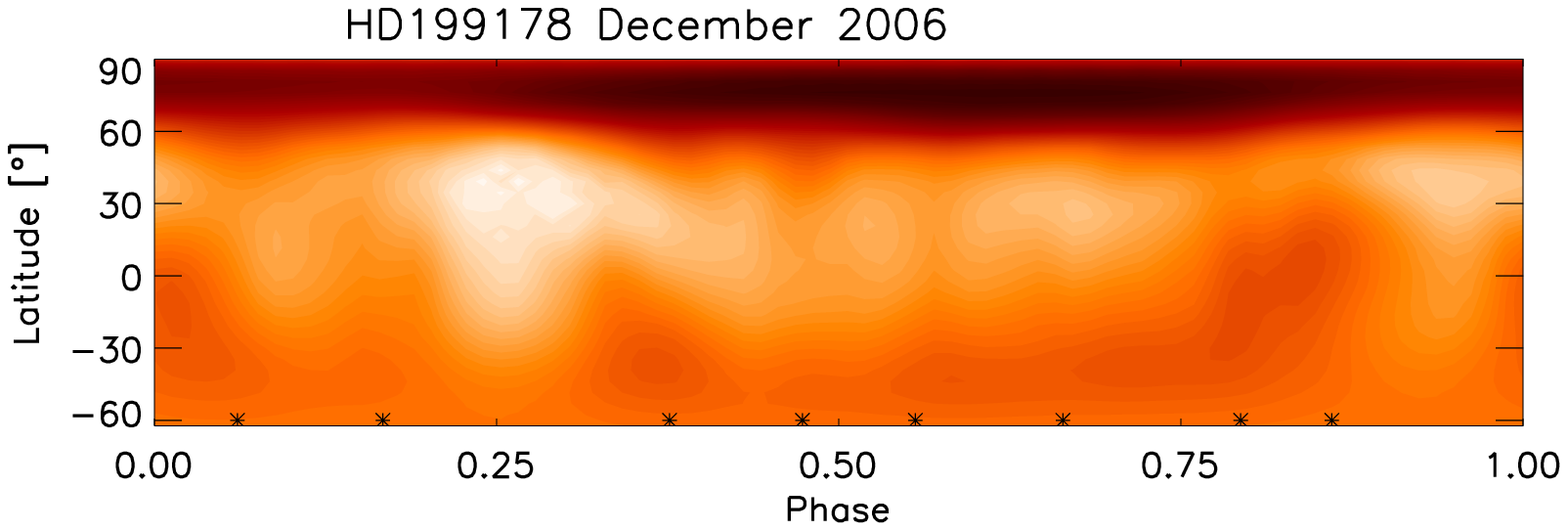}
\includegraphics[width=5.5cm,clip]
{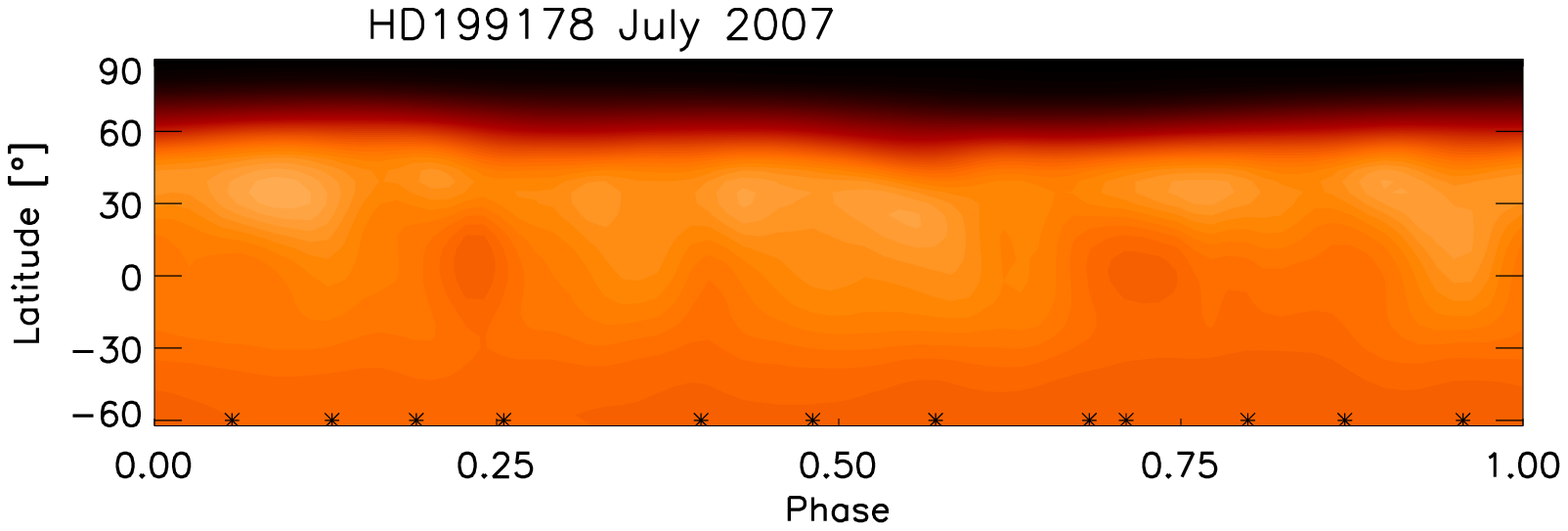}

\vspace{-2.25cm}

\includegraphics[width=5.5cm,clip]
{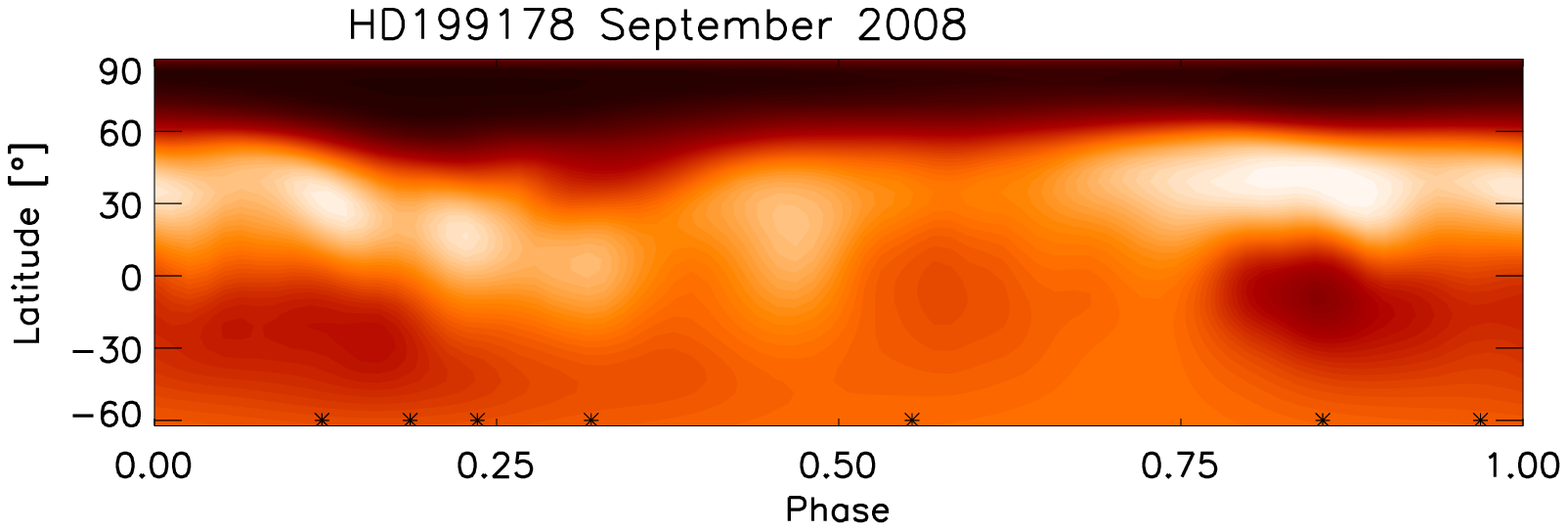}
\includegraphics[width=5.5cm,clip]
{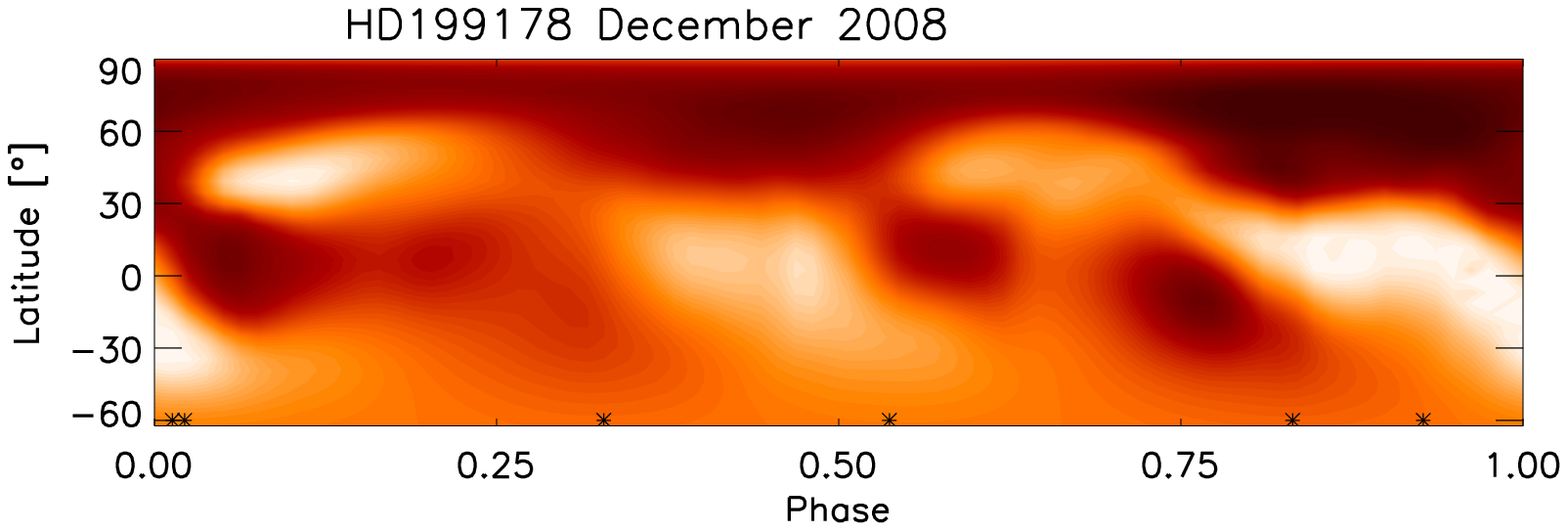}
\includegraphics[width=5.5cm,clip]
{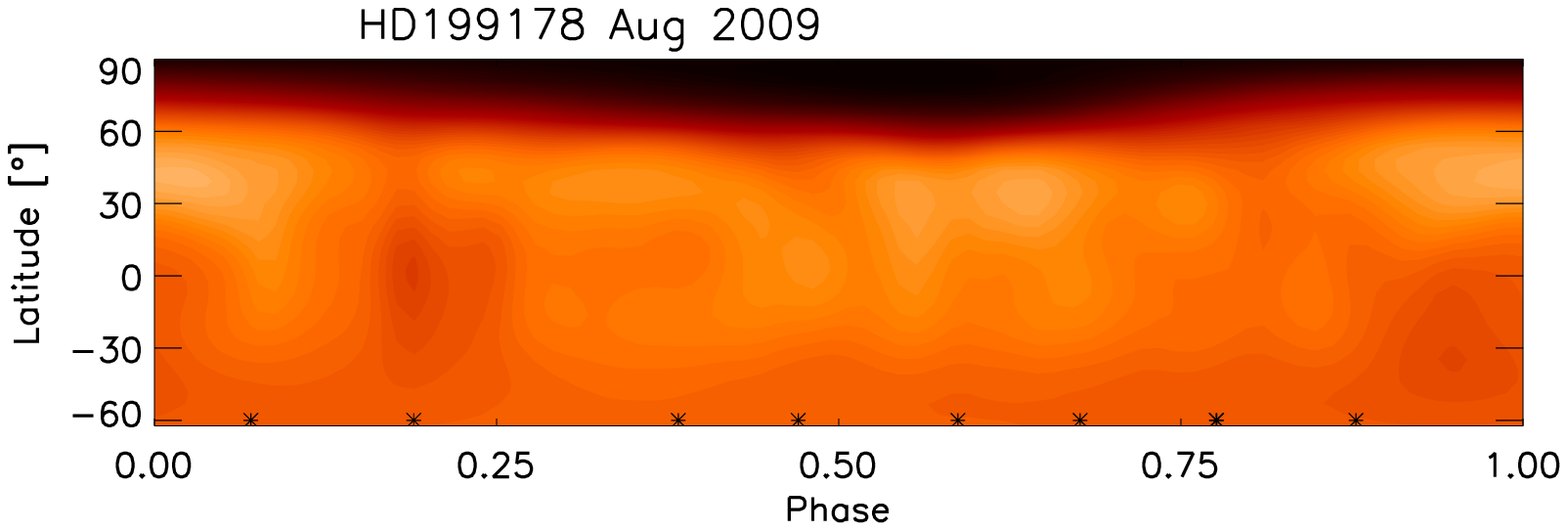}

\vspace{-2.25cm}

\includegraphics[width=5.5cm,clip]
{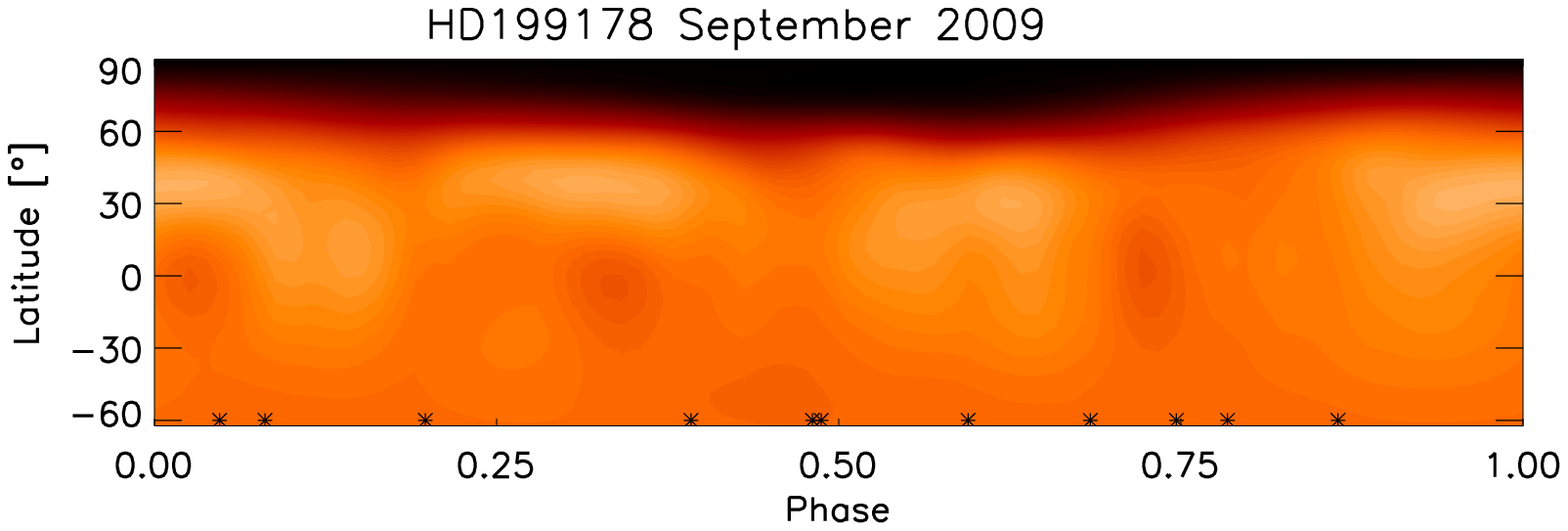}
\includegraphics[width=5.5cm,clip]
{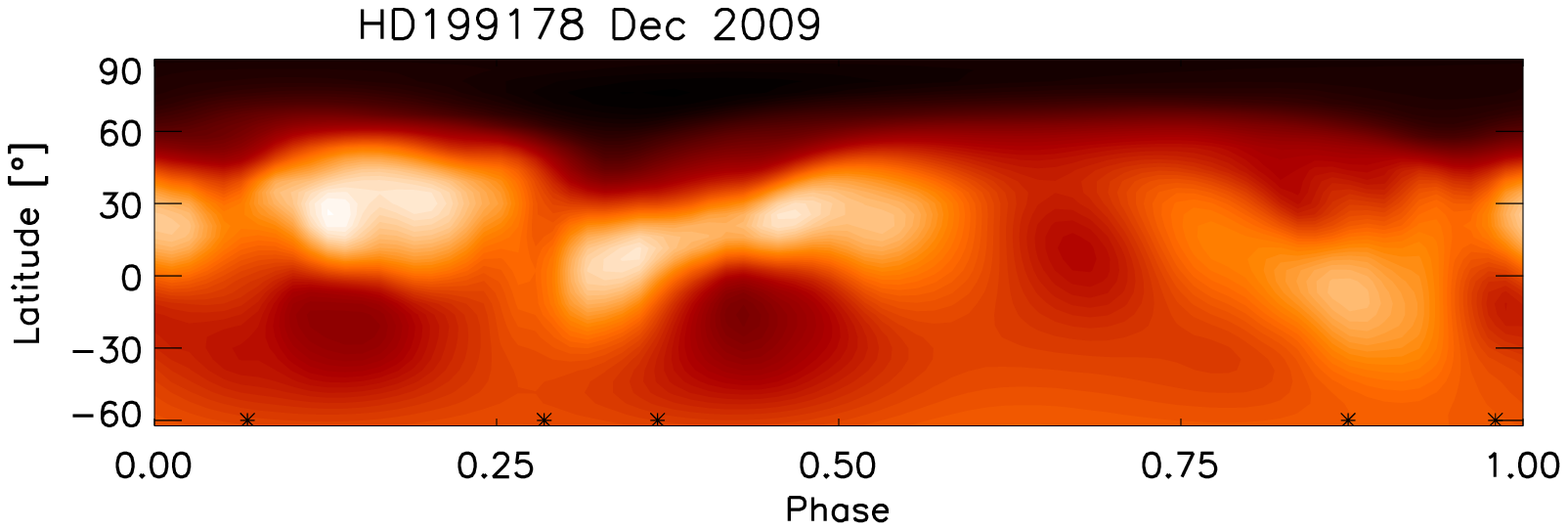}
\includegraphics[width=5.5cm,clip]
{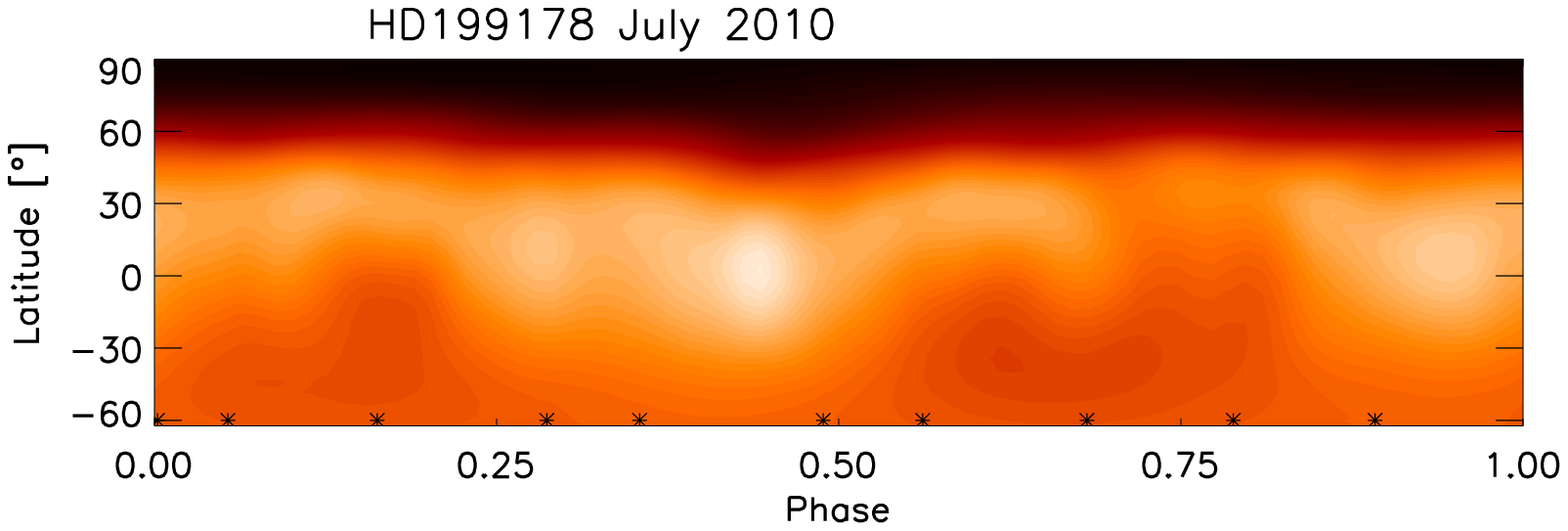}

\vspace{-2.25cm}

\includegraphics[width=5.5cm,clip]
{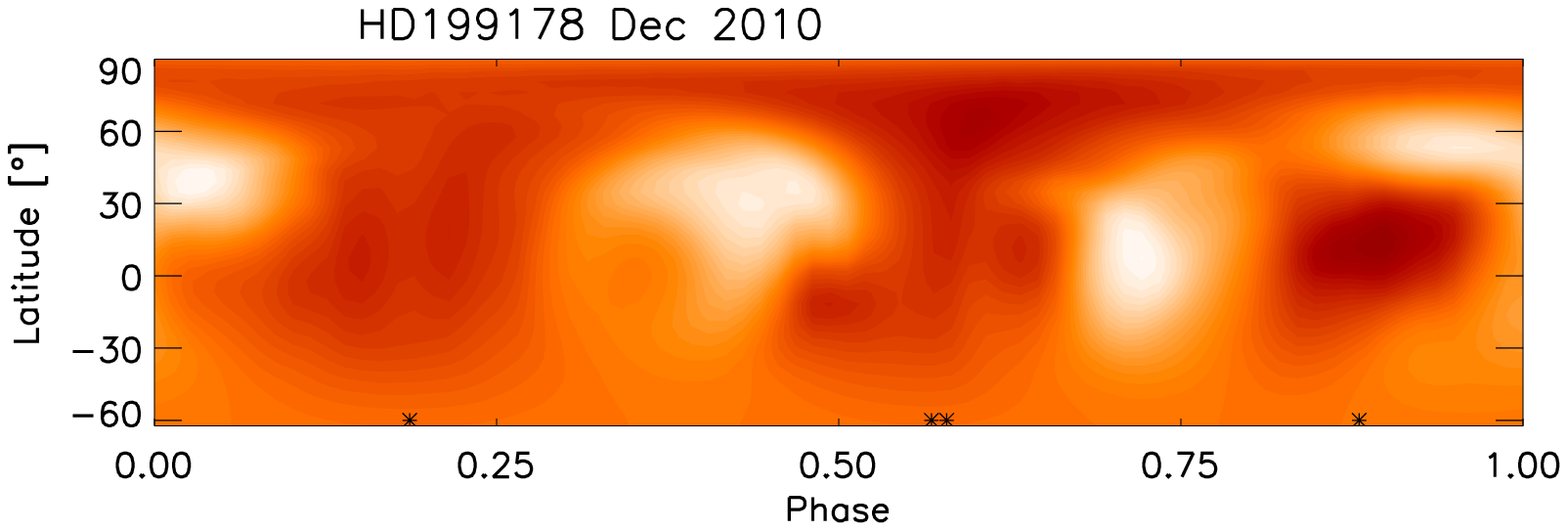}
\includegraphics[width=5.5cm,clip]
{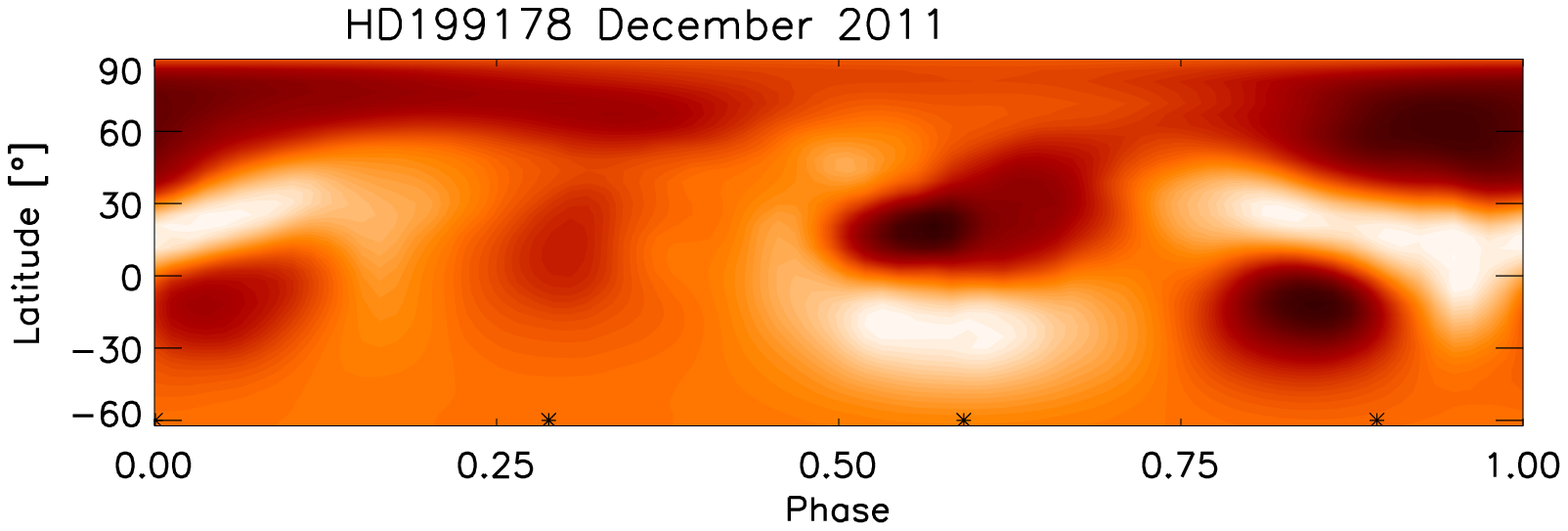}
\includegraphics[width=5.5cm,clip]
{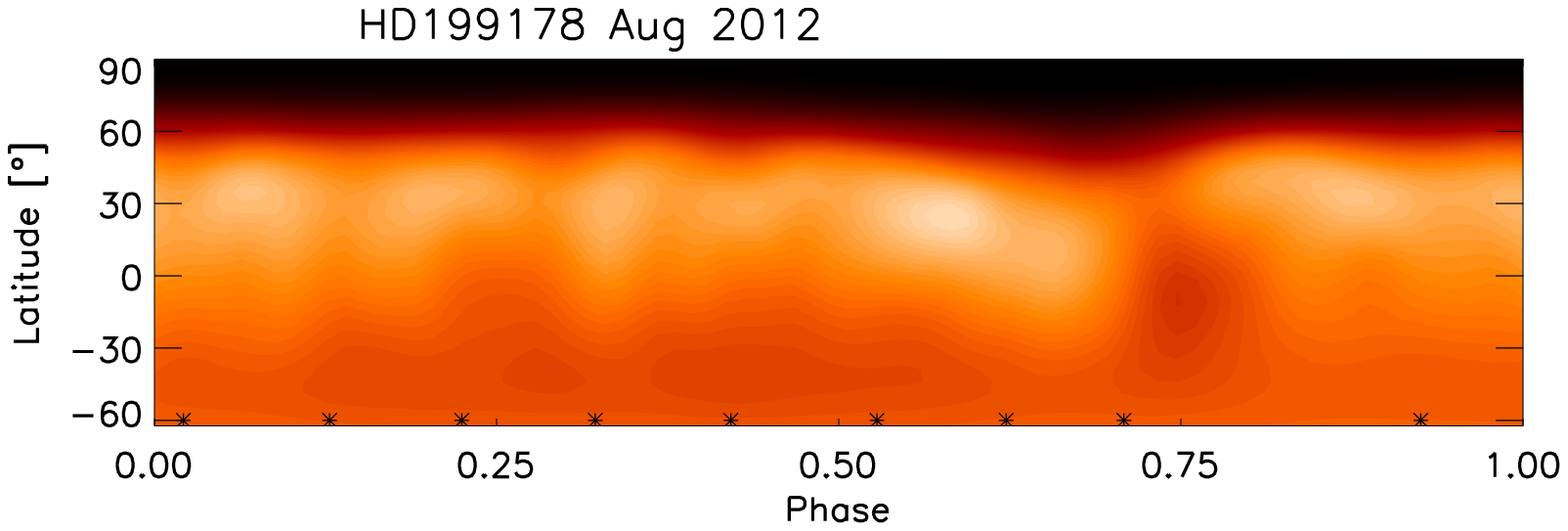}

\vspace{-2.25cm}

\includegraphics[width=5.5cm,clip]
{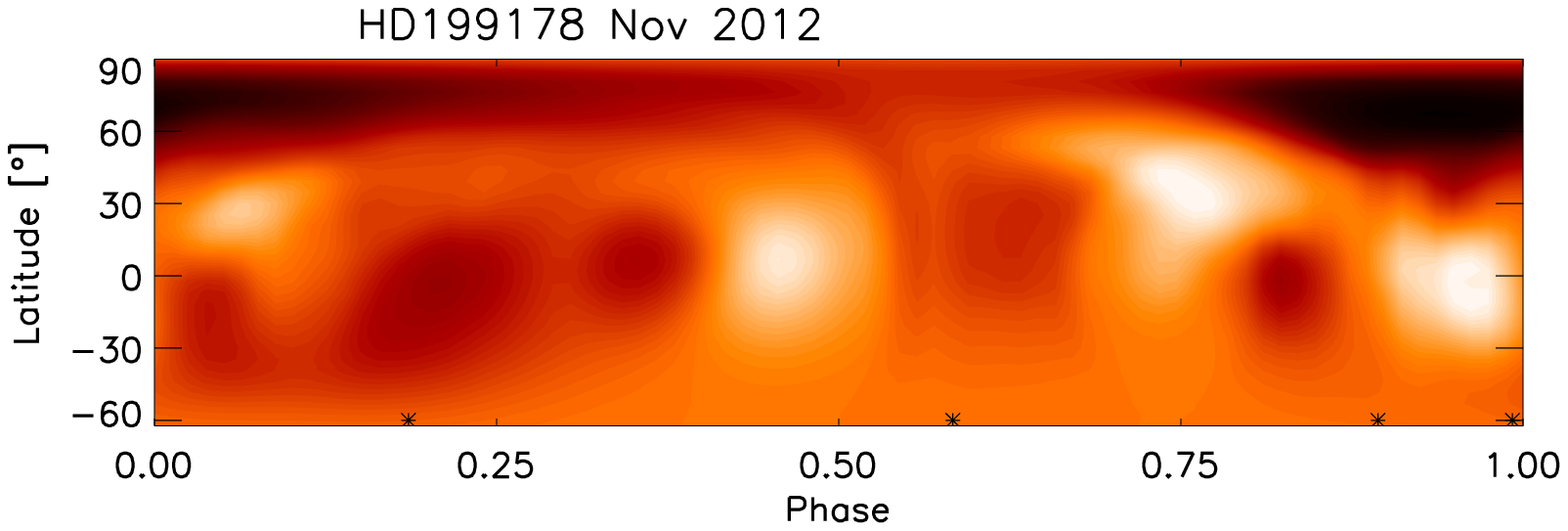}
\includegraphics[width=5.5cm,clip]
{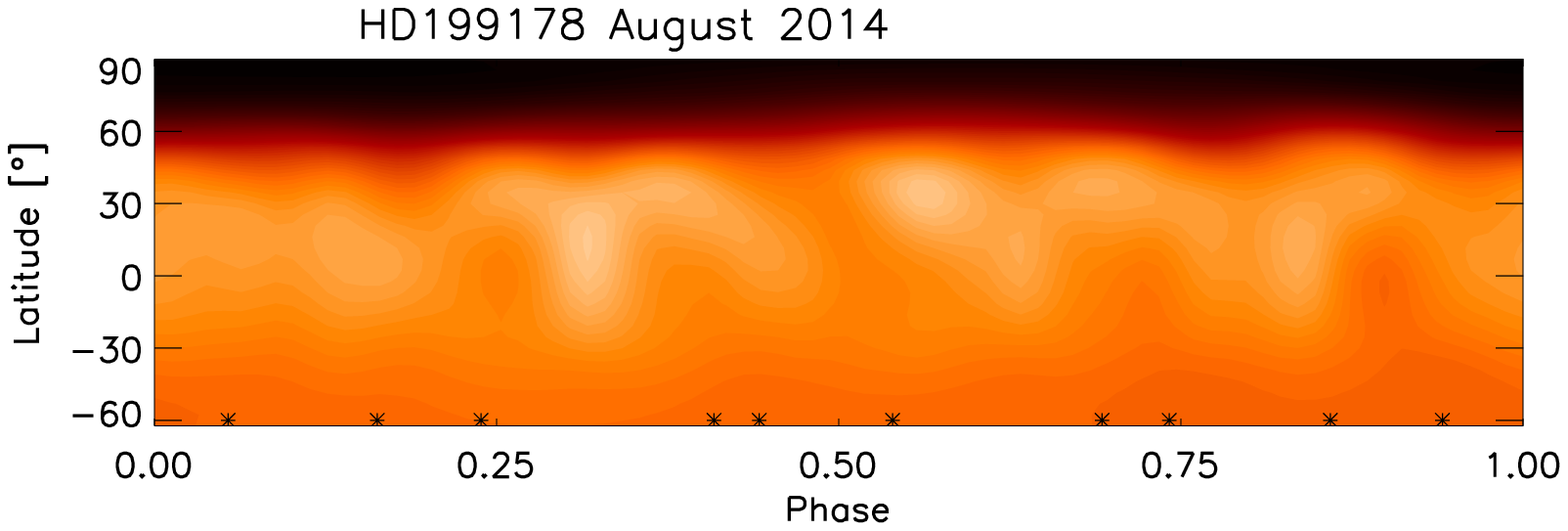}
\includegraphics[width=5.5cm,clip]
{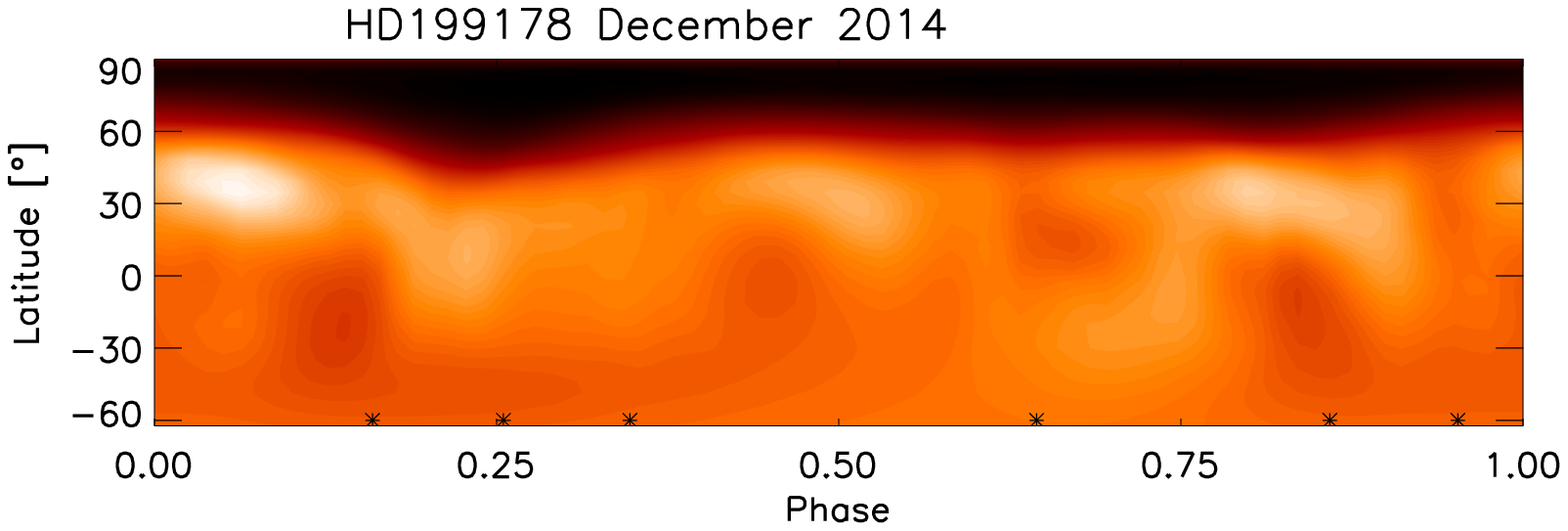}

\vspace{-2.25cm}

\includegraphics[width=5.5cm,clip]
{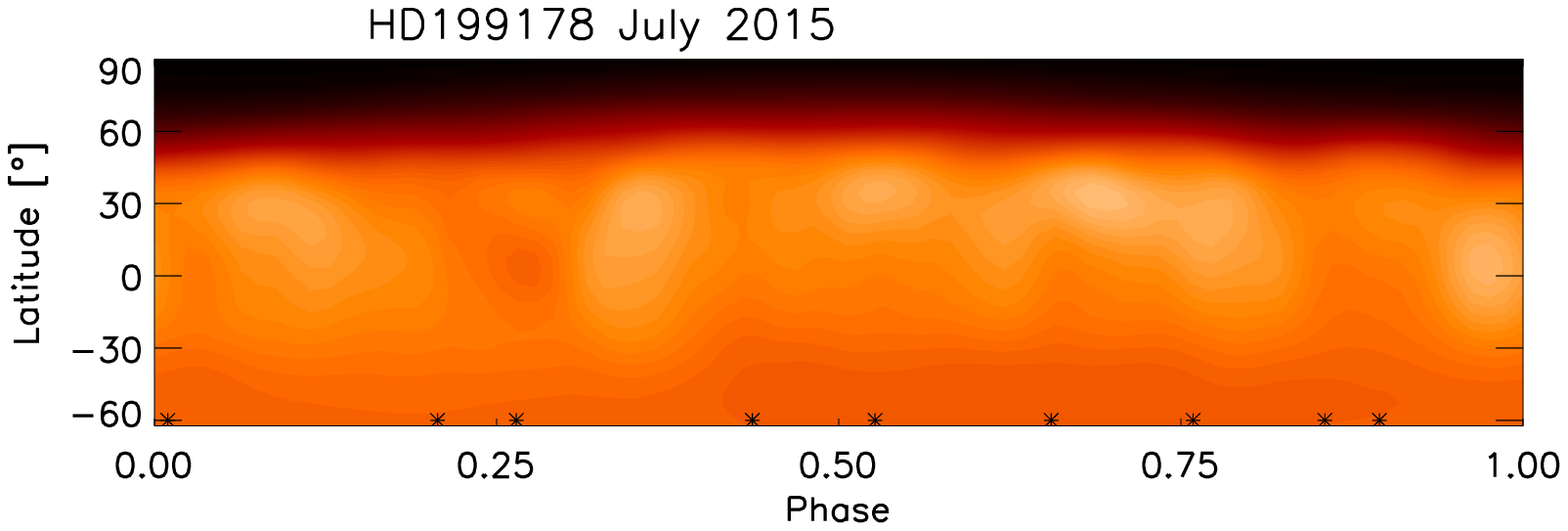}
\includegraphics[width=5.5cm,clip]
{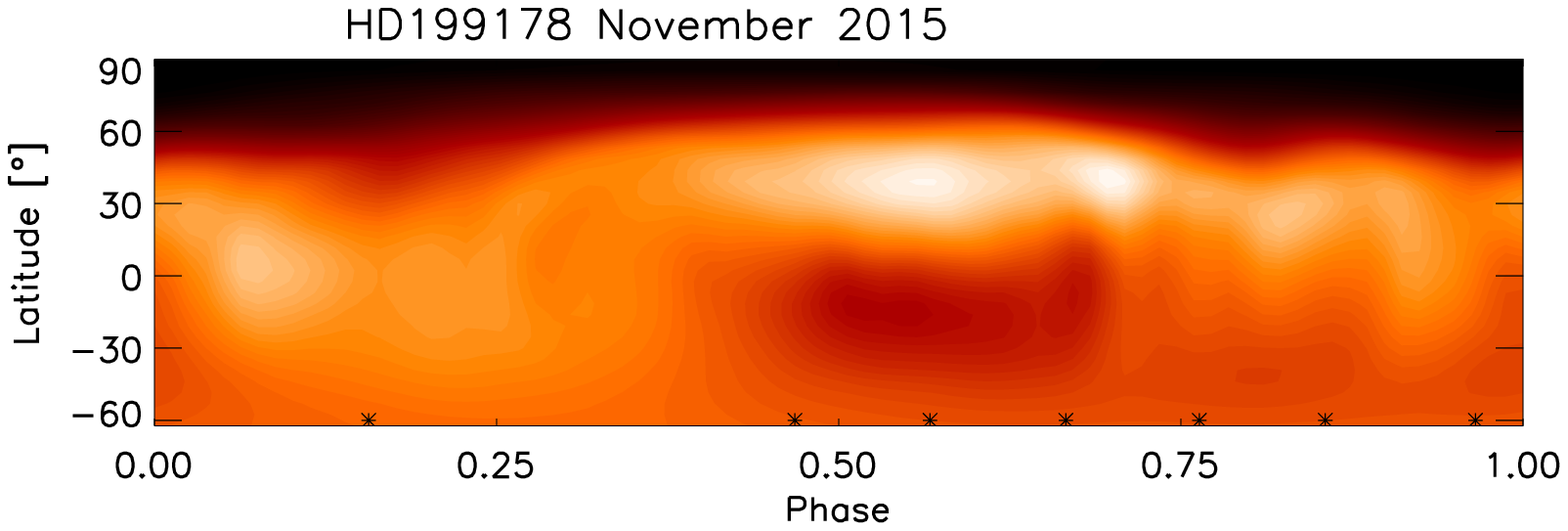}
\includegraphics[width=5.5cm,clip]
{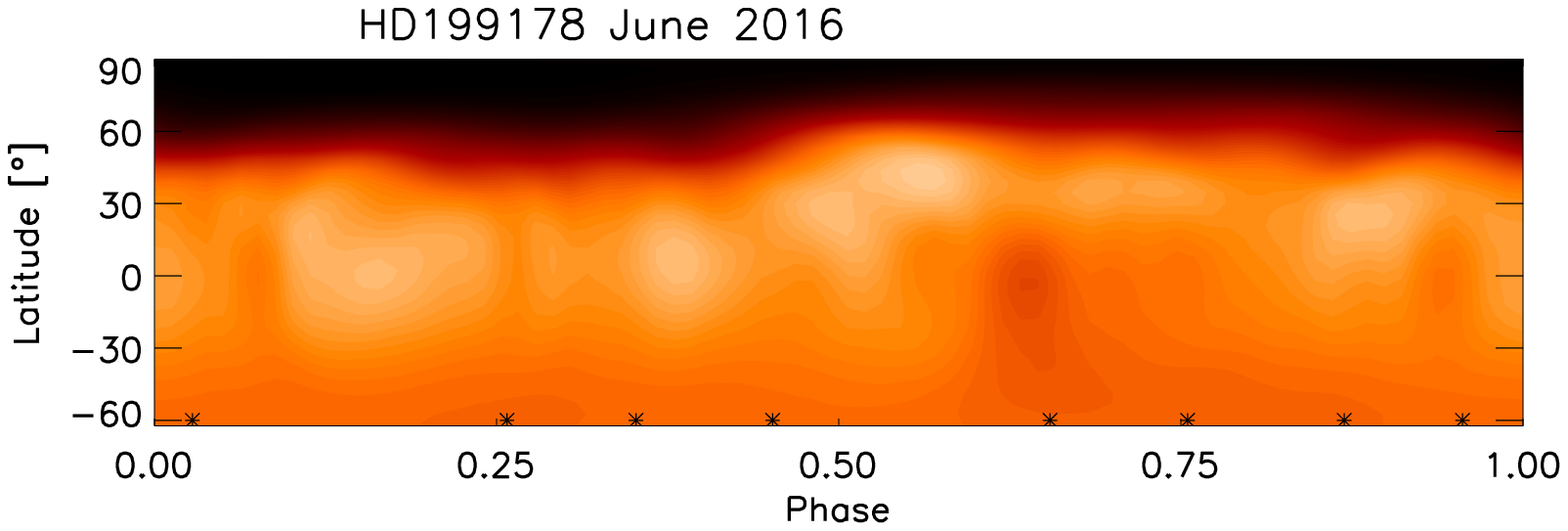}

\vspace{-2.25cm}

\includegraphics[width=5.5cm,clip]
{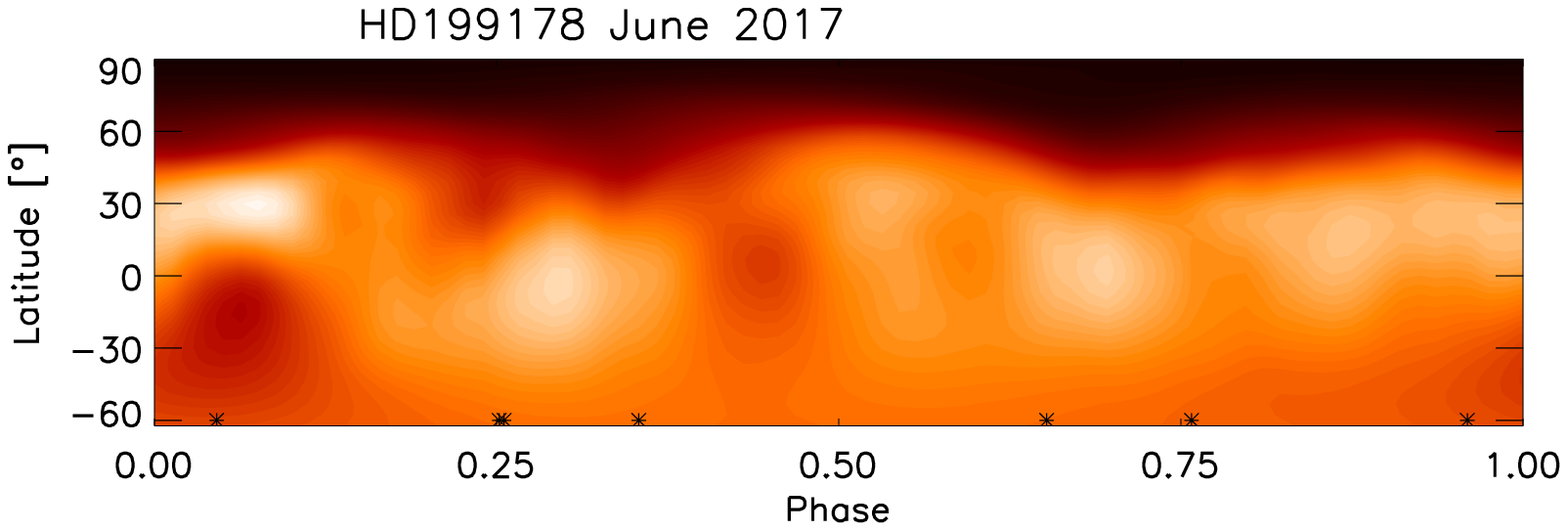}
\includegraphics[width=5.5cm,clip]
{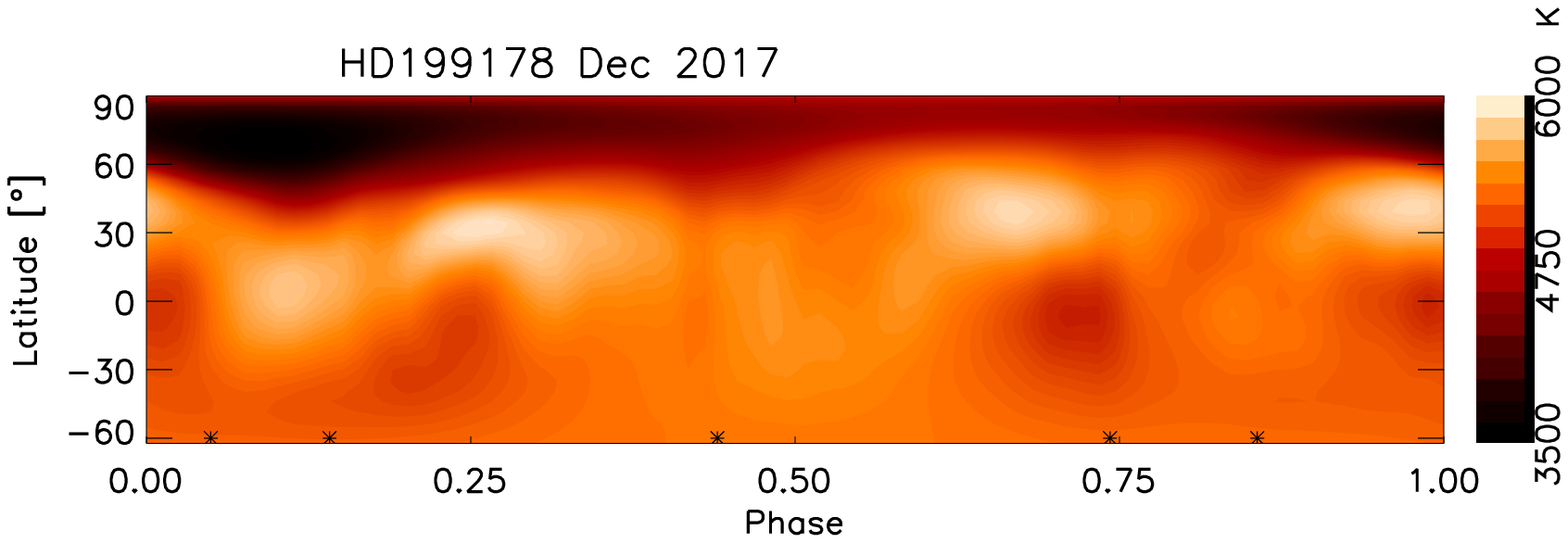}

\vspace{-2.4cm}

      \caption{Doppler imaging surface temperature maps. The observed phases are indicated with `*' 
and the same temperature scale is used in all images.}
         \label{di}
   \end{figure*}

The August 2009 and September 2009 observations are actually simultaneous; there is  a difference in $\langle$HJD$\rangle$ of only one day. However, the images were derived using different spectral set-ups. This was a robust test for
differences in the resulting DI caused by using different spectral lines and
phase coverage. The polar spots in these images are almost identical,
except for slight differences around phases near  $\phi = 0$. This difference
can be explained by more observations near this phase in the case of 
September 2009. Otherwise, the only visible differences in these images are the weak lower latitude features, which are thus most probably artefacts.

\section{Results}

We see high-latitude spot structures in all DIs (Fig. \ref{di}). 
If maps based on
observations with $f_\phi < 50$\%  are regarded as unreliable, we can also 
conclude that all reliable images show a large polar spot structure. 
This structure is varying in time, and the rotationally modulated signal is 
dominated by annexes extending towards lower latitudes. This polar spot
structure is usually limited to latitudes $\theta > 60 \degree$; sometimes the annex 
extends to $\theta \approx 40 \degree$. There are also lower latitude
weaker features, particularly during November 1998, September 2008, November 
2015, and June 2017, which could contribute to the light curve modulation. 
However, these cases are all images with $f_\phi < 70$\% and the lower 
latitude cool spot structures have nearby hot structures, indicating that they
may be artefacts \citep[see][]{Hackman2001}.

Images near in time allow us to study  short-term changes in the spot 
configuration. In 1994 (July and August) and 1998 (October and November)
we have images separated by about one month. In both cases we see some small 
differences, but the overall spot configuration remains unchanged. Images with 
a longer time difference (e.g. May and July 1999) already show
significant differences that cannot be explained merely by phase shifts
caused by rotation period uncertainties.

\begin{figure}
\includegraphics[width=9cm,clip]
{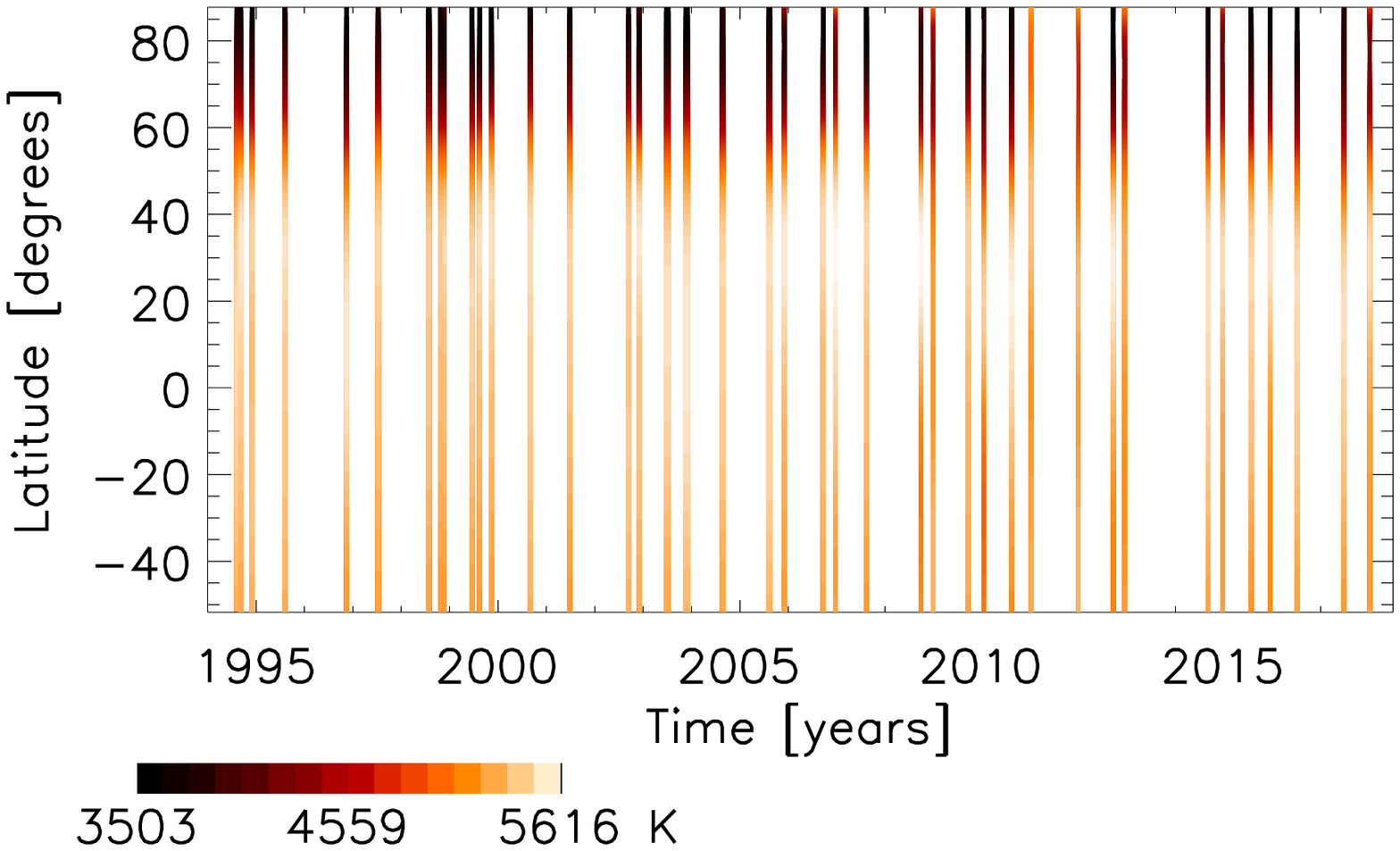}

      \caption{DIs averaged over longitudes vs. dates: Time-latitude 
diagram.}
         \label{lattime}
   \end{figure}

\begin{figure}
\includegraphics[width=9cm,clip]
{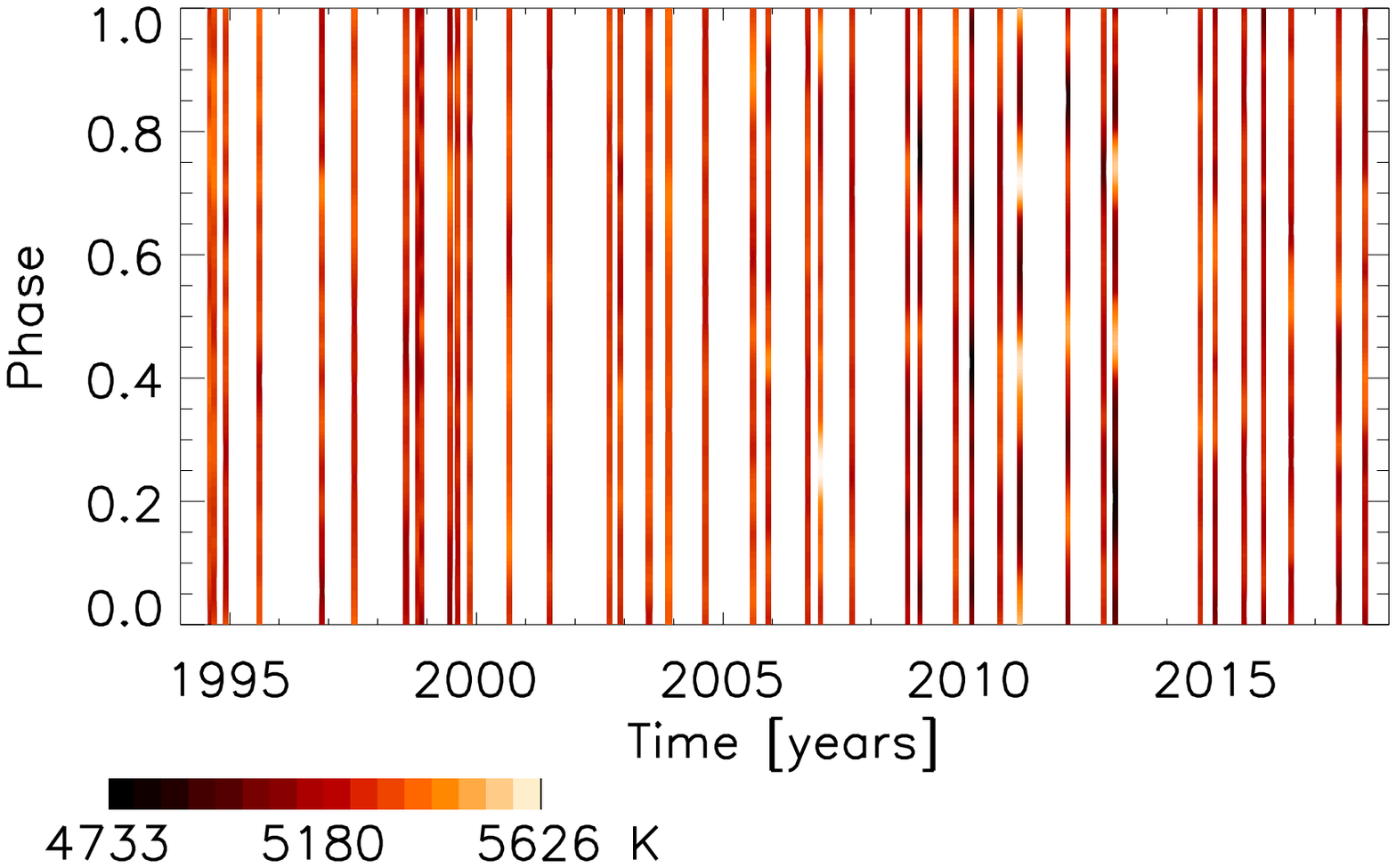}

\caption{DIs averaged over latitudes vs. dates: Time-longitude diagram.}
\label{phasetime}
\end{figure}

To study the long-term spot evolution we present the temporal variation
in  the longitudinally and latitudinally averaged temperatures (Figs. 
\ref{lattime} 
and \ref{phasetime}). The former  is an analogue to the butterfly 
diagram, while the latter  can be used to detect stable or drifting 
active longitudes. These structures have been revealed  in observations
\citep[see e.g.][]{Lindborg2011,Hackman2011} and in numerical simulations 
\citep{Cole2014,Viviani2018}.

In Fig. \ref{phasetime} the phases
correspond to the frame rotating according to the ephemeris of Eq. \ref{ephem}.
Figure \ref{lattime} indicates a smooth behaviour, when disregarding the 
most unreliable images based on insufficient phase coverage ($f_\phi < 50$\%).
Figure \ref{phasetime} gives no indication of active longitudes.

\begin{figure}
\includegraphics[width=7cm,clip]
{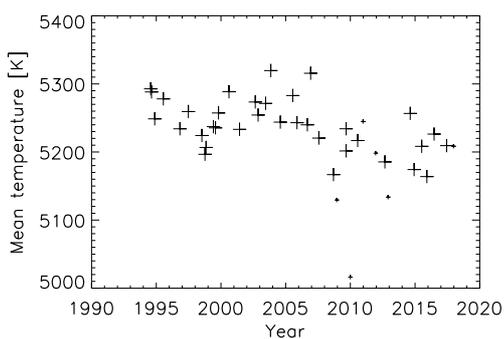}

      \caption{Mean surface temperature $T_\mathrm{mean}$ of the DI maps. Reliable maps 
($f_\phi \ge 0.5$) are represented with larger symbols than the unreliable
ones.}
         \label{meant}
   \end{figure}

\begin{figure}
\includegraphics[width=7cm,clip]
{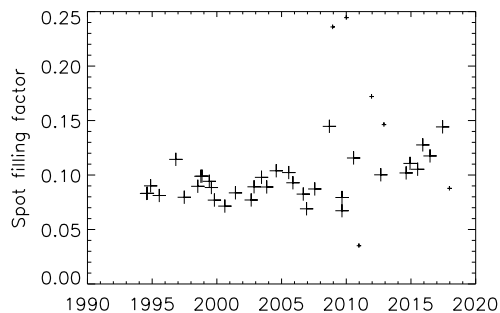}

\caption{Spot filling factors. Reliable maps  ($f_\phi \ge 0.5$) are 
represented with larger symbols than the unreliable ones.}
\label{spfill}
\end{figure}

We also calculated the mean surface temperatures $T_\mathrm{mean}$ and spot 
filling factors $f_\mathrm{spot}$ for each DI. We used $T_\mathrm{spot}=4800$\,K as 
the spot threshold. This is 500\,K lower than the assumed unspotted surface 
$T_\mathrm{eff}=5300$ K.
The results are plotted in Figs. \ref{meant} and \ref{spfill}. The
linear Pearson's correlation between $T_\mathrm{mean}$ and $f_\mathrm{spot}$
of the reliable maps ($f_\phi > 0.5$) was
$r(T_\mathrm{mean},f_\mathrm{spot})$$\approx$$-0.60$. It should be pointed out
that the chosen threshold is, in a sense, arbitrary, but the correlation did
not change significantly when using slightly different values. For example,  using
$T_\mathrm{spot}=5000$\,K gave $r(T_\mathrm{mean},f_\mathrm{spot})$$\approx$$-0.60$
and $T_\mathrm{spot}=4500$\,K yielded 
$r(T_\mathrm{mean},f_\mathrm{spot})$$\approx$$-0.55$.

\section{Conclusions}

We have presented an extremely long series of 41 DI surface temperature maps
of \object{HD 199178}, including all but one year during 1994--2017. The
dominating spot structure is a large spot/spot group covering the region
around the rotational pole and extending down to latitudes 40$\degree$ -- 
70$\degree$ depending on the season. This polar structure is clearly 
a stable phenomenon. It is usually off-centred from the rotational pole, thus 
causing the rotationally modulated variations in the observed line profiles of 
photospheric absorption lines. Some of our  images show spot activity at 
lower latitudes, but in all cases this can be caused by artefacts due to limited
phase coverage.

Some of our DIs are  nearly simultaneous with the ZDIs presented by 
\cite{Petit2004}. Assuming their surface magnetic radial fields are correct,
then any magnetic origin of the large cool polar structure
must come from smaller spot structures with opposite magnetic polarities. 
In this  case nearby opposite polarities would cancel out in a ZDI, leading to 
difficulties in characterising the magnetic field associated with the largest 
cool spots.
Another problem connected to correlating surface dark spots with magnetic 
fields is that the polarisation signal is weighted by the flux. 
\cite{Rosen2012} showed that this may lead to an underestimation of
the magnetic field in a cool spot.

Comparing images taken in the same years indicates that the spot
configuration is stable for about a month, but significant differences can 
occur after two months. This is in line with the time scales of 
significant light curve changes for \object{FK Comae} \citep{Hackman2013}.
By plotting the averages of the DIs over longitudes and latitudes (Figs.
\ref{lattime} and \ref{phasetime}) we studied the long-term evolution of
the spot configuration. The latitudinal spot distribution behaves smoothly
while the longitudinal distribution shows no indication of active longitudes.
This pattern differs somewhat from those derived for three other single 
rapidly rotating stars: FK Com \citep{Hackman2013}, \object{LQ Hya} 
\citep{Cole2015}, and \object{V899 Her} \citep{Willamo2018}. In these cases 
there were coherent structures visible for a few years, while HD 199178 seems 
more erratic in this sense.

The negative correlation between the mean 
temperature $T_\mathrm{mean}$ and spot filling factor $f_\mathrm{spot}$ shows 
that the variability is dominated by cool spots. The variations
in  $T_\mathrm{mean}$ and $f_\mathrm{spot}$ are not regular enough
to draw conclusions on any activity cycles. There is, 
however, a tendency of a decrease in $T_\mathrm{mean}$ since 2007 as 
well as a slight increase in $f_\mathrm{spot}$ since 2010.

By comparing two images derived from almost simultaneous data (August and 
September 2009) we demonstrate that the use of different wavelength regions will
not cause any significant systematic differences in the resulting DI maps.
This comparison also shows that the weaker contrast lower latitude structures
may be artefacts.

Despite previously reported differential rotation \citep[see e.g.][]{Petit2004},
we calculated our DI maps without differential rotation.
Since the DIs do not show any
reliable lower latitude spots, it is hard to even roughly estimate
the differential rotation of the star. The adopted differential
rotation parameter $\alpha$=0 also produced good fits for all seasons, and
tests with other values did not significantly improve the goodness of the fit.

Concerning the derivation of rotational parameters, we demonstrate that the
generally used approach of searching for the fastest or best convergence of the
DI solution may be misleading. In particular, the possible differential 
rotation of HD 199178 cannot be determined with this approach. 

\begin{acknowledgements}
This work has made use of the VALD database which is  operated at Uppsala University, the Institute of Astronomy RAS in Moscow, and the University of Vienna.
Part of the observations were obtained at the National Astronomical 
Observatory, Rozhen, Bulgaria. J.J.L. and M.J.K. were partially supported by 
the Academy of Finland Centre of Excellence ReSoLVE (project 307411).
O.K. acknowledges support
from the Knut and Alice Wallenberg Foundation, the Swedish Research Council,
and the Swedish National Space Board. The work of T.W. was 
financed by the Emil Aaltonen Foundation.
We thank the referee Dr. Pascal Petit for the valuable and constructive
suggestions on how to improve the paper. 

\end{acknowledgements}

\bibliography{hackman}

\begin{appendix}
\section{Observed and modelled spectra}
\label{spectra}

\begin{figure*}
\centering
\label{spec1}

\includegraphics[width=6cm,clip]{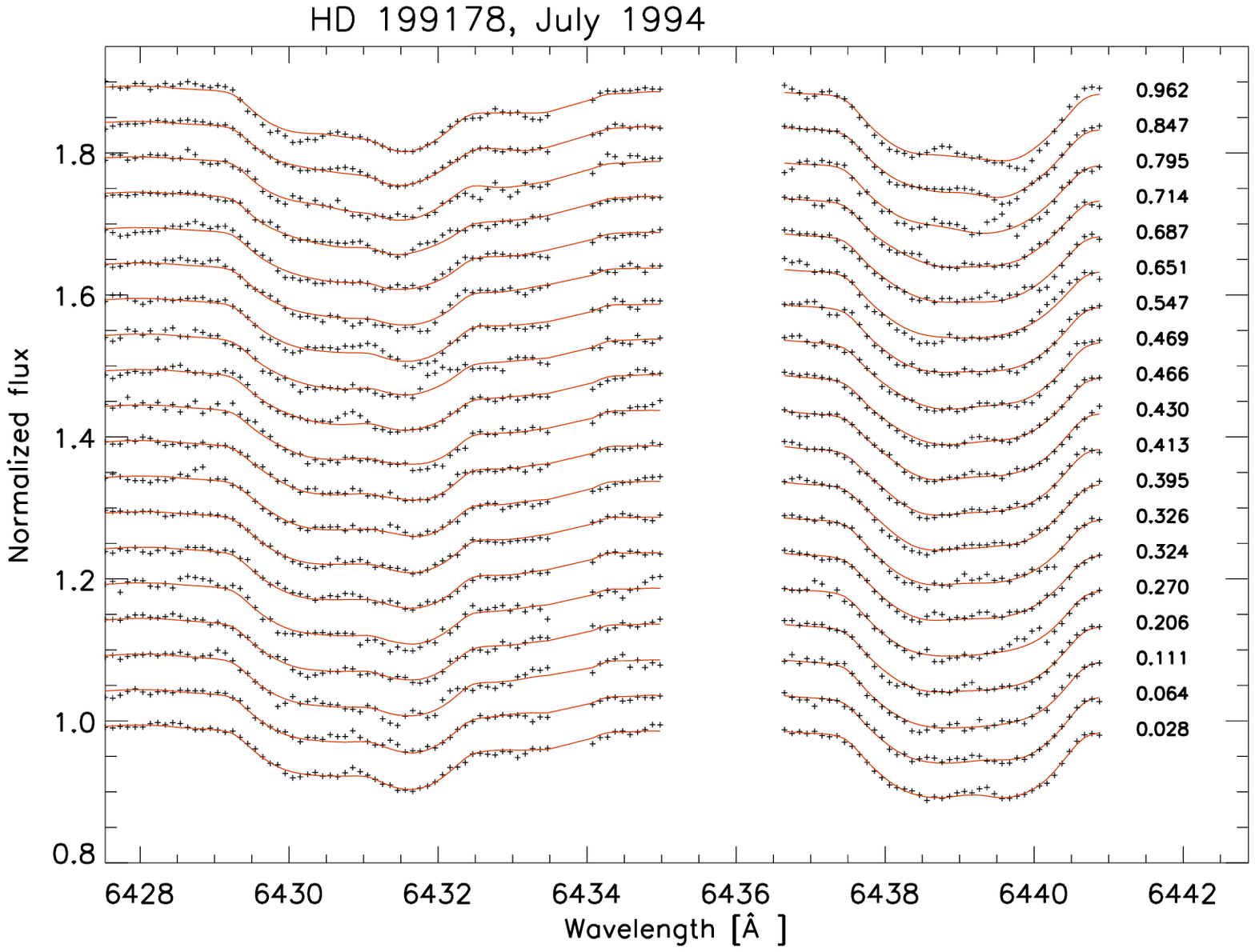}
\includegraphics[width=6cm,clip]{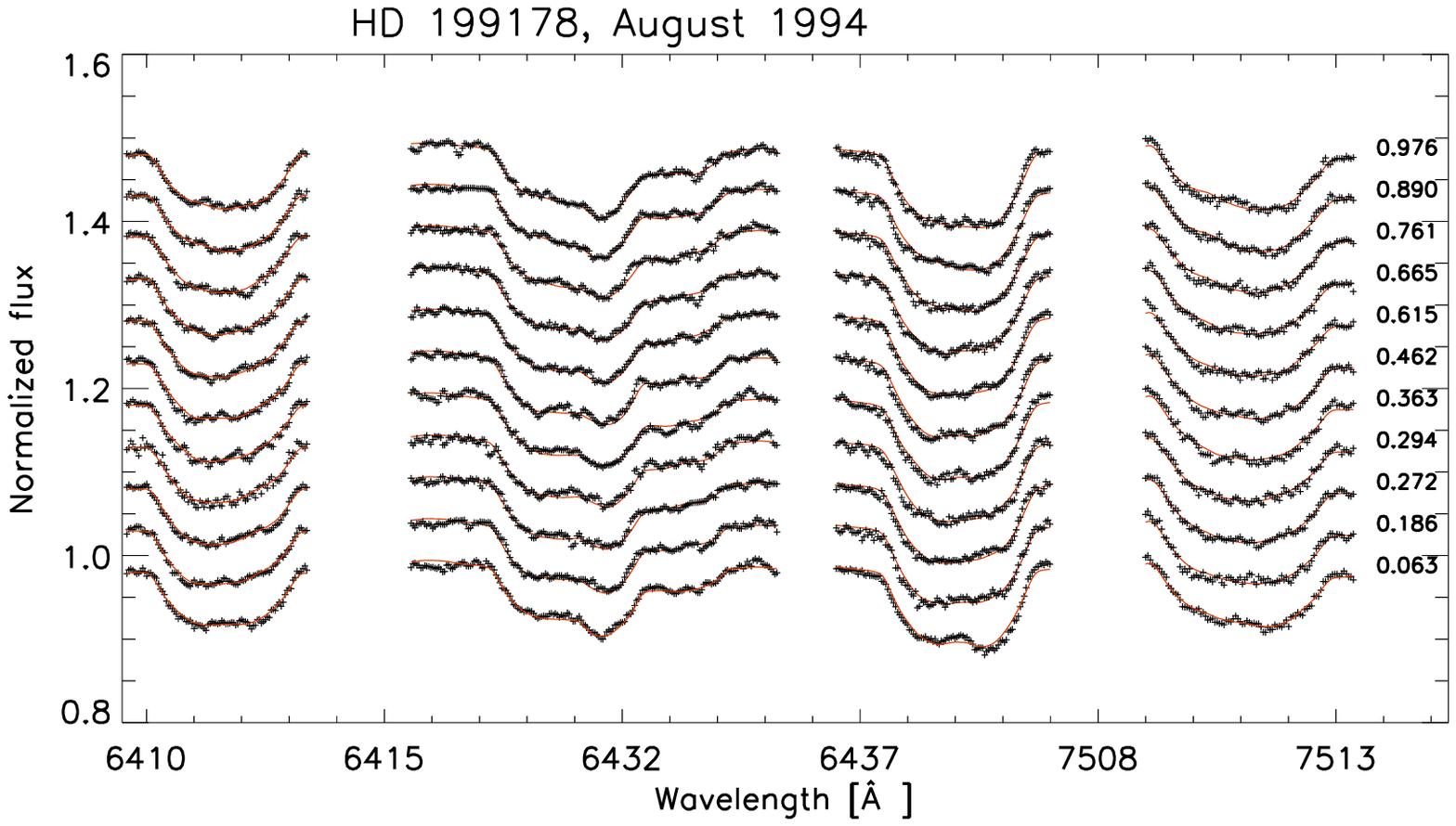}
\includegraphics[width=6cm,clip]{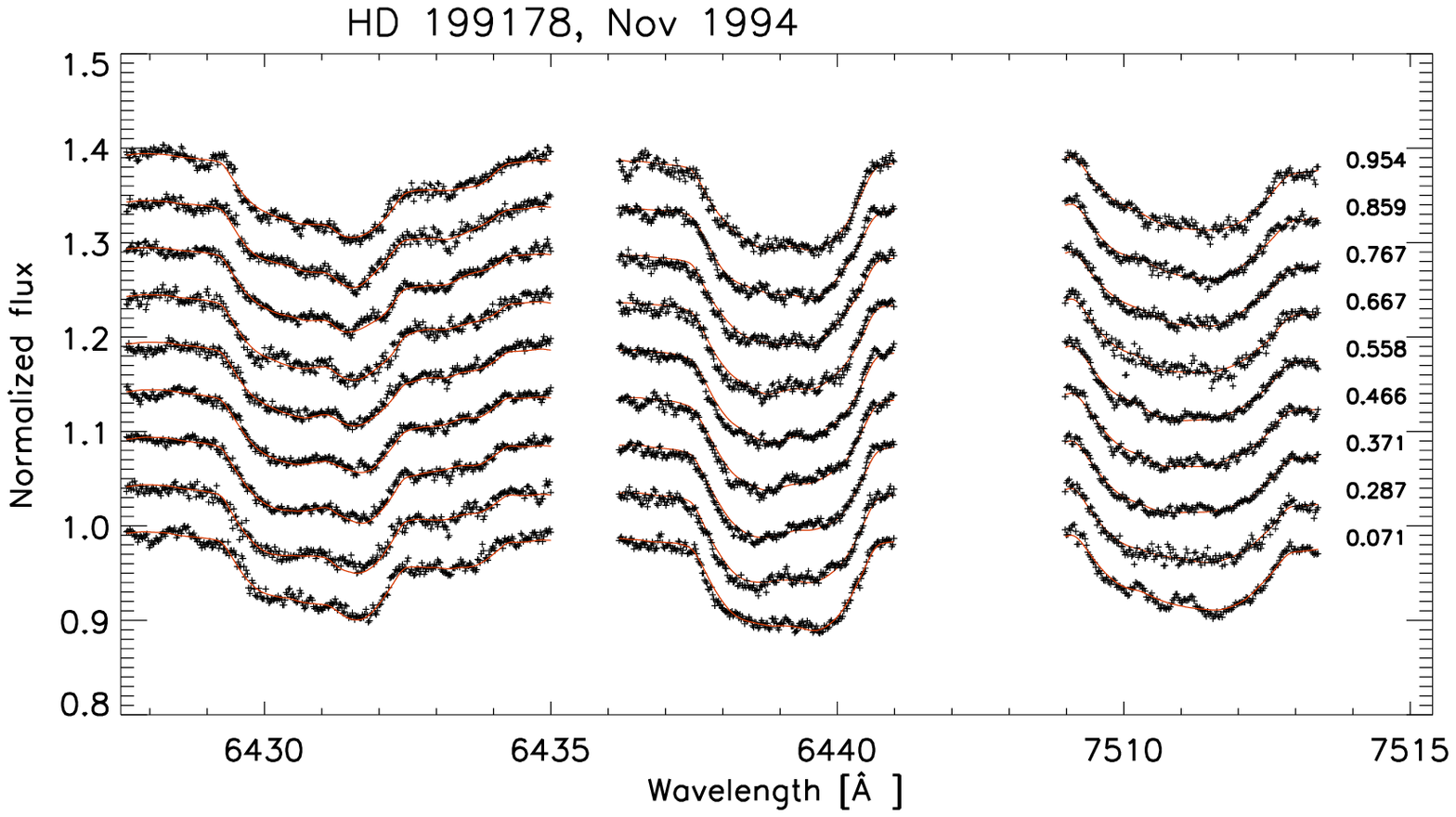}

\vspace{-1cm}

\includegraphics[width=6cm,clip]{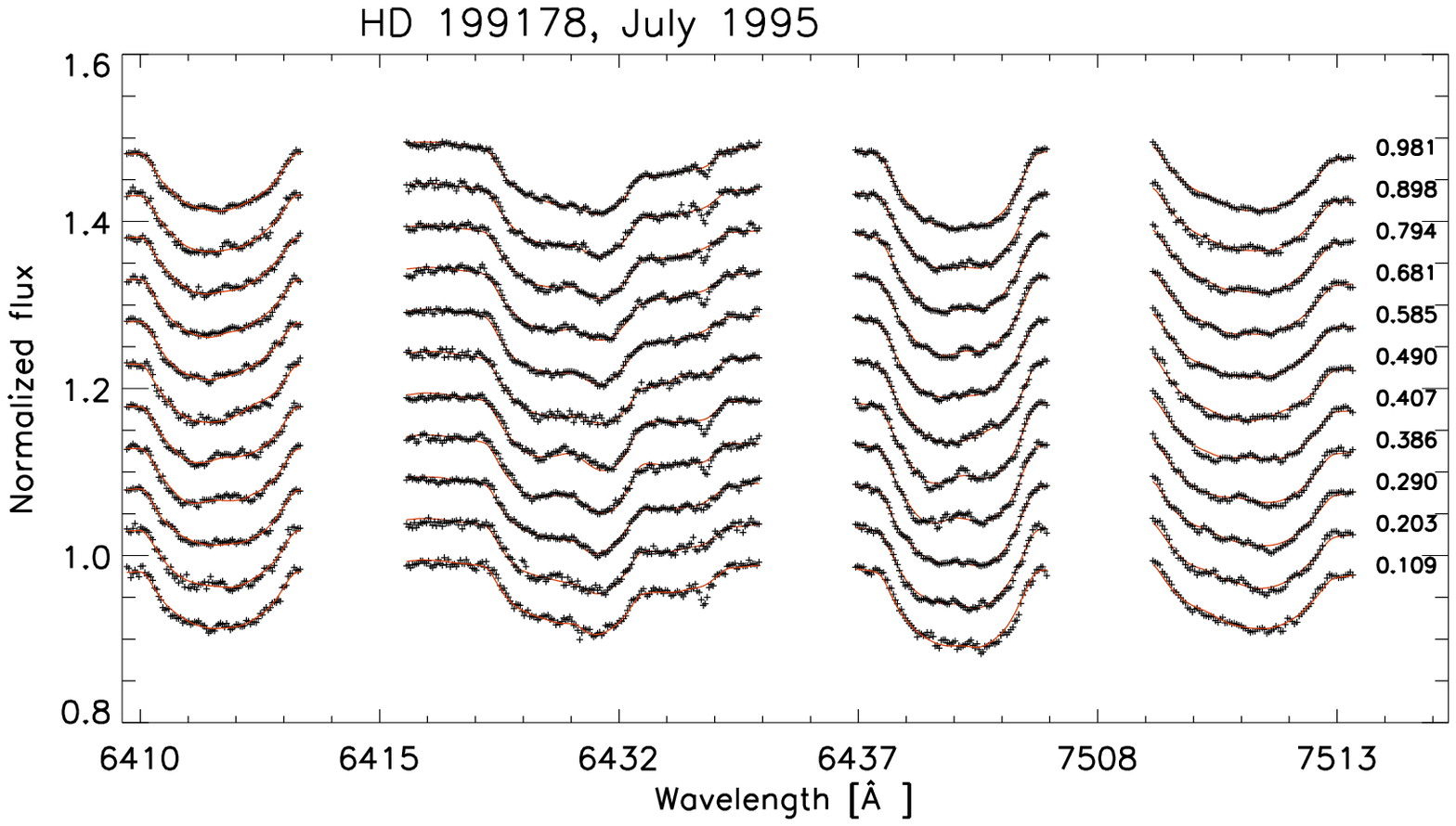}
\includegraphics[width=6cm,clip]{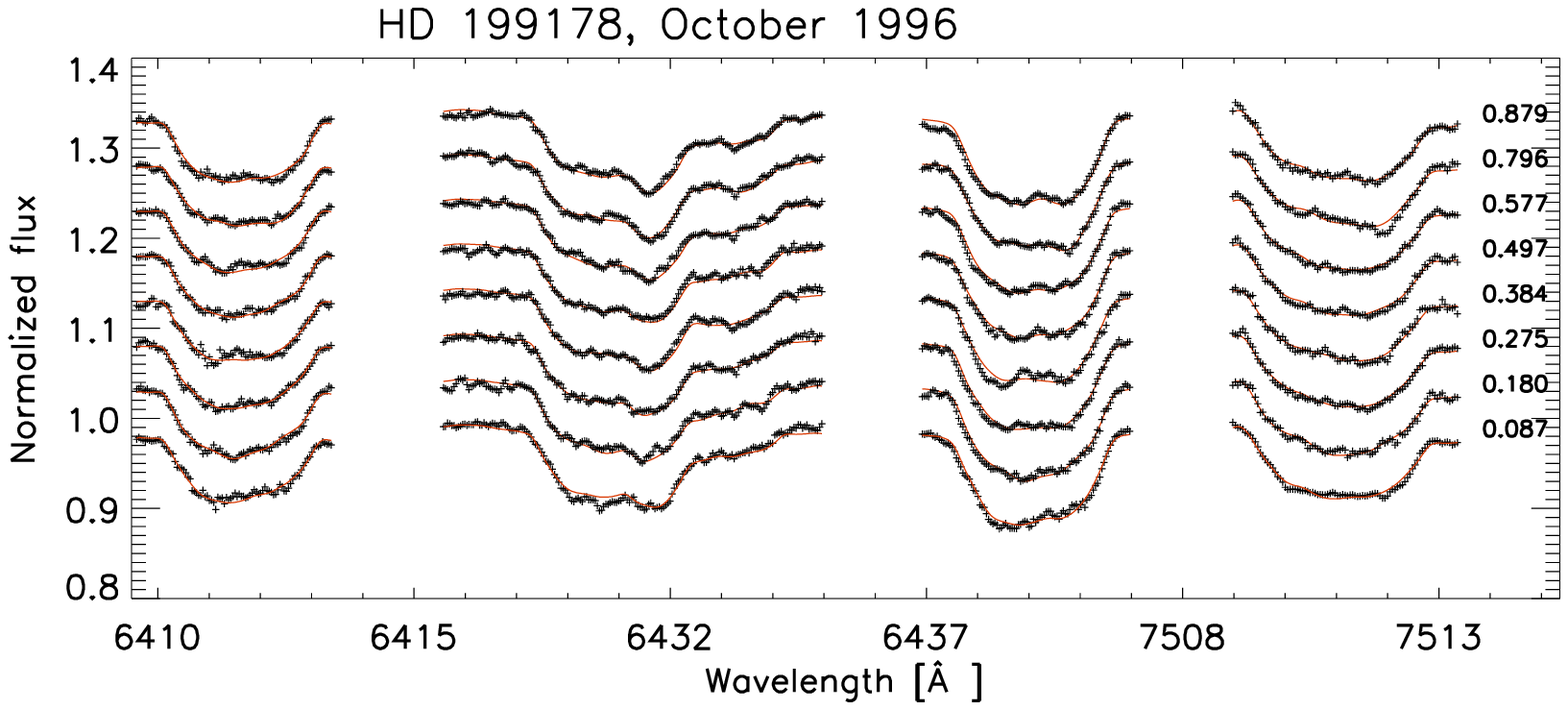}
\includegraphics[width=6cm,clip]{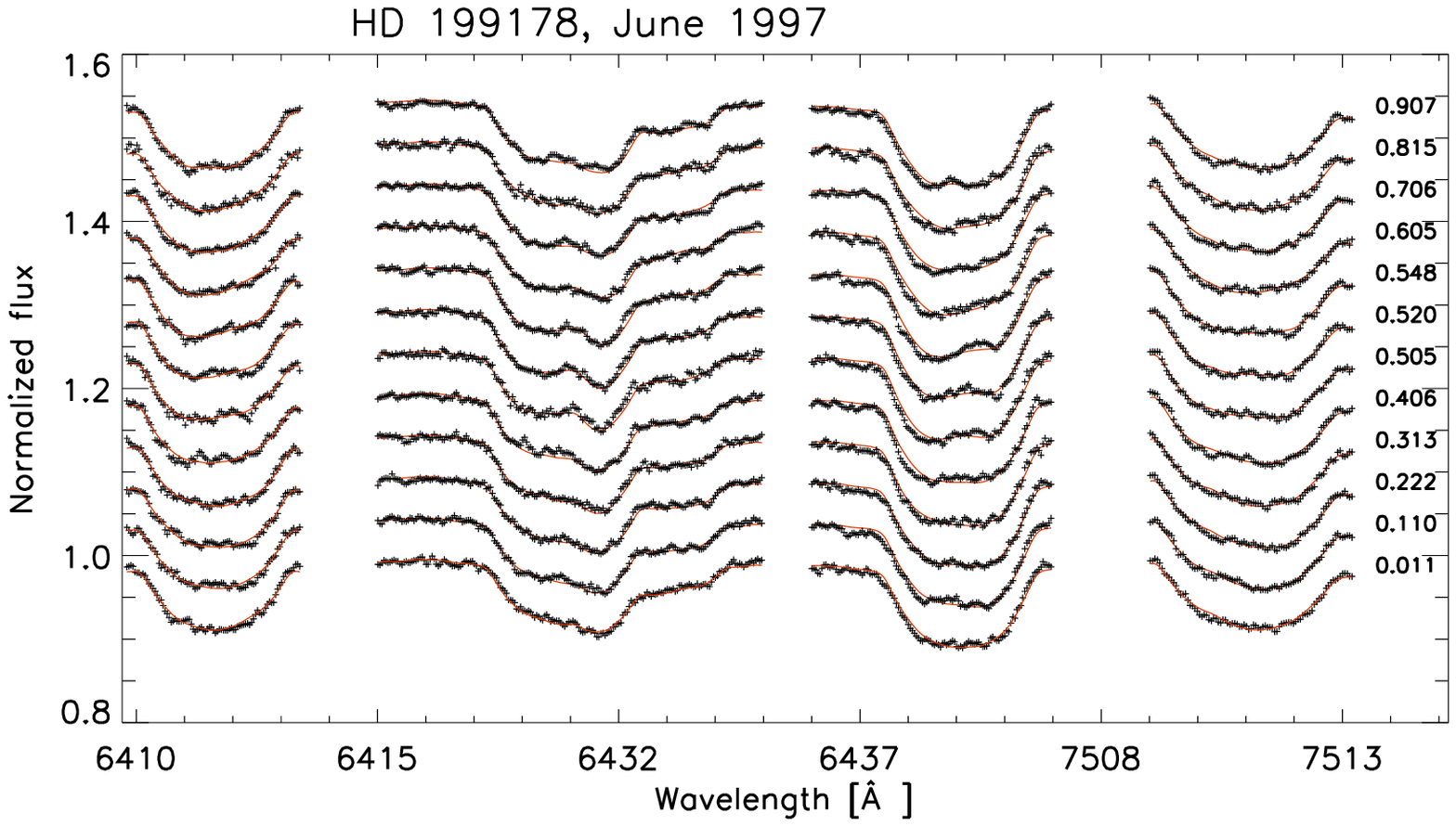}

\vspace{-1cm}

\includegraphics[width=6cm,clip]{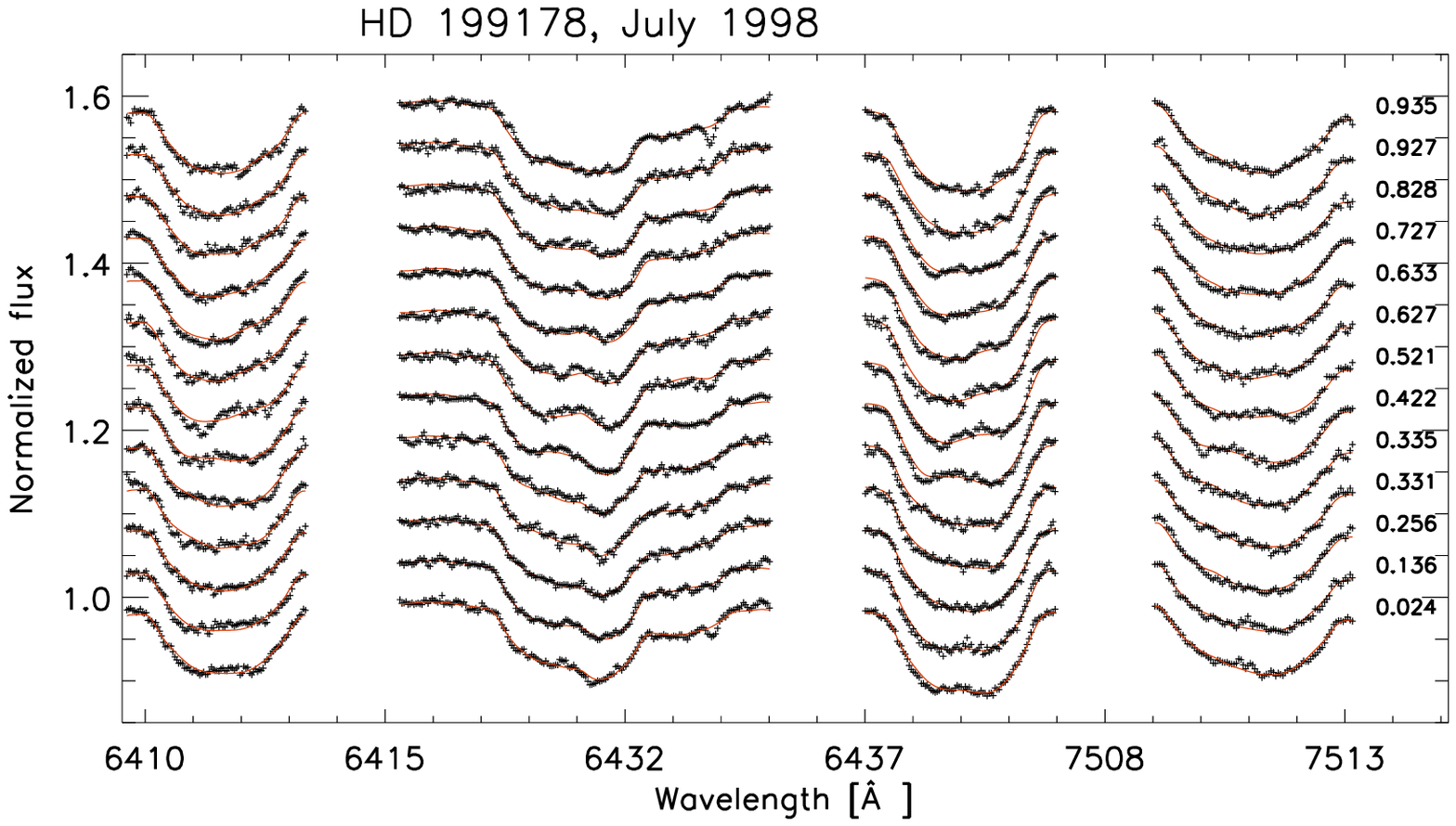}
\includegraphics[width=6cm,clip]{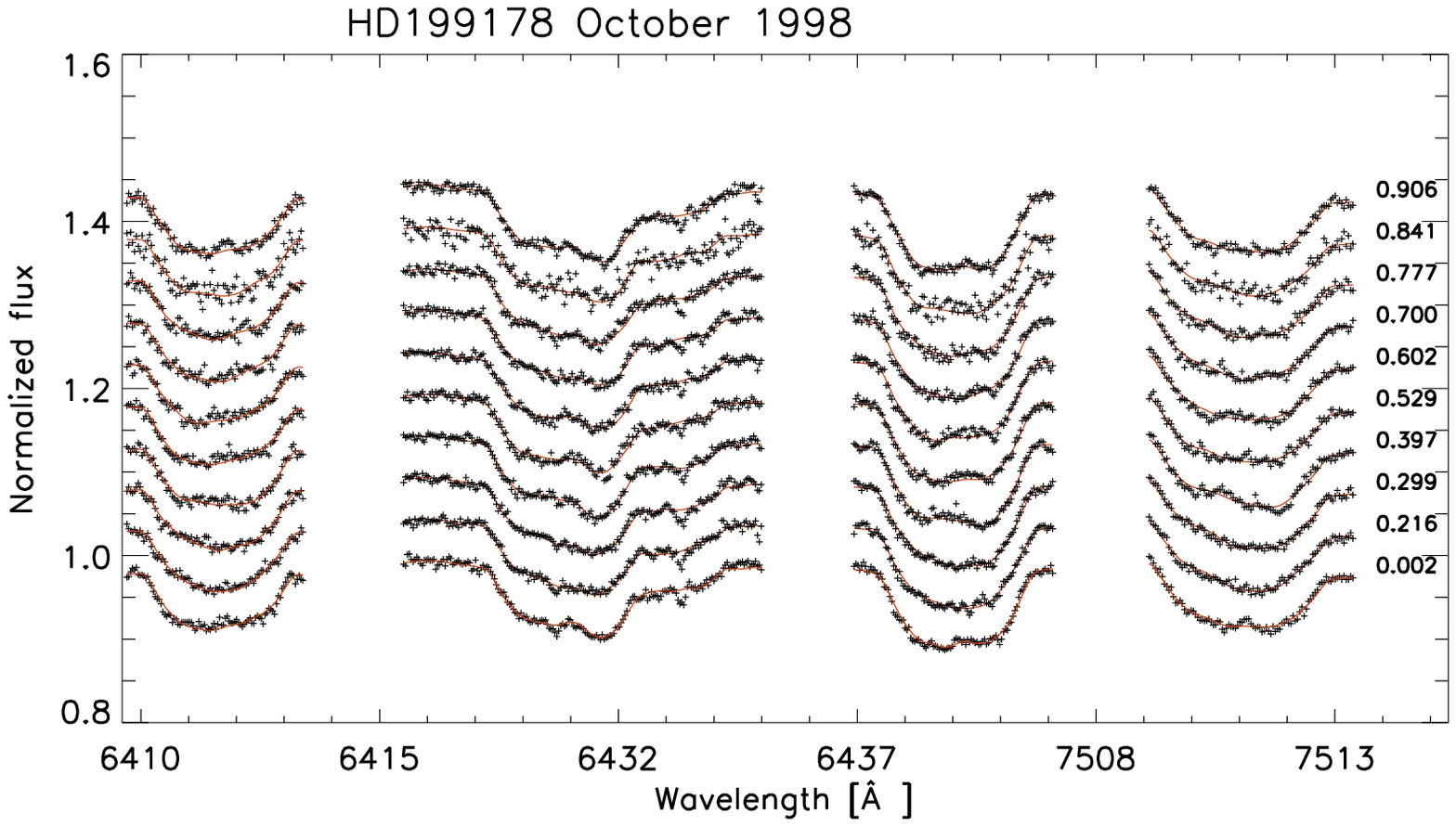}
\includegraphics[width=6cm,clip]{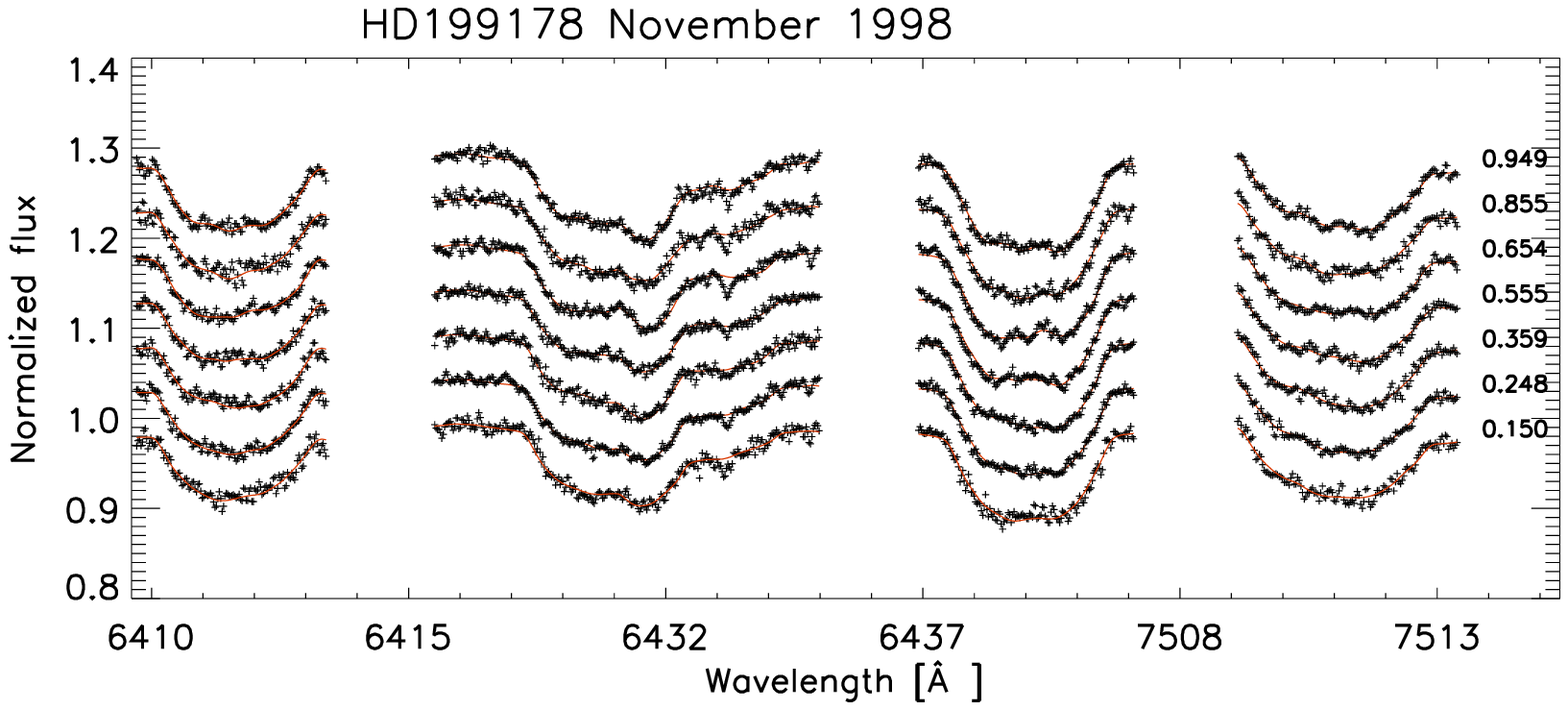}

\vspace{-1cm}

\includegraphics[width=6cm,clip]{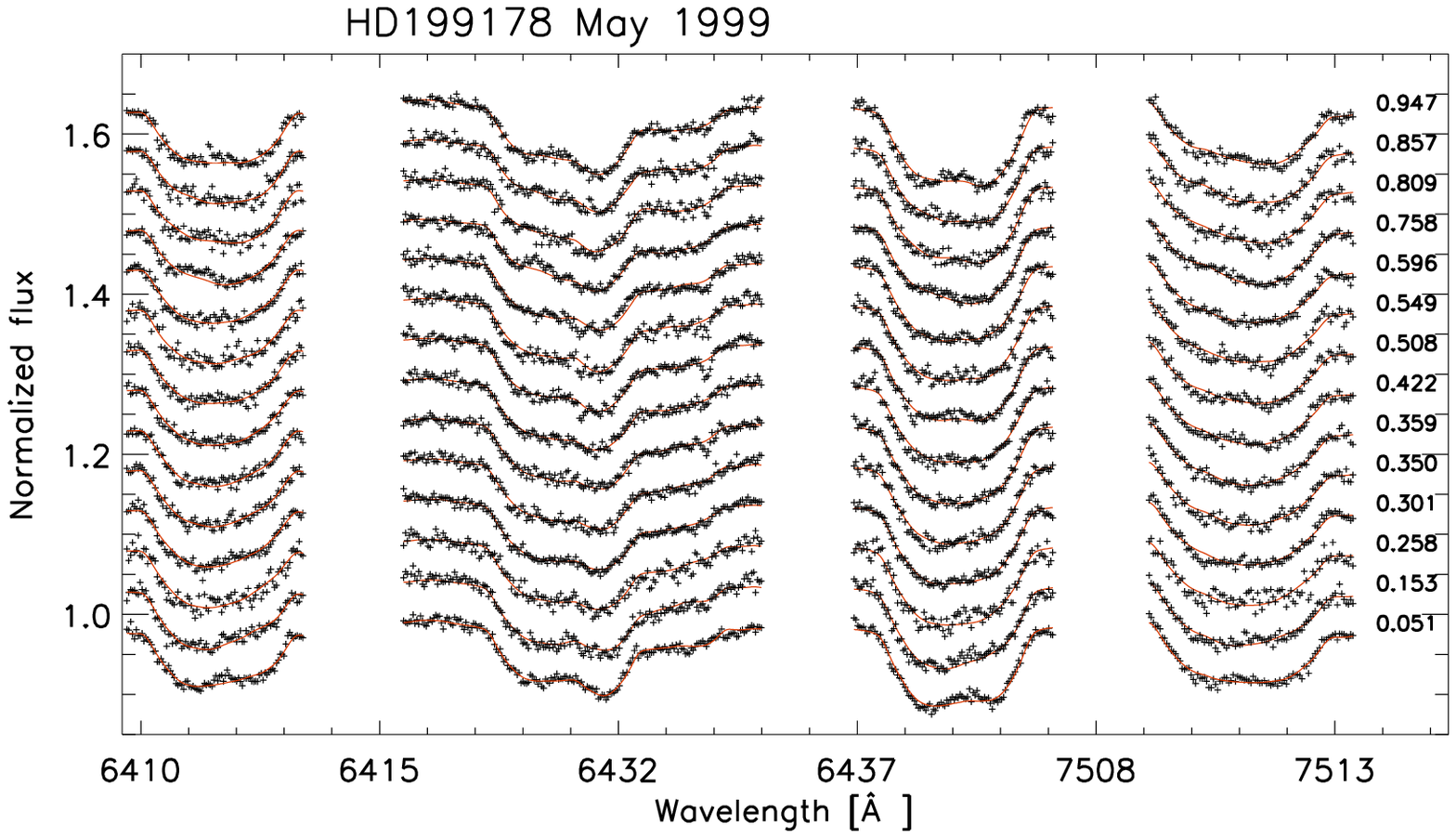}
\includegraphics[width=6cm,clip]{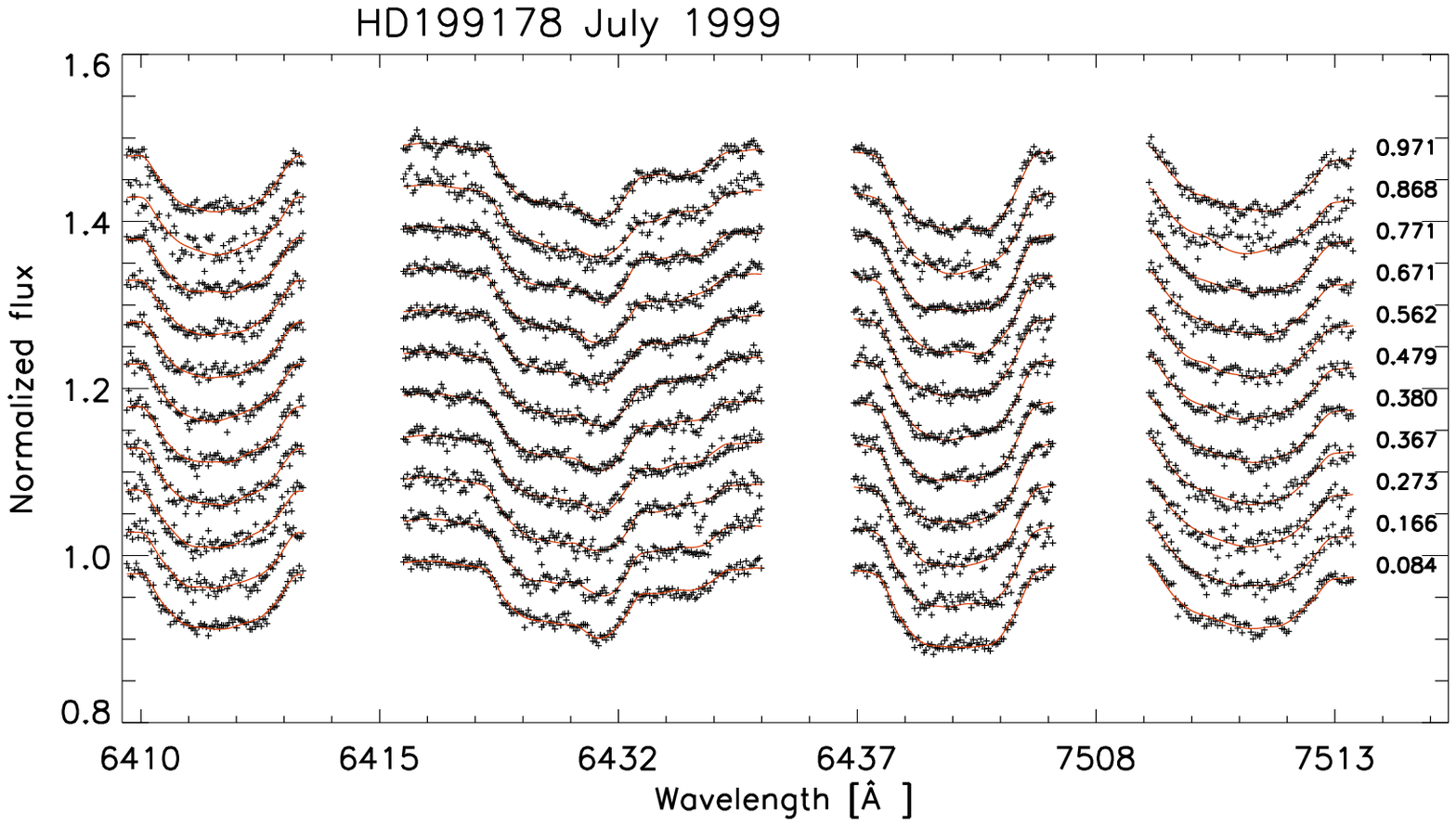}
\includegraphics[width=6cm,clip]{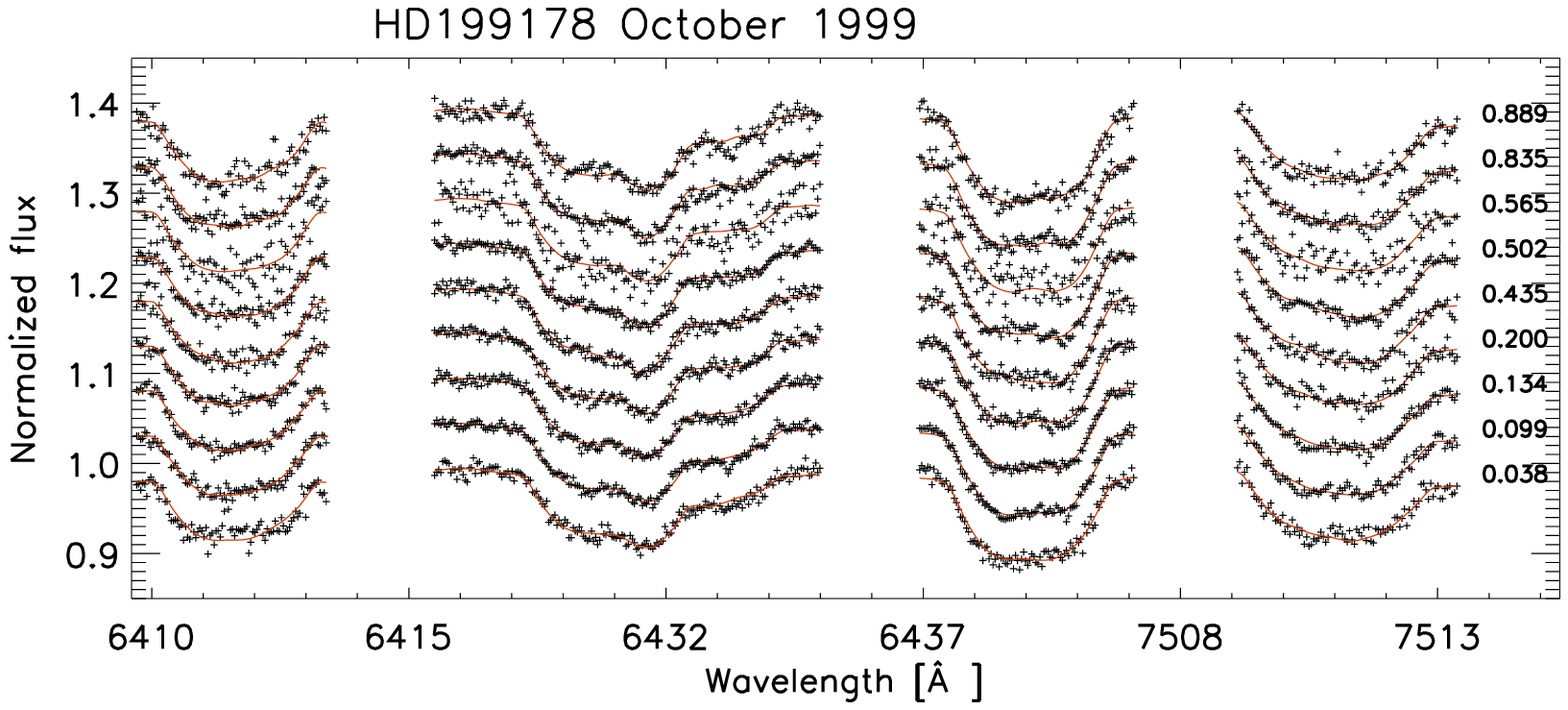}

\vspace{0.5cm}

\includegraphics[width=6cm,clip]{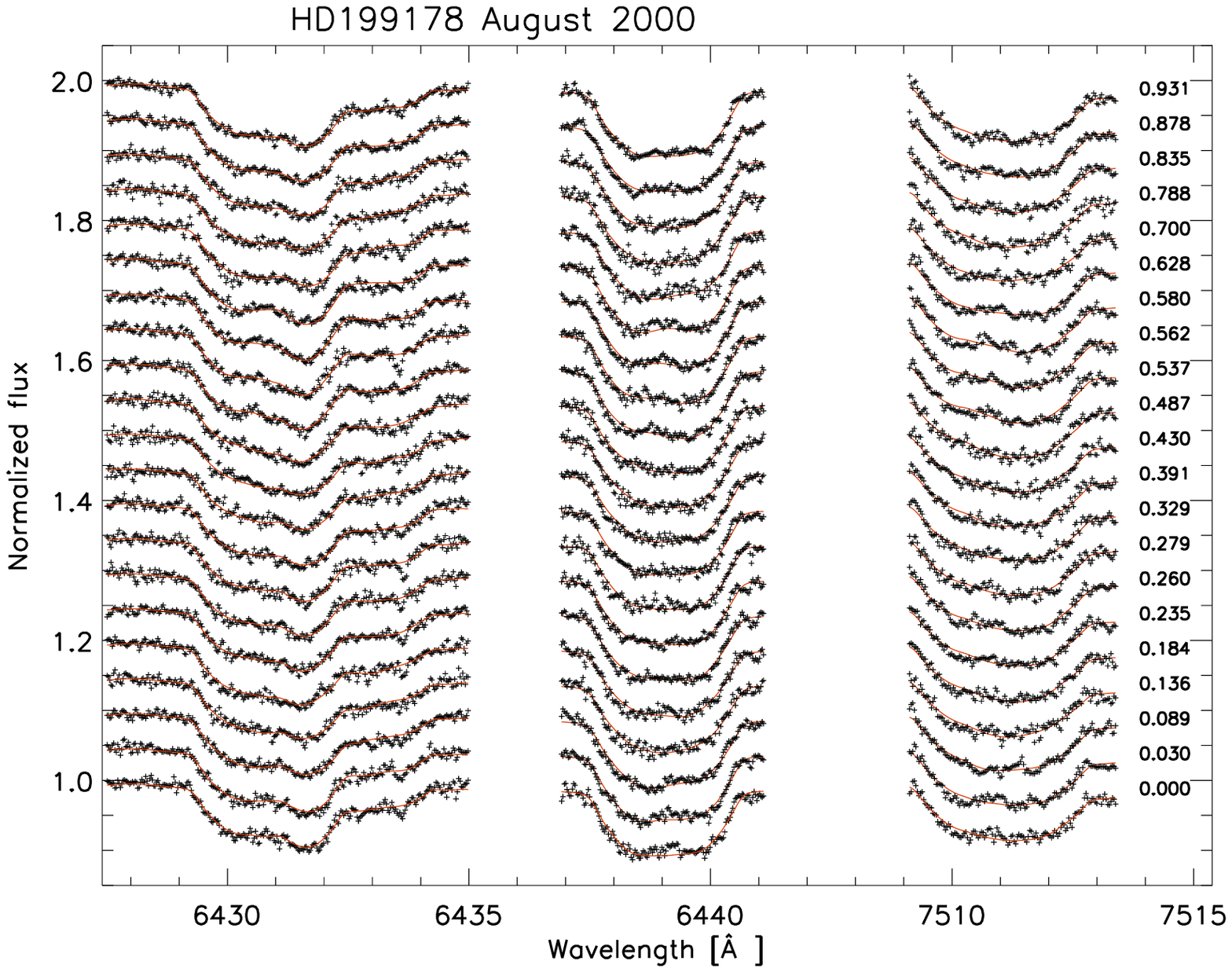}
\includegraphics[width=6cm,clip]{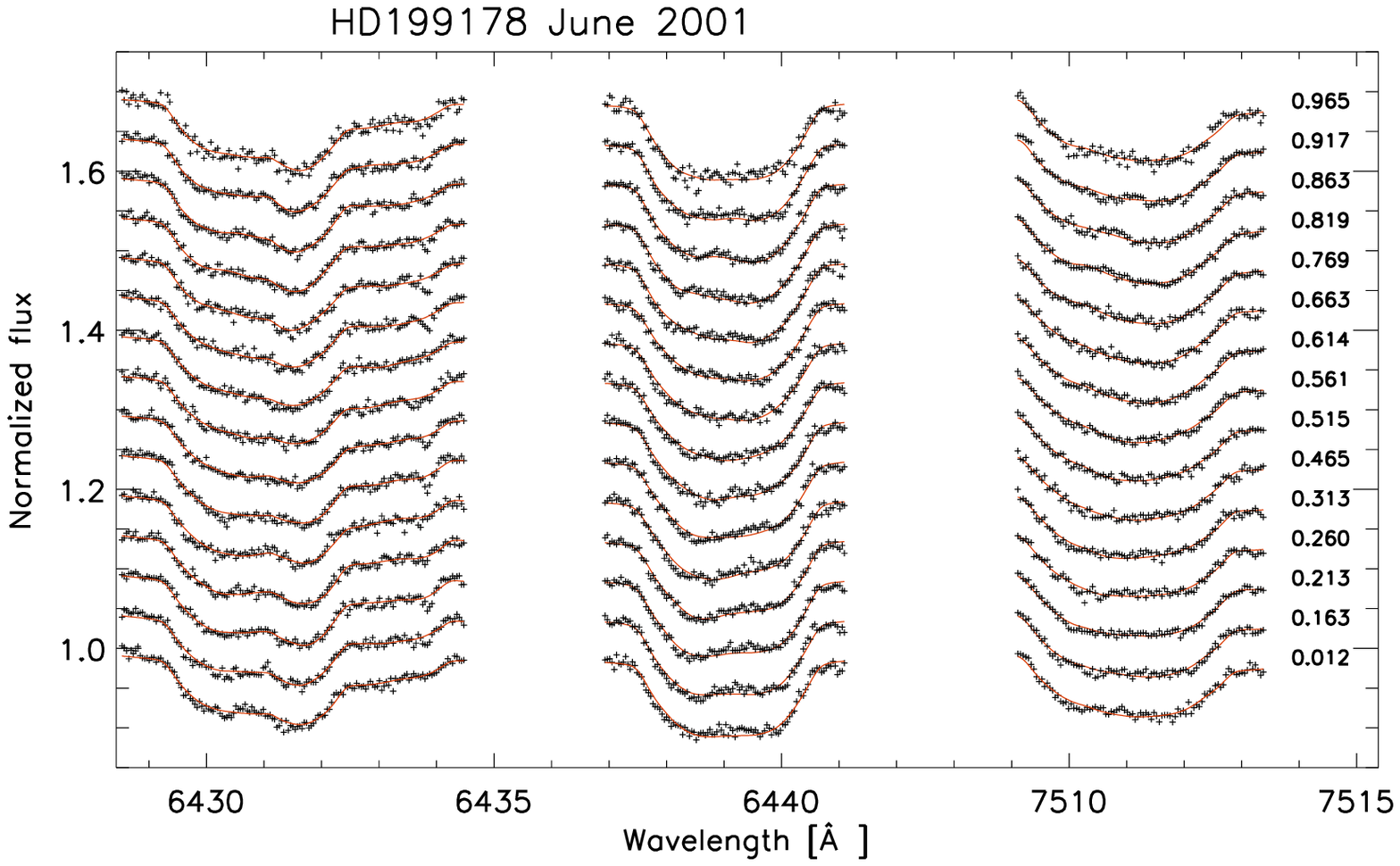}
\includegraphics[width=6cm,clip]{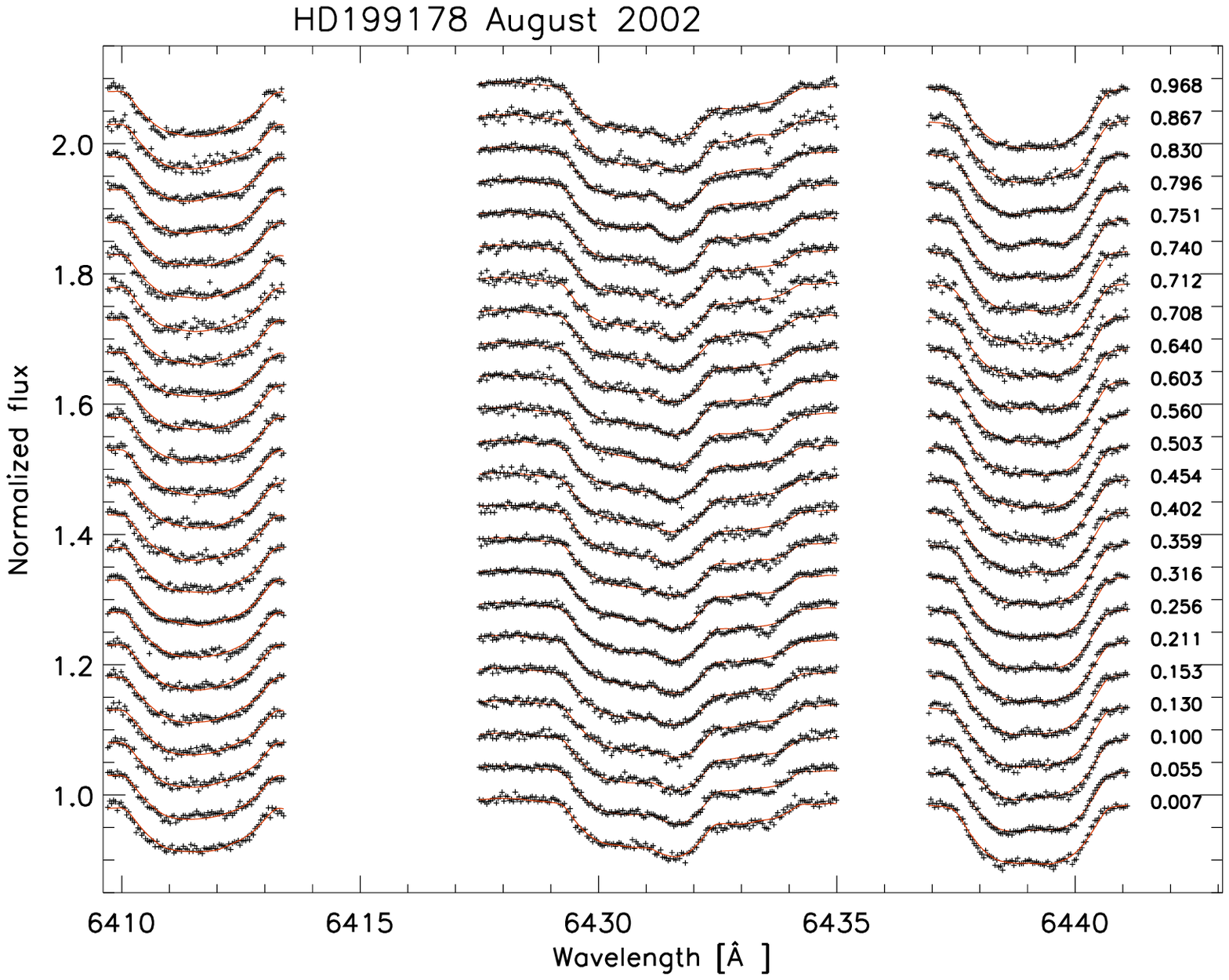}


\includegraphics[width=6cm,clip]{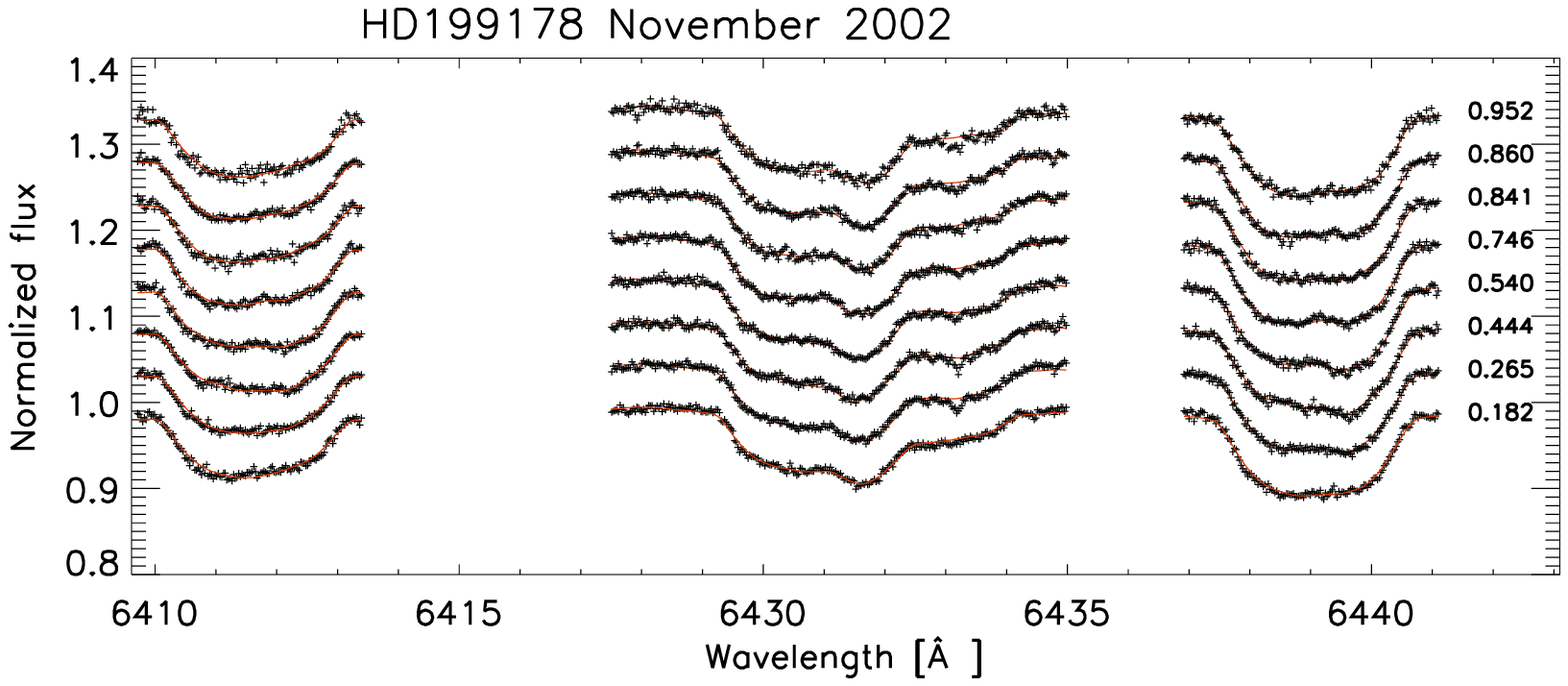}
\includegraphics[width=6cm,clip]{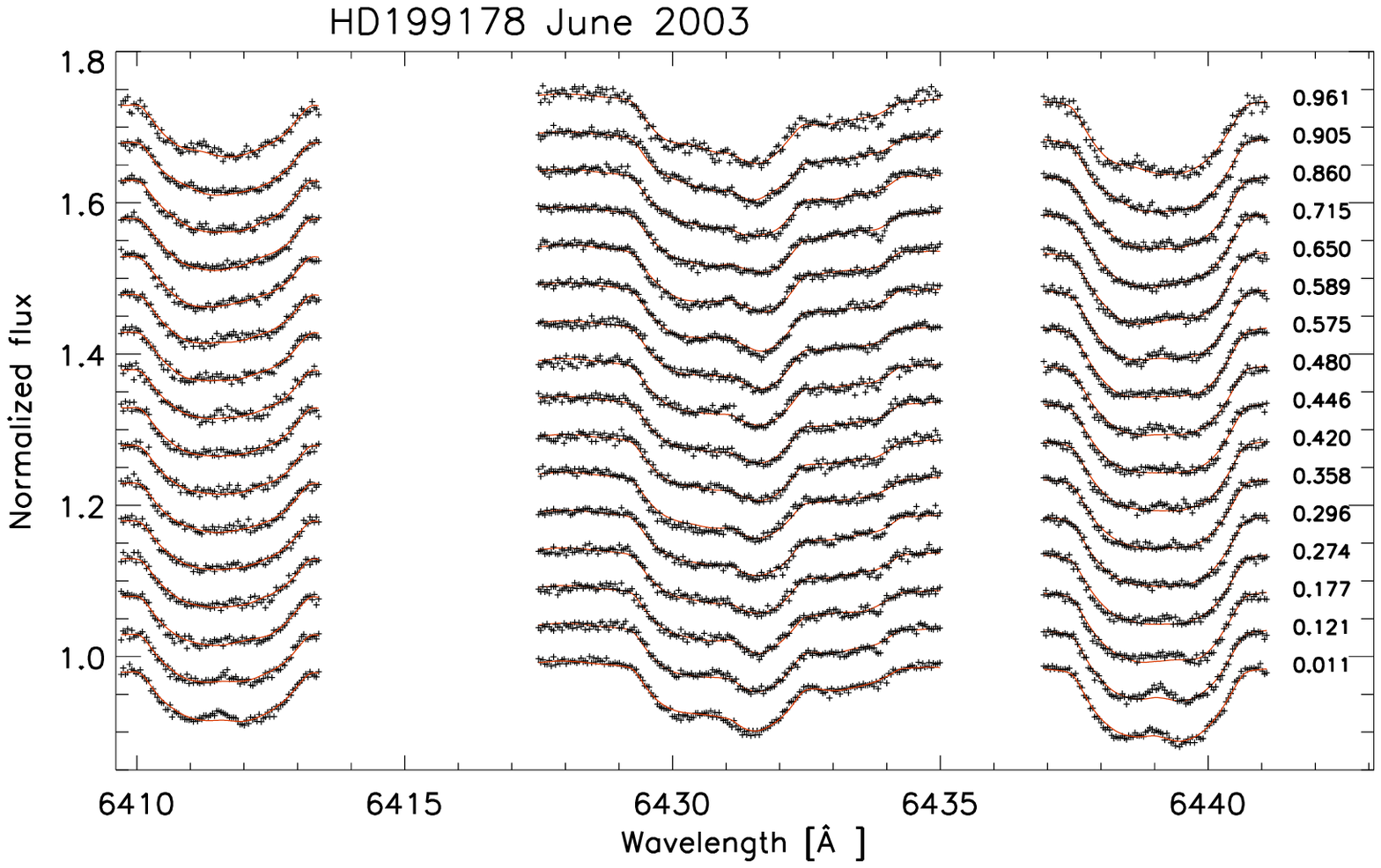}
\includegraphics[width=6cm,clip]{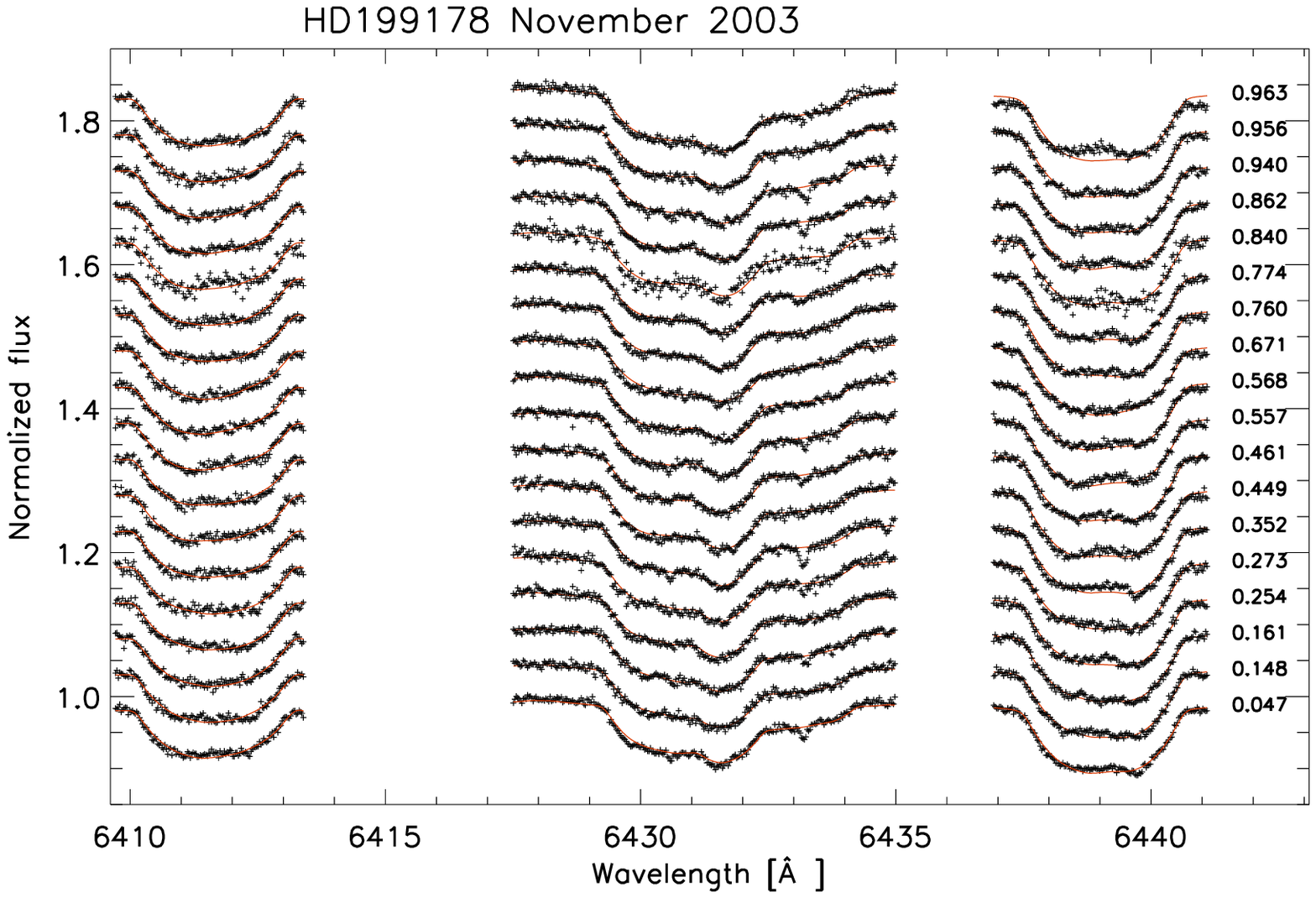}

\caption{Observed (plus signs) and modelled (red line) spectra. Different 
wavelength regions are plotted in the same frames, the
wavelength step being constant. The rotational phase is given to the
right of each spectrum.}

\end{figure*}

\begin{figure*}
\centering

\vspace{-1cm}

\includegraphics[width=6cm,clip]{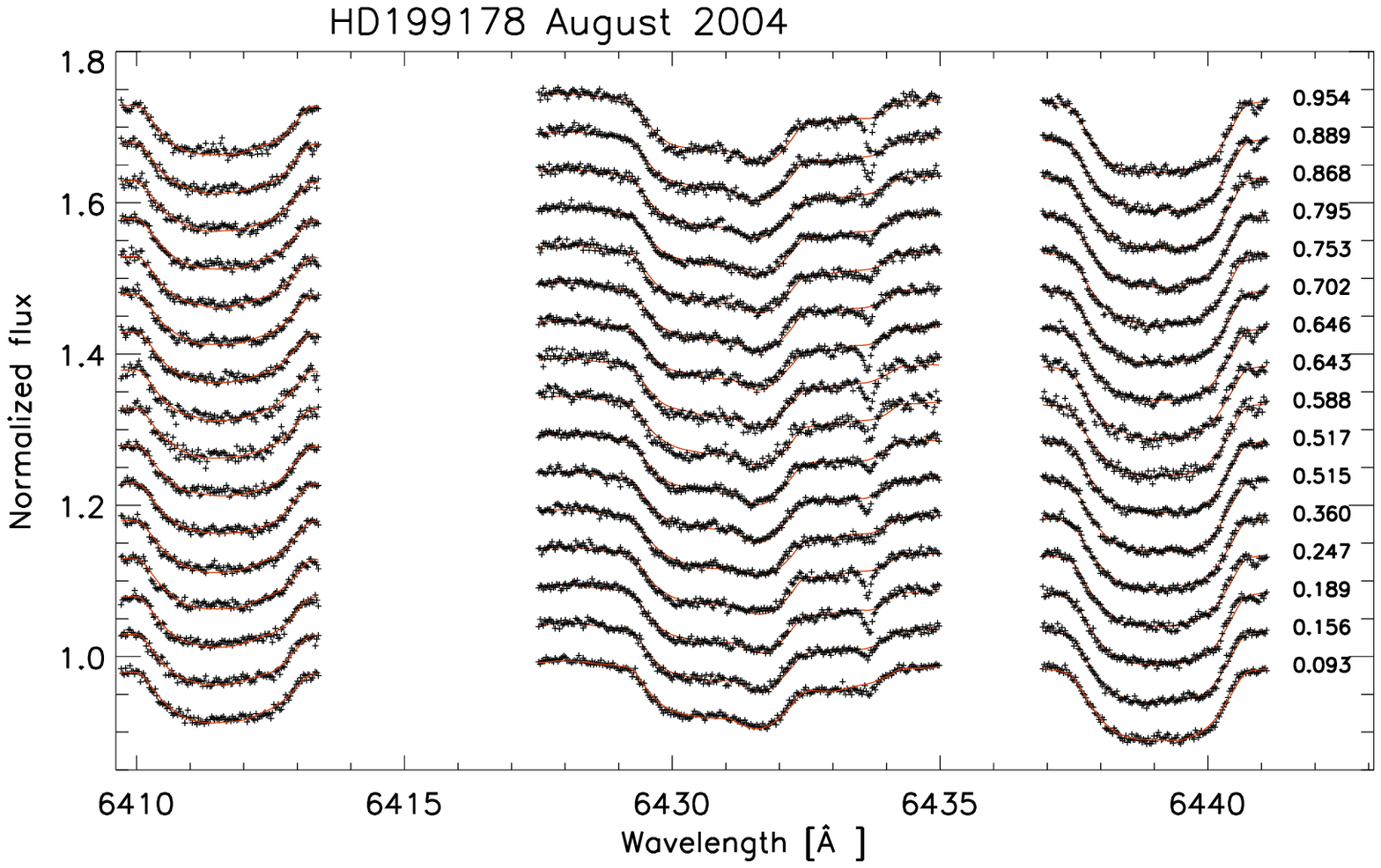}
\includegraphics[width=6cm,clip]{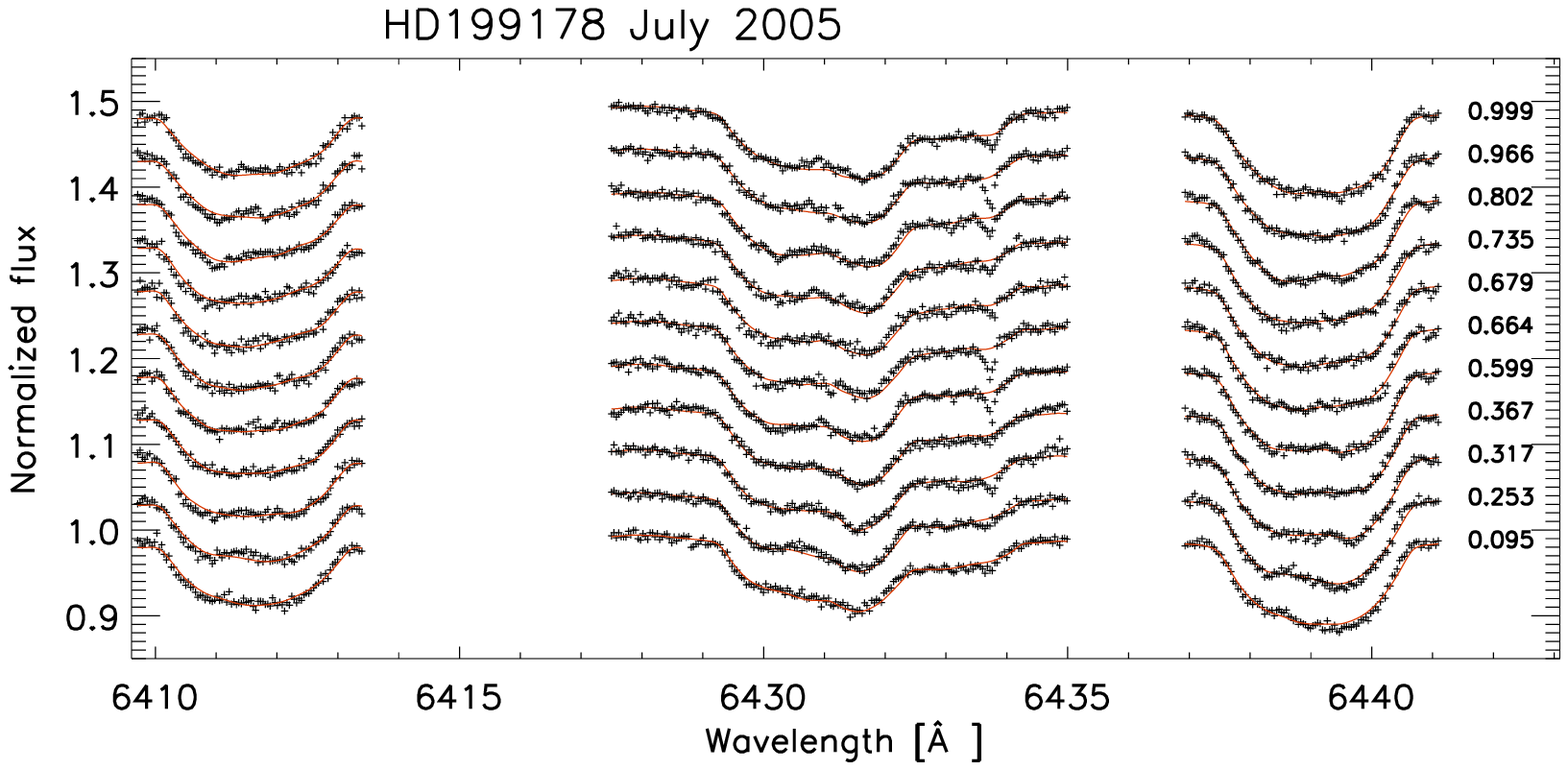}
\includegraphics[width=6cm,clip]{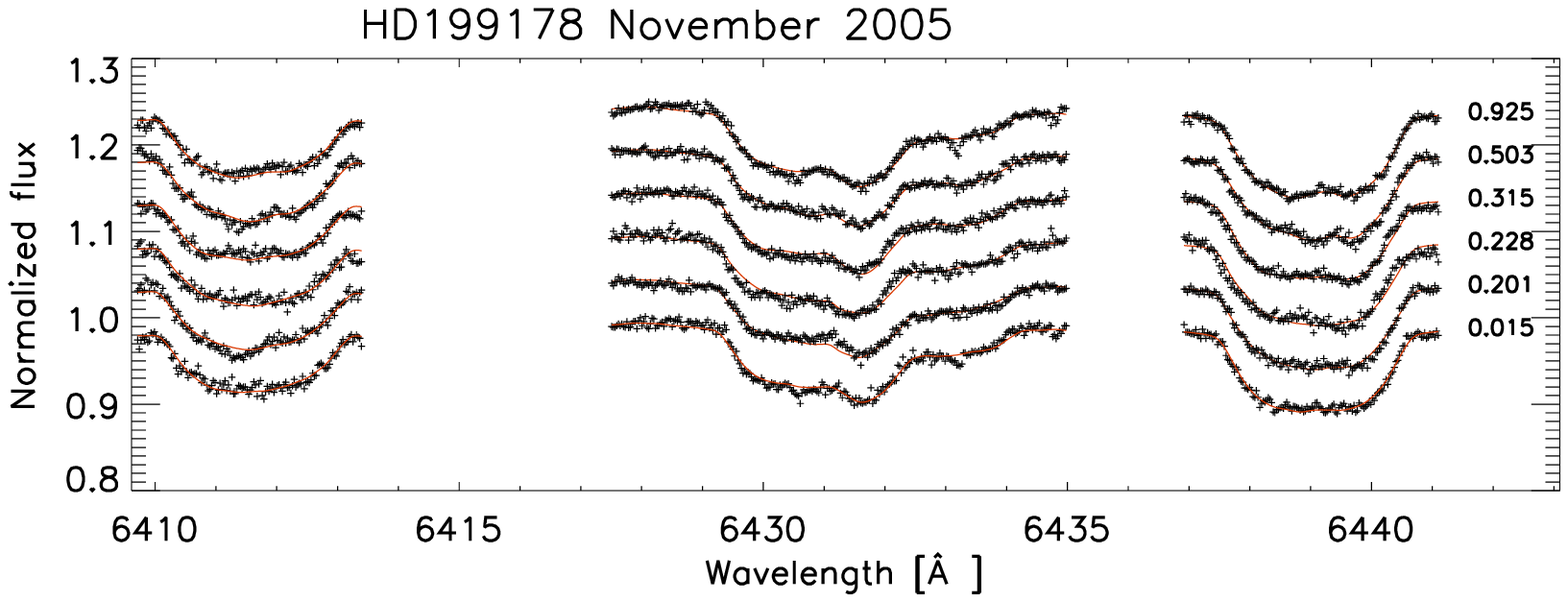}

\vspace{-0.5cm}

\includegraphics[width=6cm,clip]{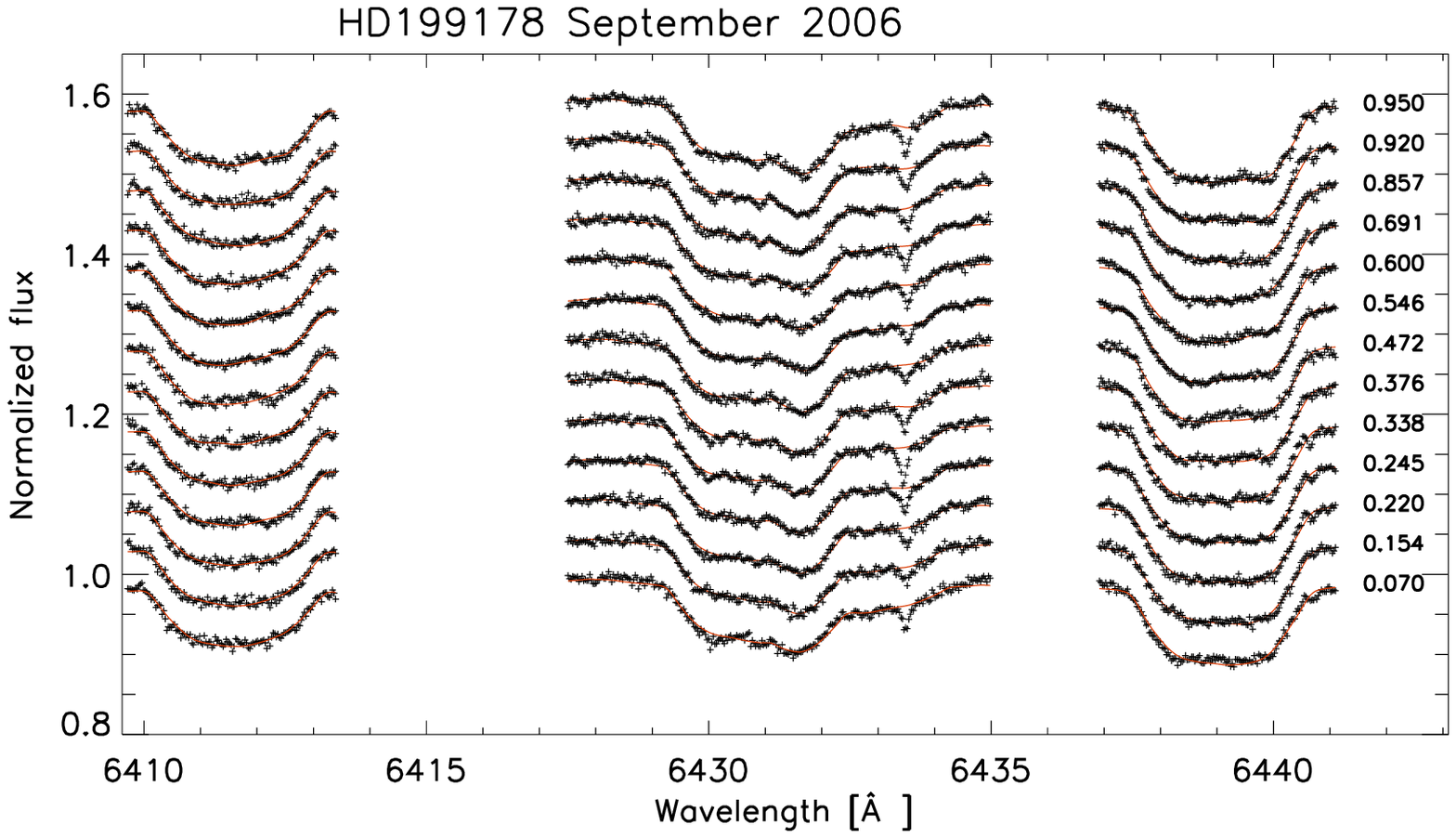}
\includegraphics[width=6cm,clip]{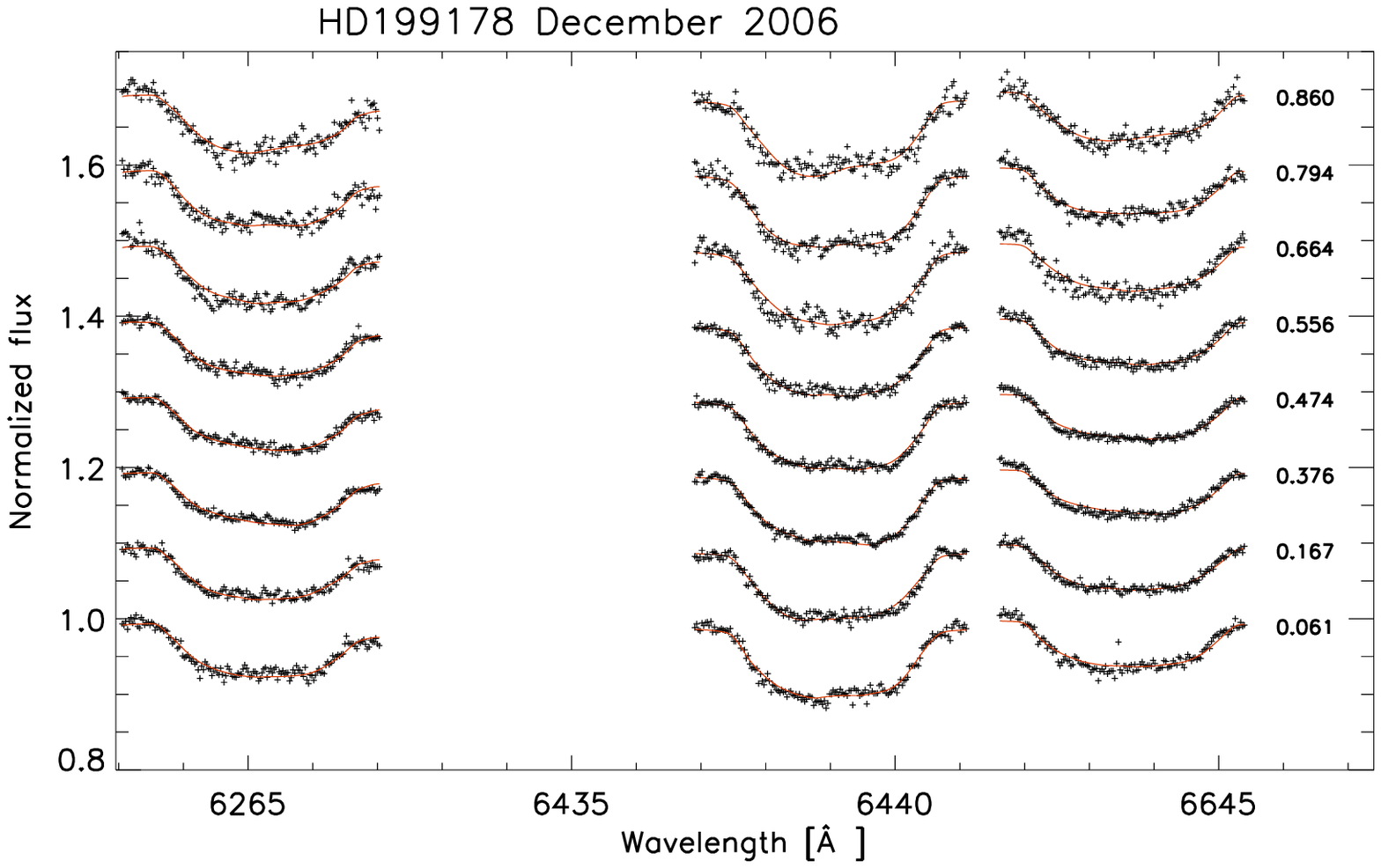}
\includegraphics[width=6cm,clip]{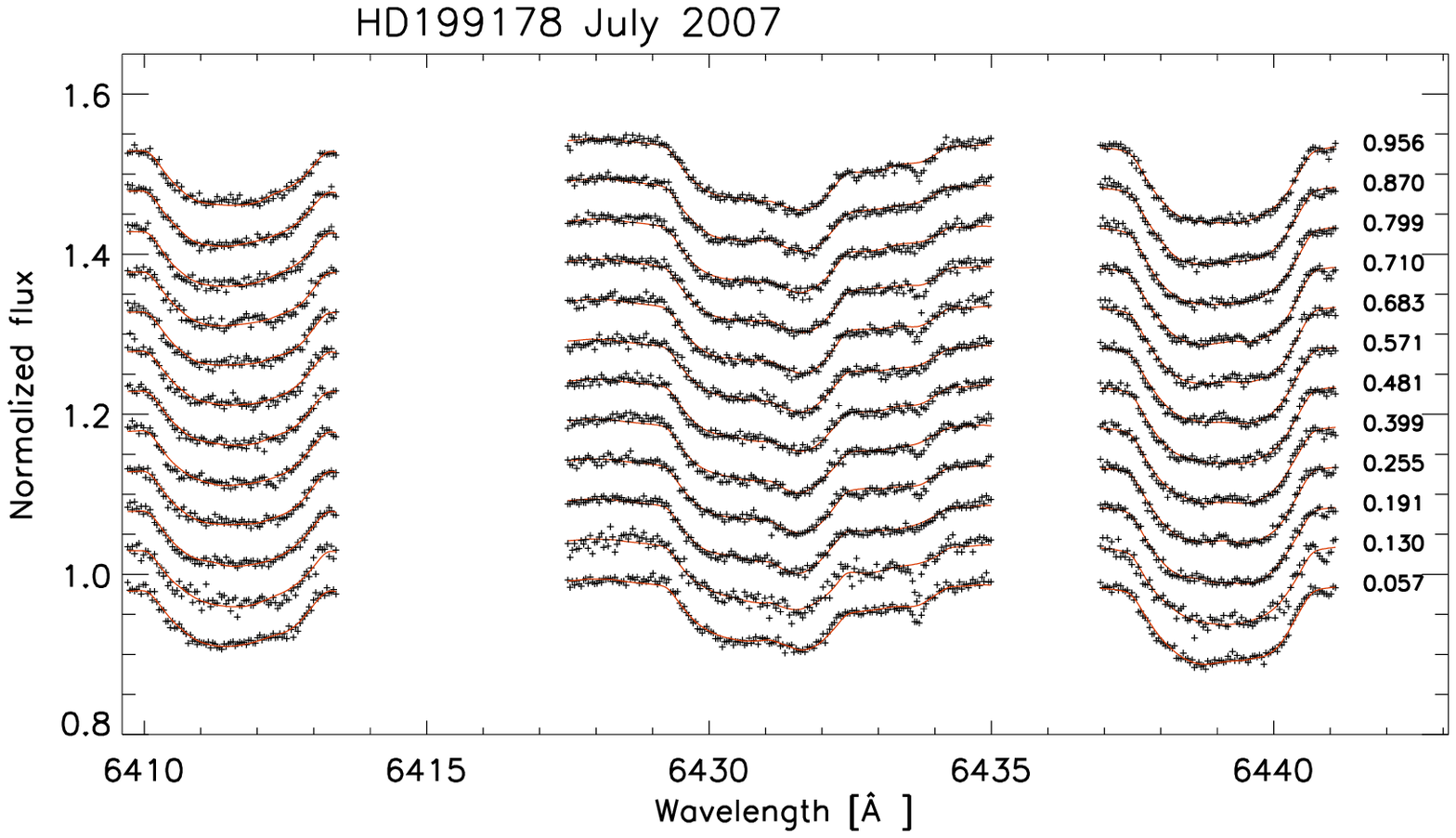}

\vspace{-1.5cm}

\includegraphics[width=6cm,clip]{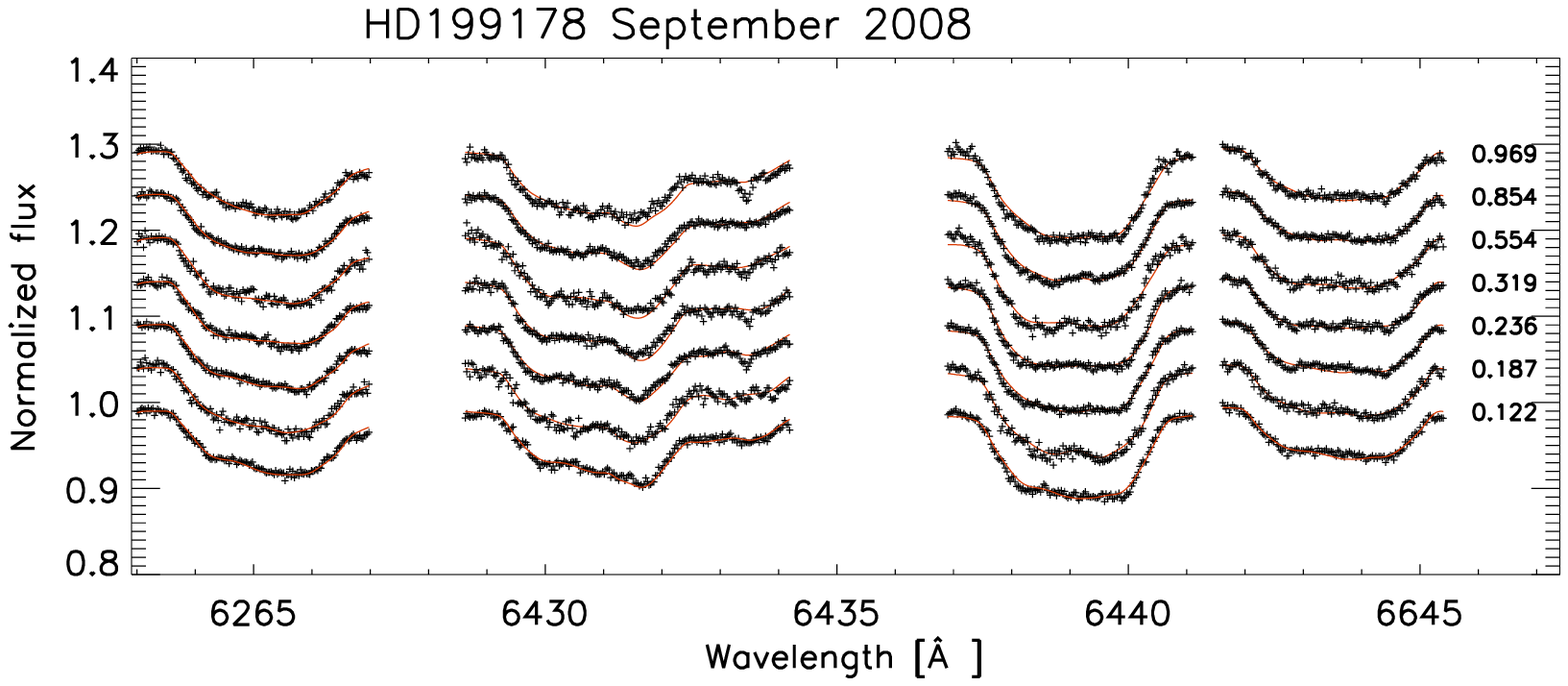}
\includegraphics[width=6cm,clip]{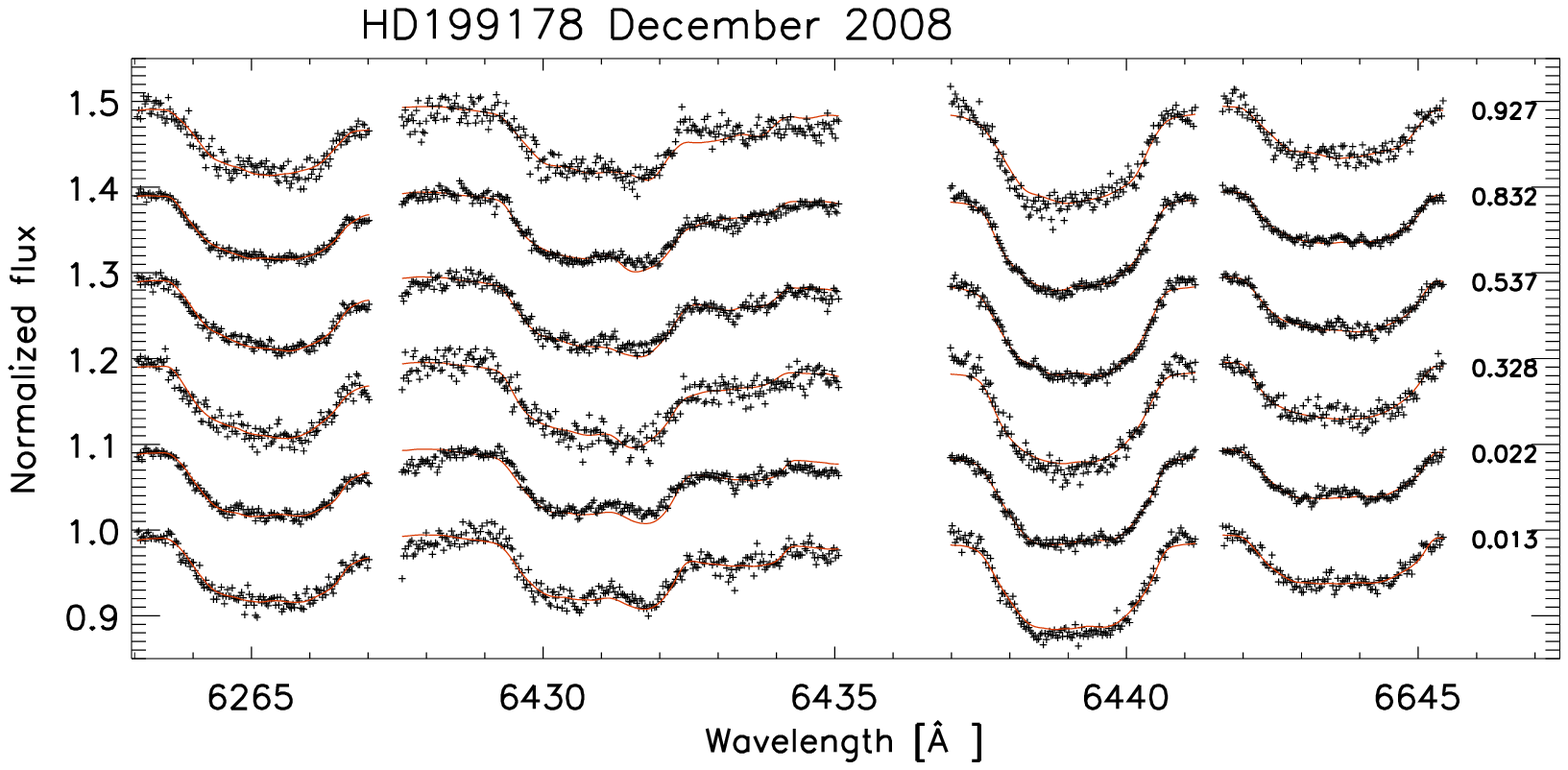}
\includegraphics[width=6cm,clip]{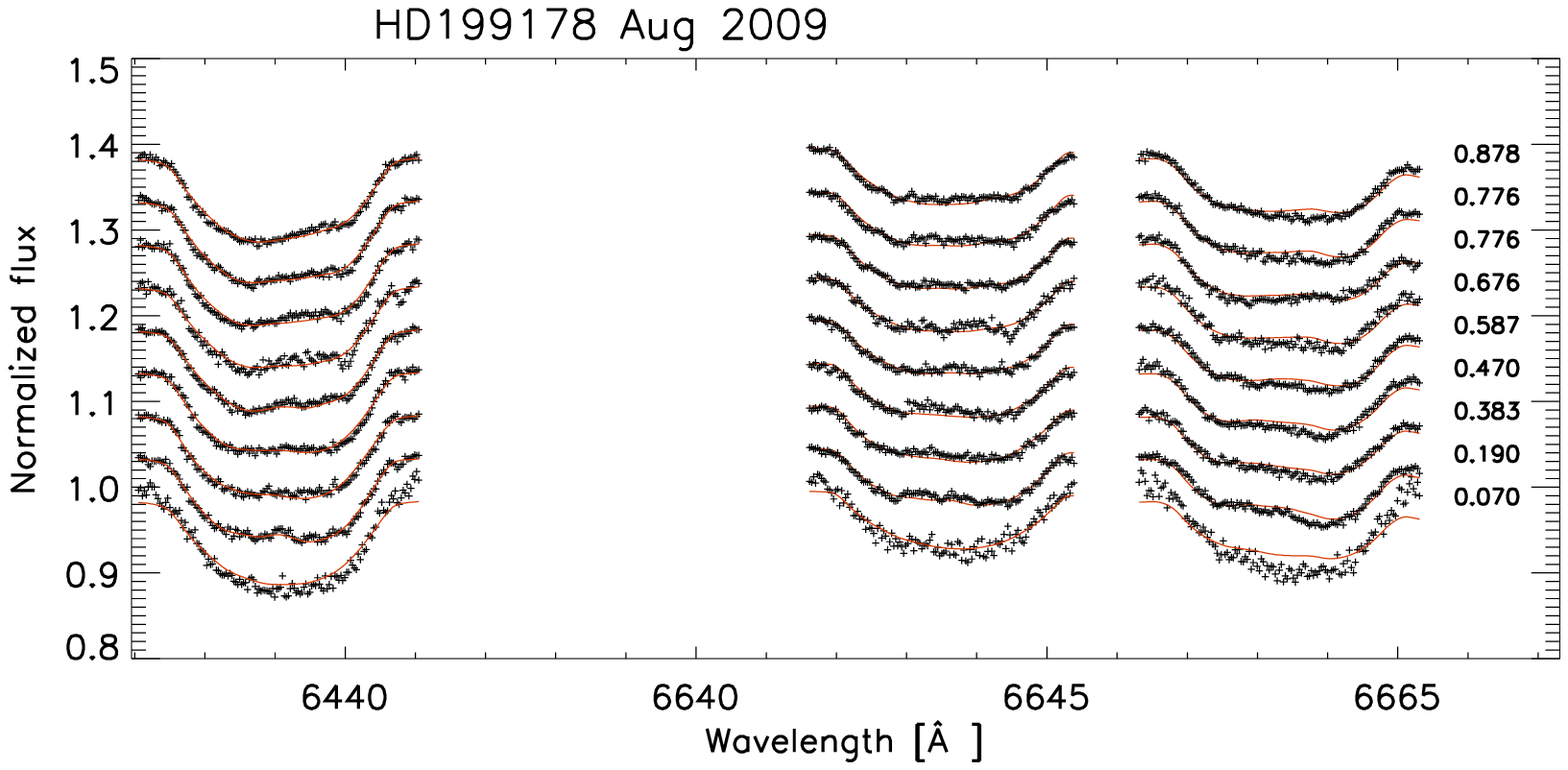}

\vspace{-1.5cm}

\includegraphics[width=6cm,clip]{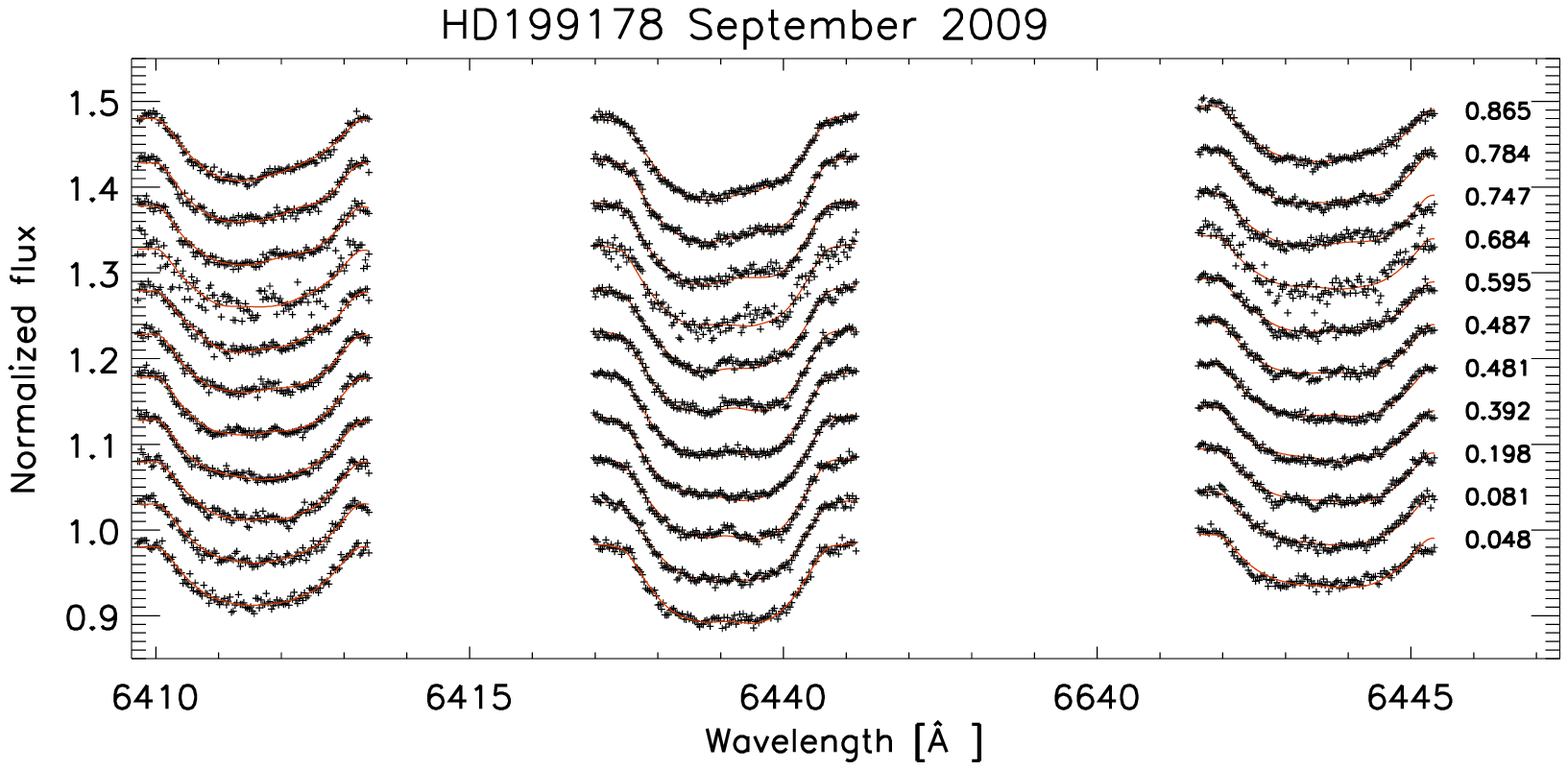}
\includegraphics[width=6cm,clip]{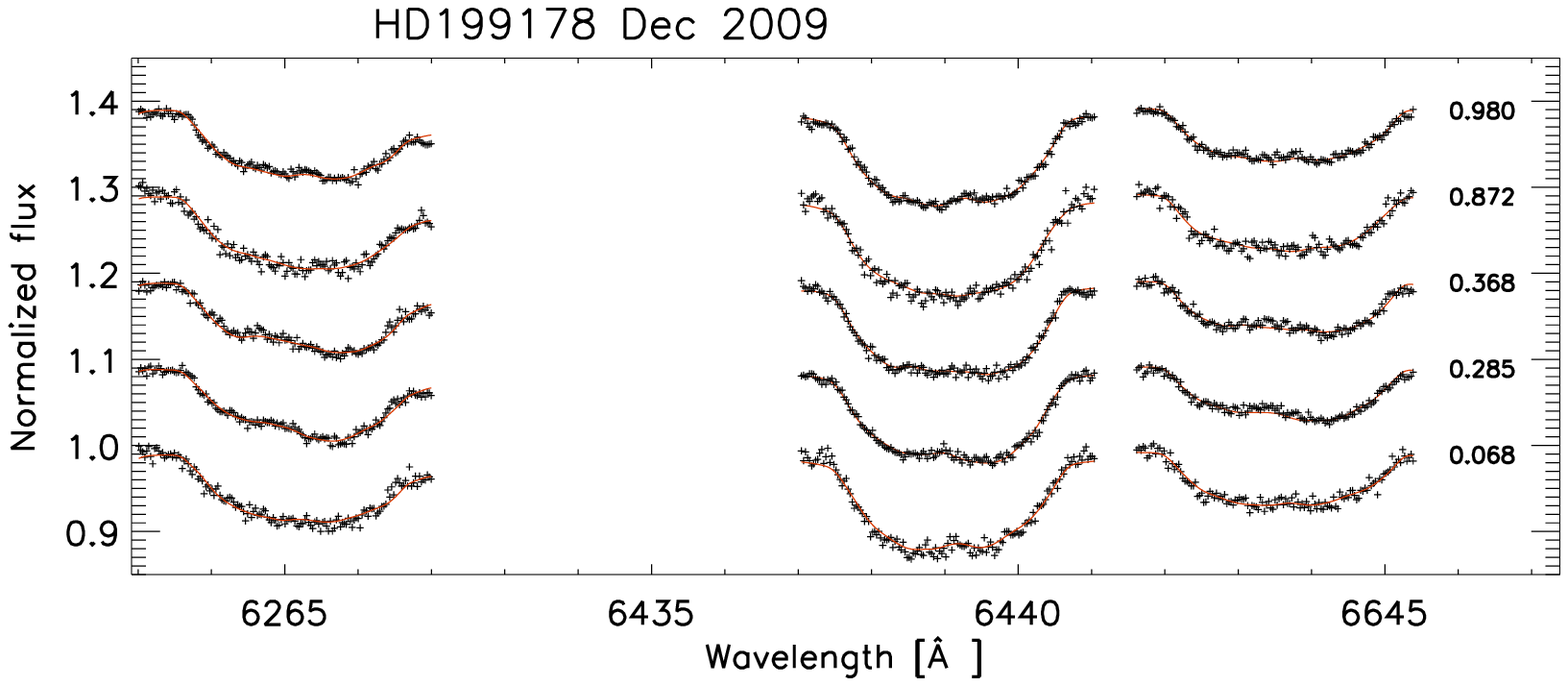}
\includegraphics[width=6cm,clip]{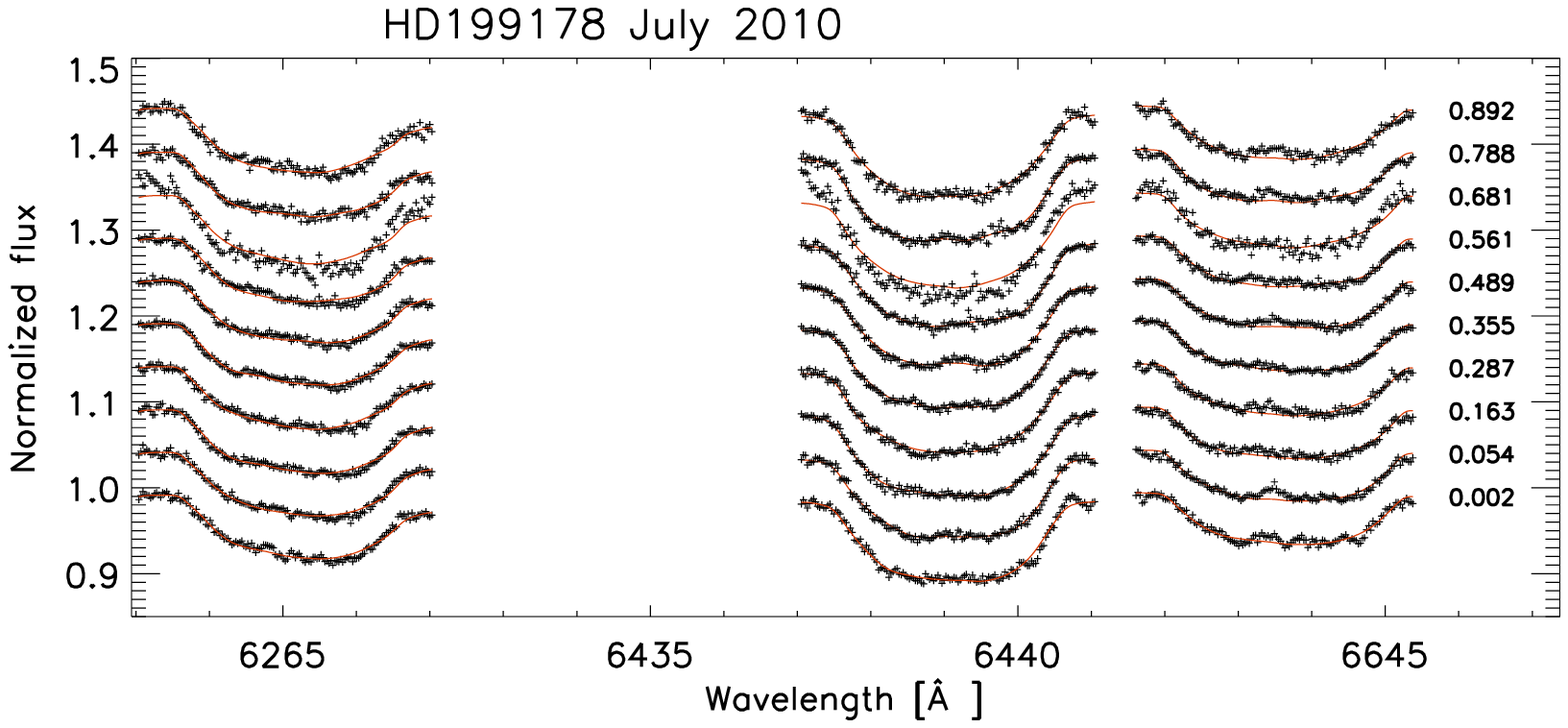}

\vspace{-1.5cm}

\includegraphics[width=6cm,clip]{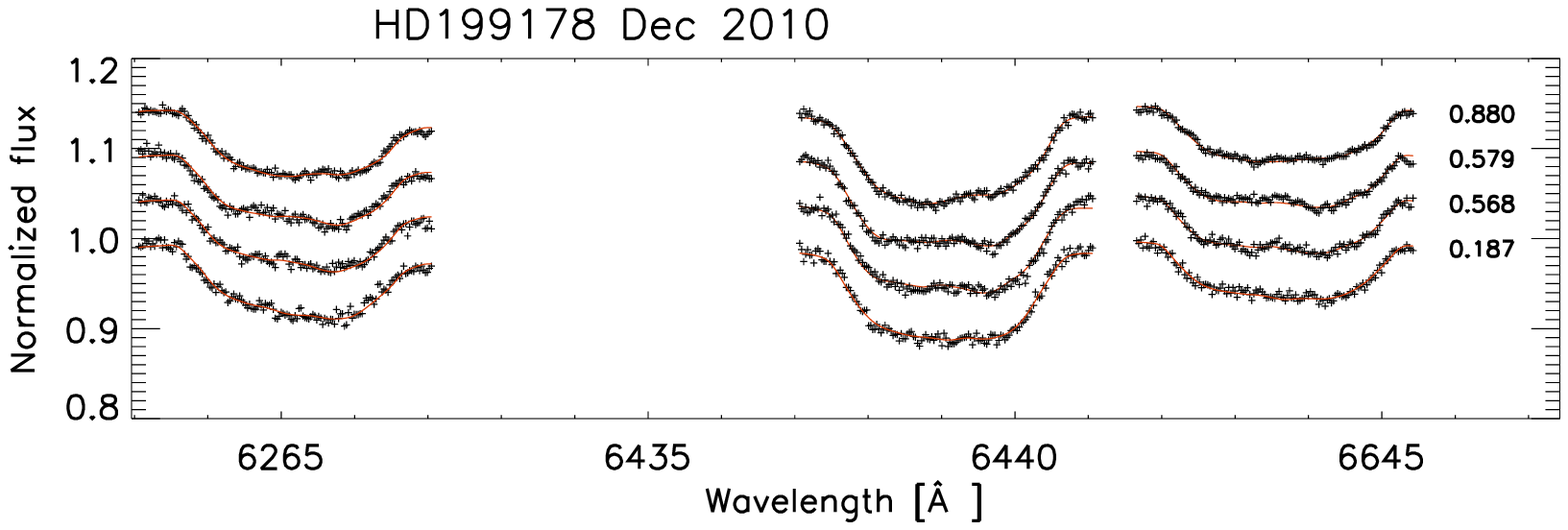}
\includegraphics[width=6cm,clip]{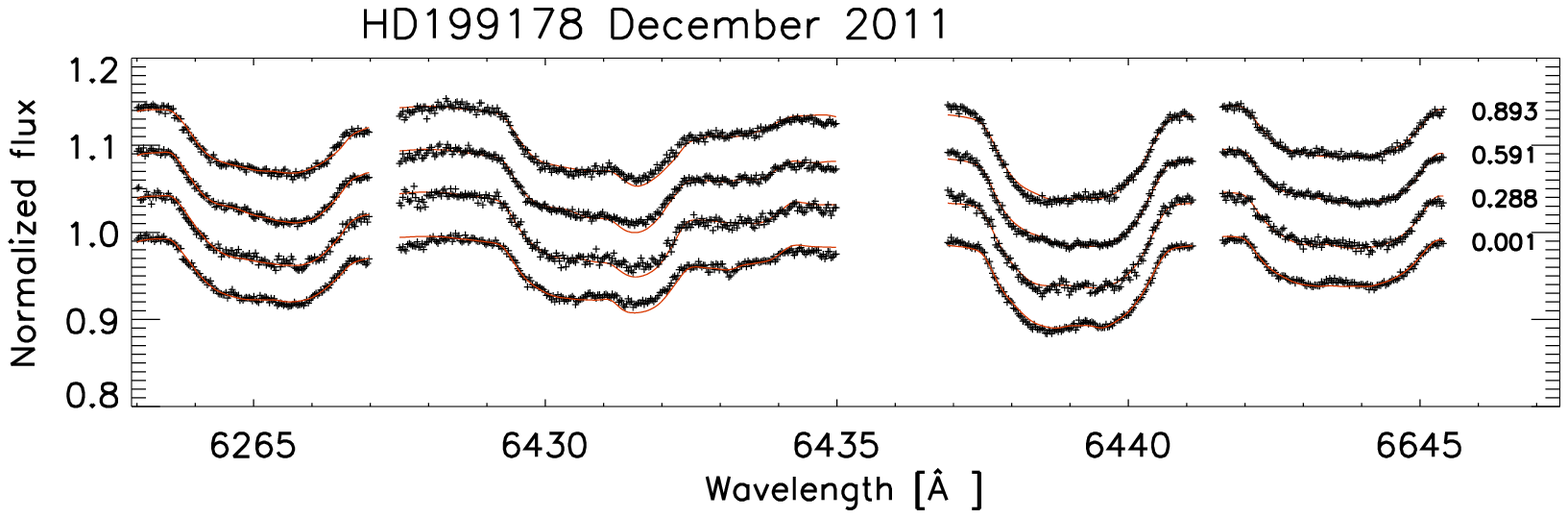}
\includegraphics[width=6cm,clip]{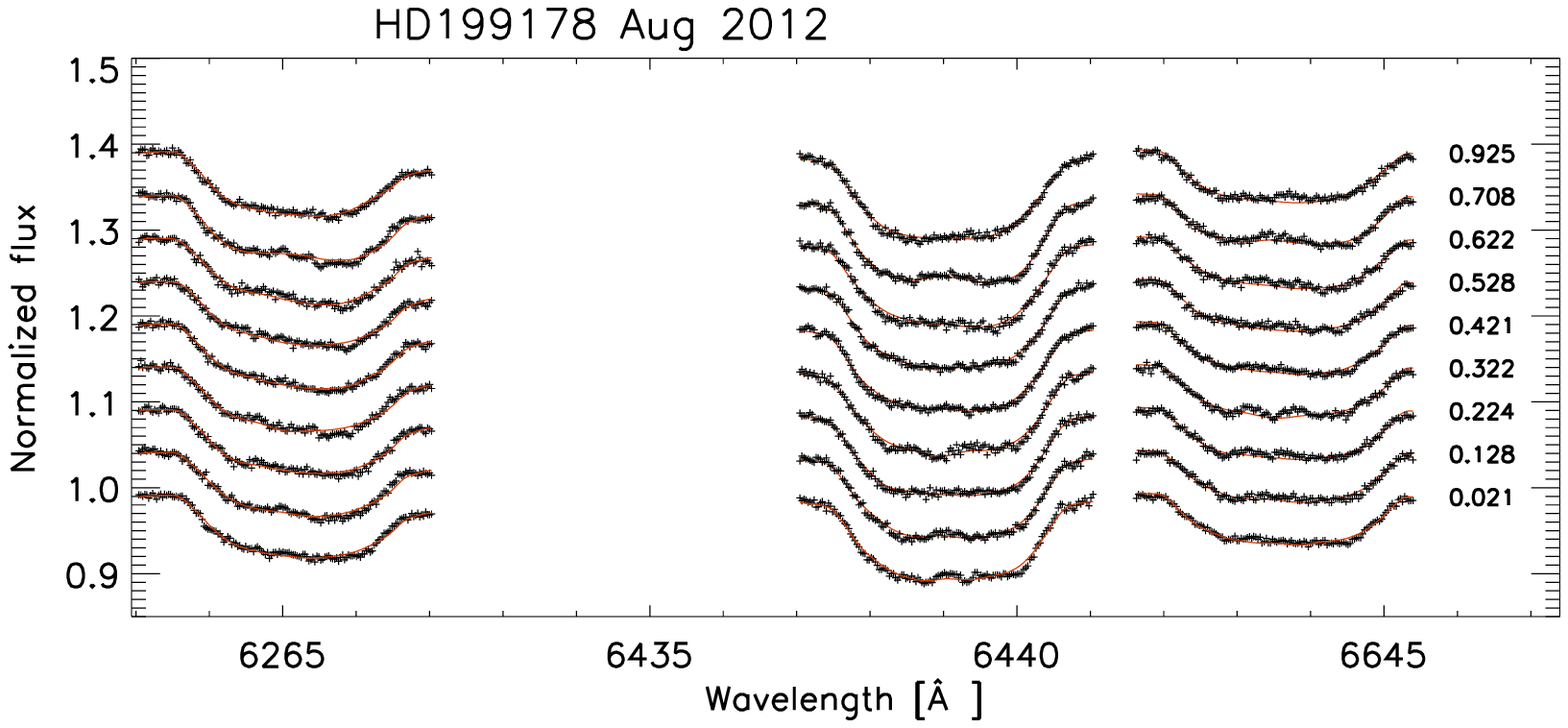}

\vspace{-1.5cm}

\includegraphics[width=6cm,clip]{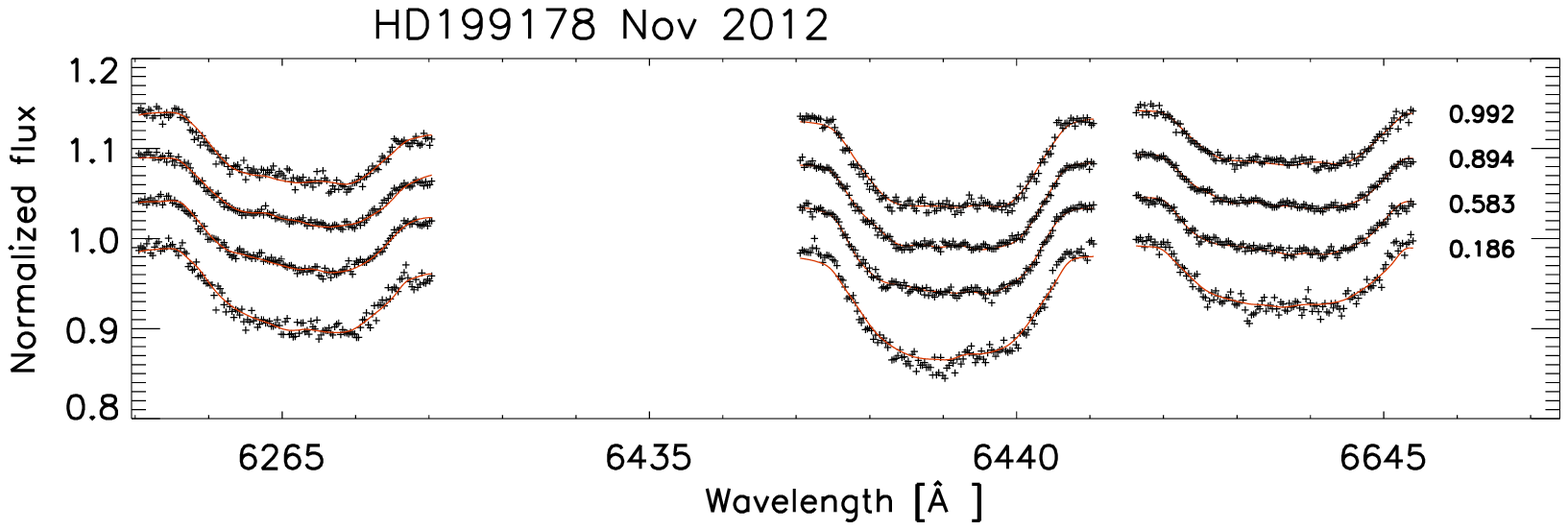}
\includegraphics[width=6cm,clip]{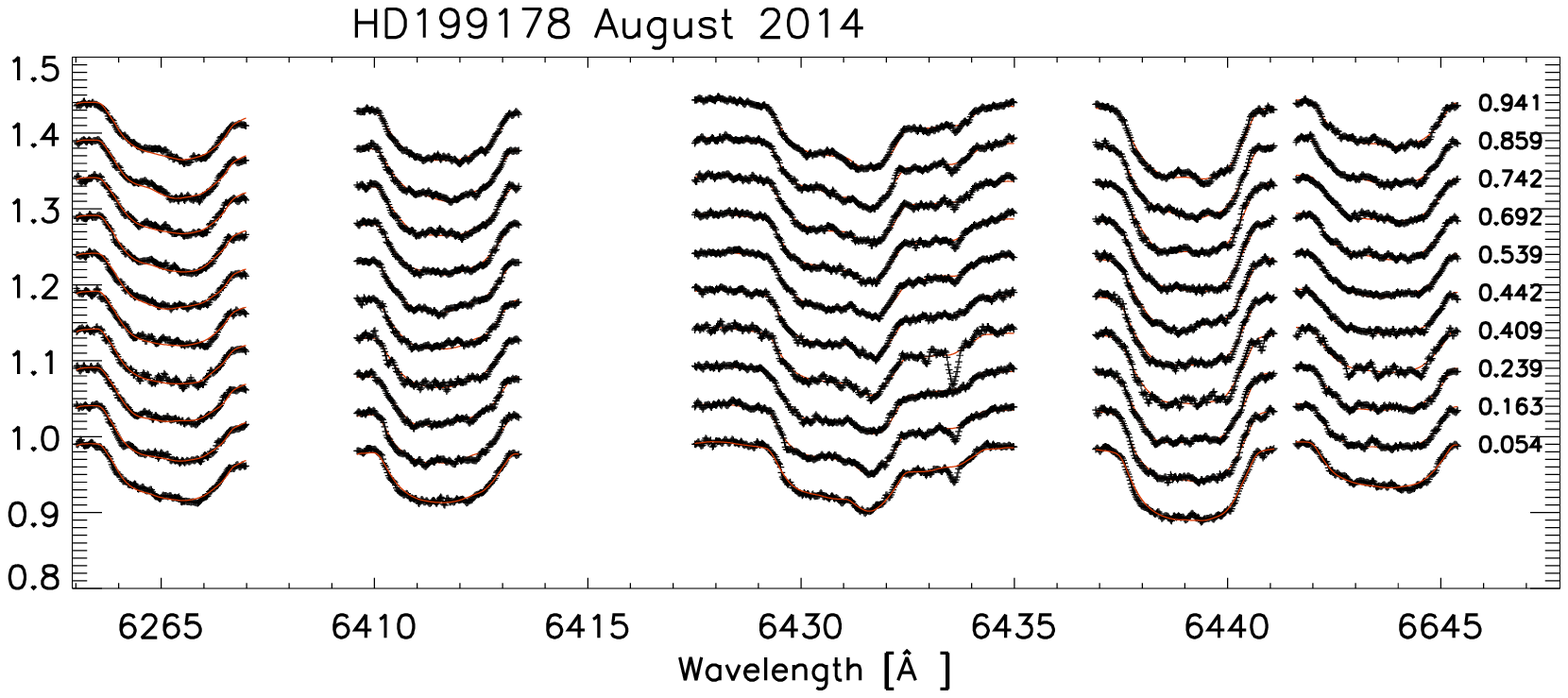}
\includegraphics[width=6cm,clip]{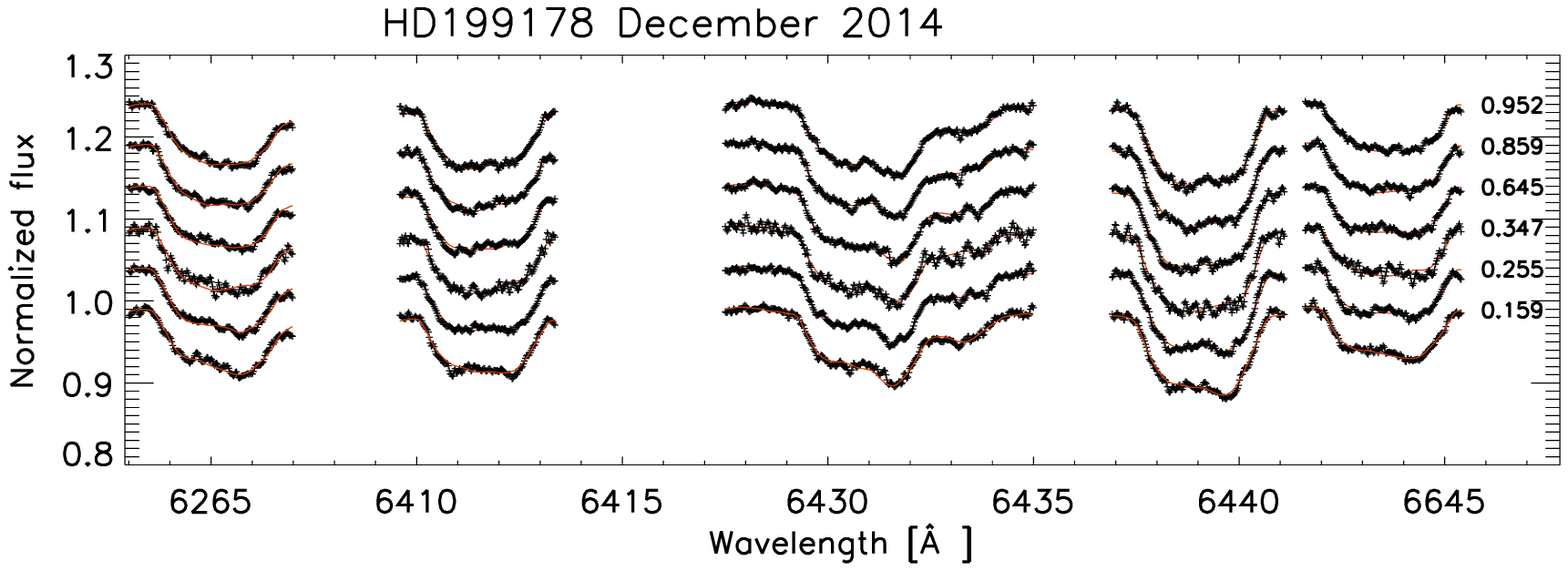}

\vspace{-1.5cm}

\includegraphics[width=6cm,clip]{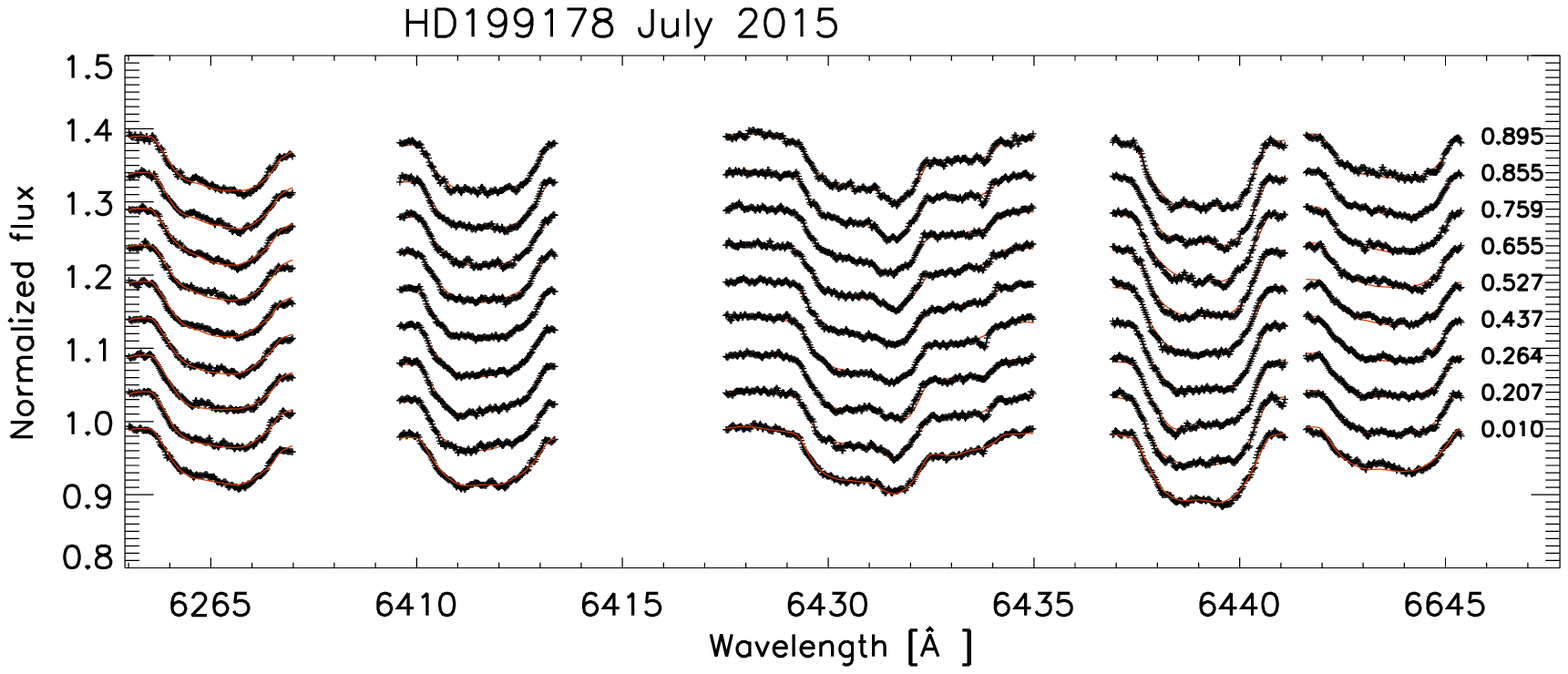}
\includegraphics[width=6cm,clip]{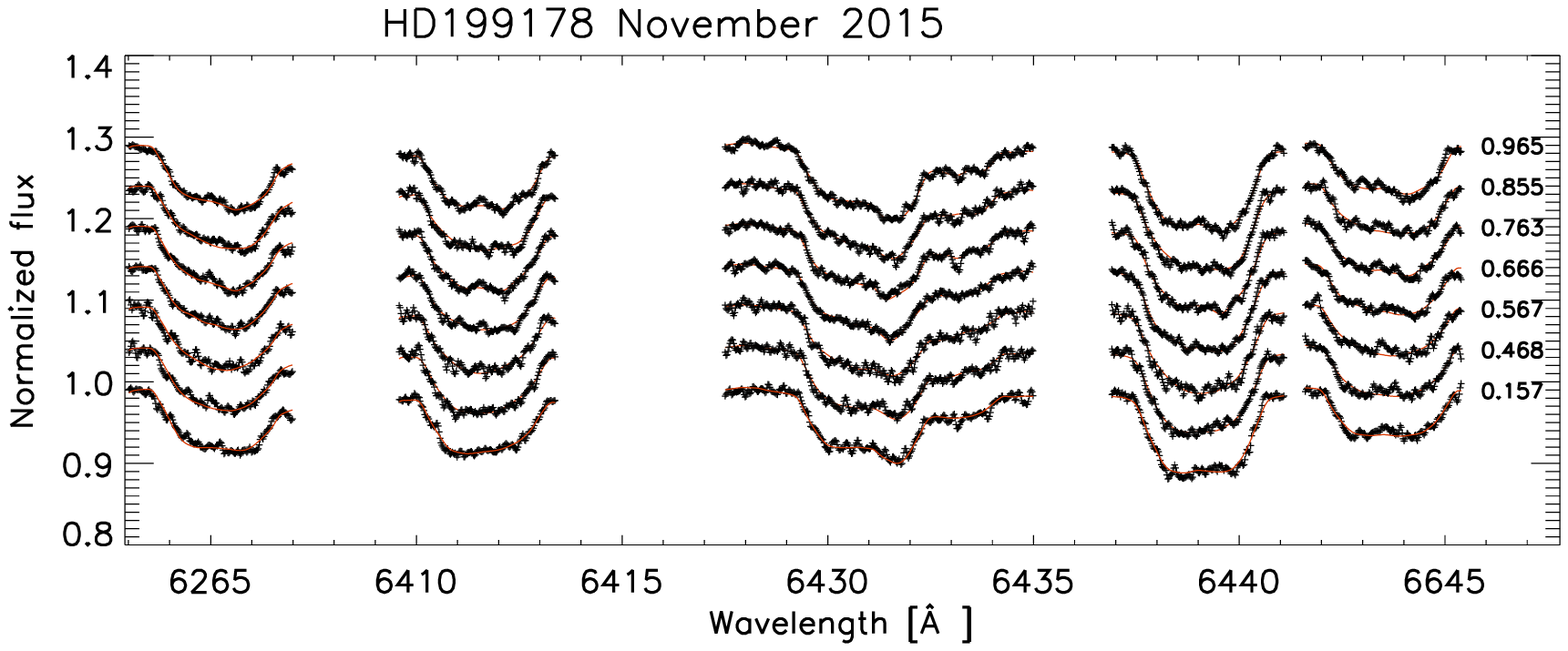}
\includegraphics[width=6cm,clip]{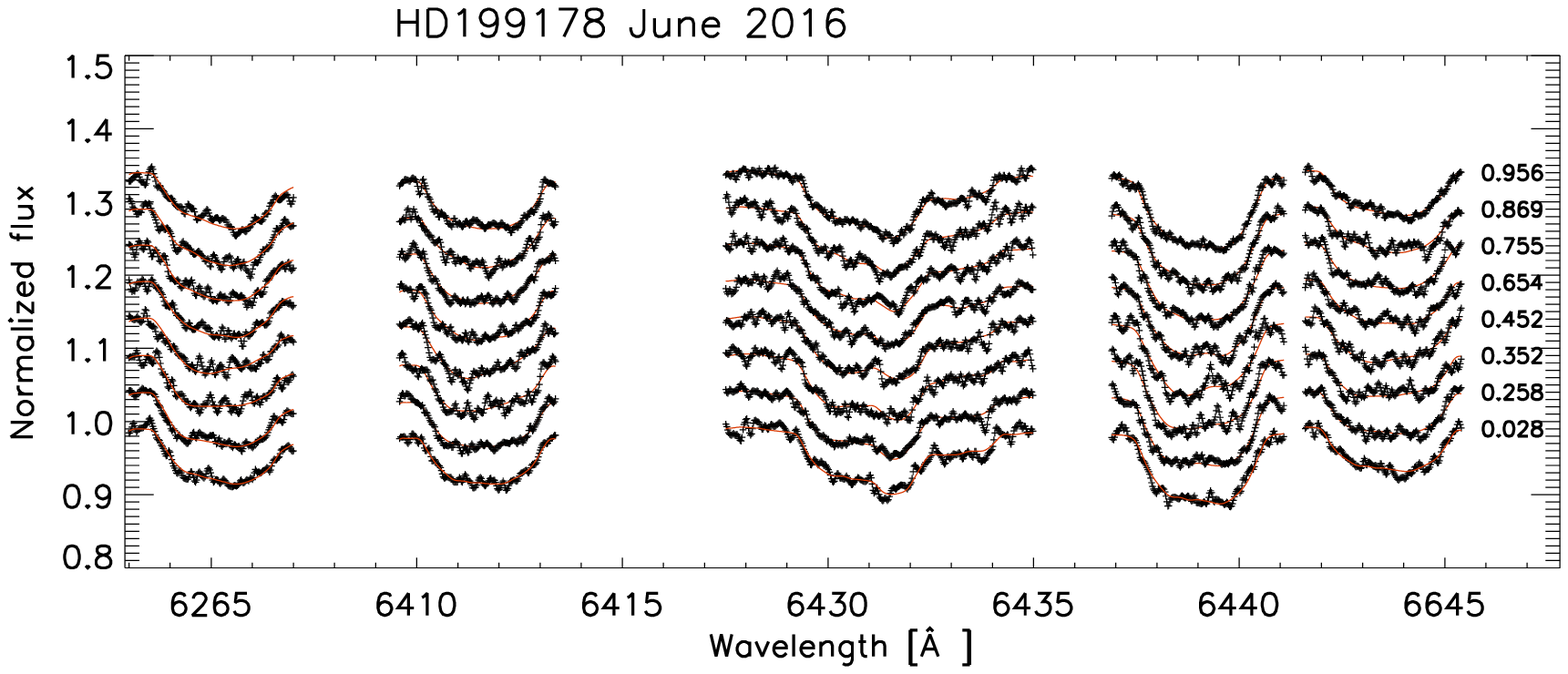}

\vspace{-1.5cm}

\includegraphics[width=6cm,clip]{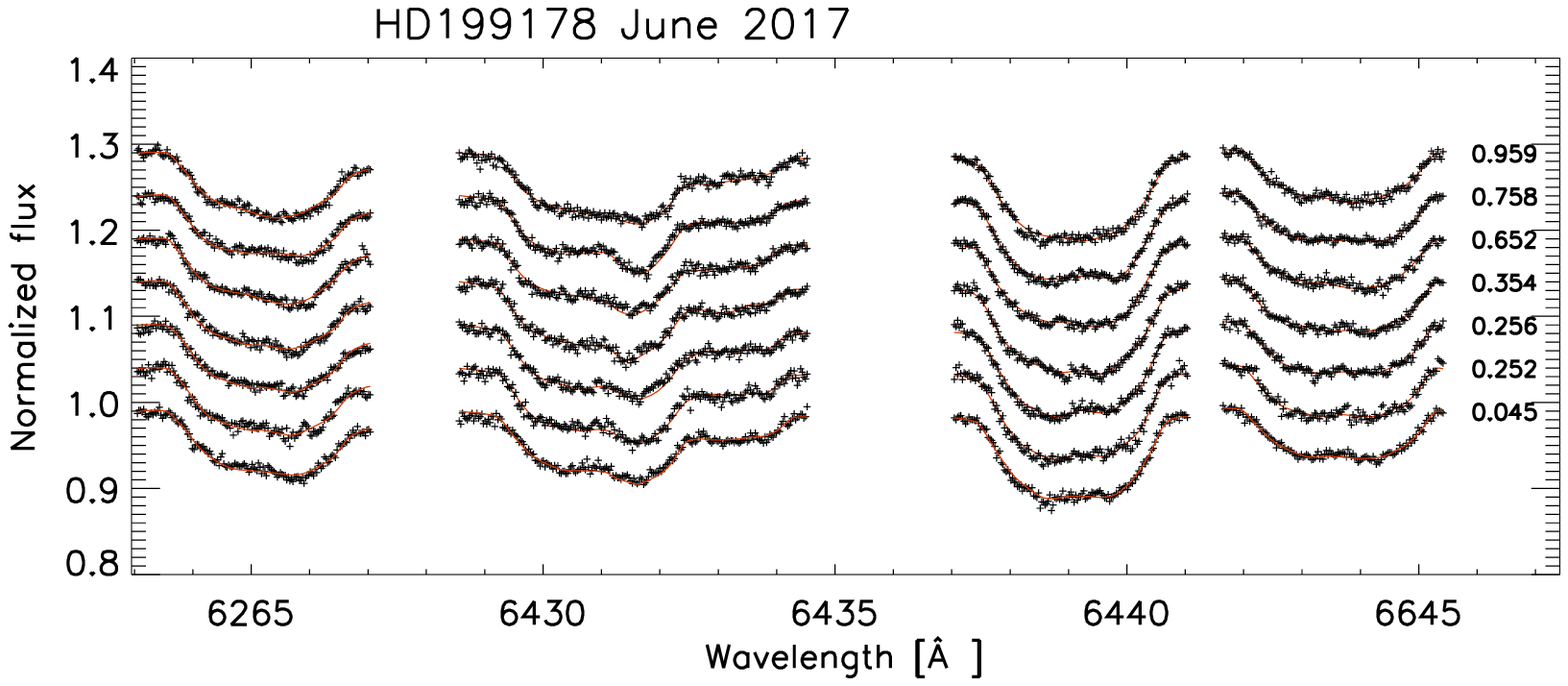}
\includegraphics[width=6cm,clip]{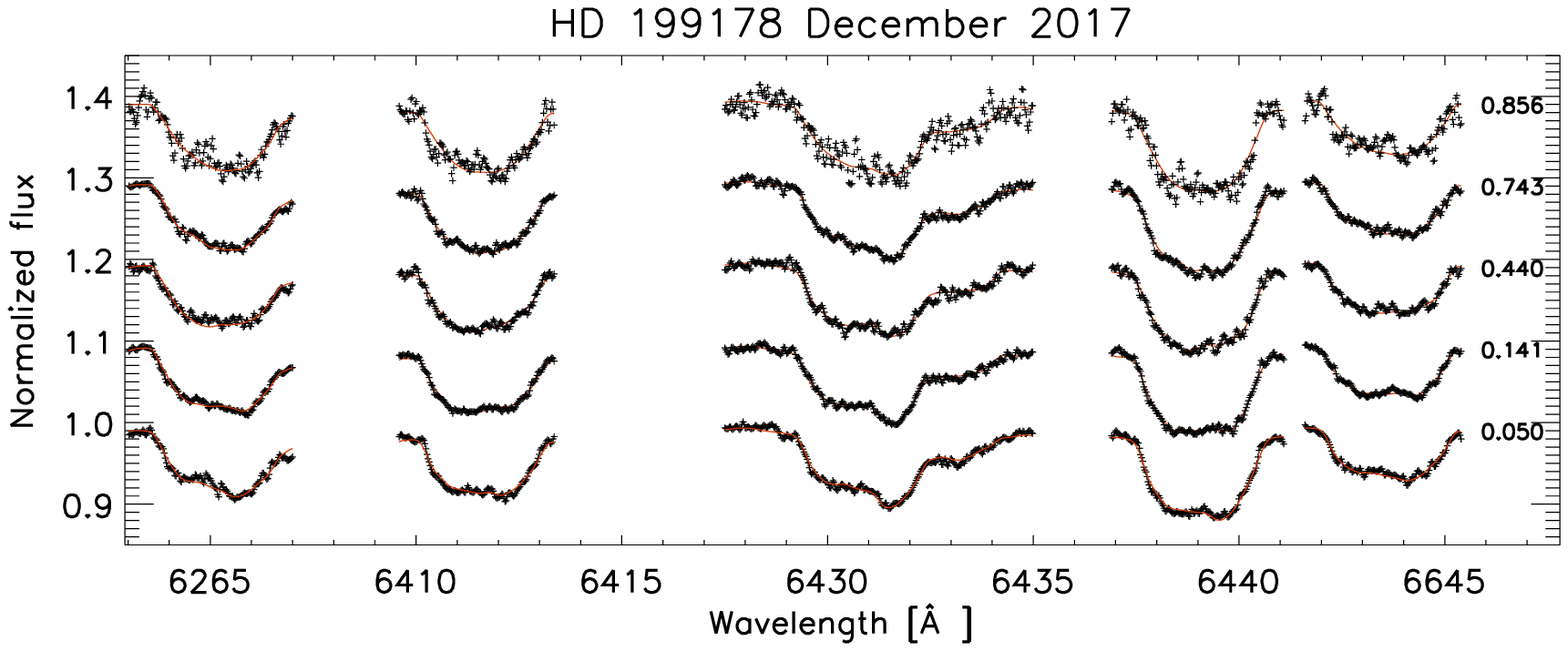}

\caption{Same as Fig. \ref{spec1}.1.}

\label{spec2}

\end{figure*}

\end{appendix}

\end{document}